\documentclass[longauth]{aa} 
\usepackage{graphicx}
\usepackage{array}

\usepackage{txfonts}
\usepackage{longtable}
\usepackage{lscape}
\usepackage{multirow}
\usepackage{bigdelim}
\usepackage{subfigure}

\usepackage{natbib}
\bibpunct{(}{)}{;}{a}{}{,} 

\begin{document}
   \title{The VLT-FLAMES Tarantula Survey\thanks {Based on observations collected at the European Southern Observatory under program ID 182.D-0222}}

   \subtitle{XI. A census of the hot luminous stars and their feedback in 30 Doradus\thanks{Complete versions of Tables \ref{tab:Candidate stars} \& \ref{tab:Feedback from stars} are available in electronic form at the CDS via anonymous ftp to cdsarc.u-strasbg.fr (130.79.128.5) or at ftp://astro.shef.ac.uk/pub/eid}}

   \author{E. I. Doran
          \inst{1}
          \and
          P. A. Crowther\inst{1}
	\and
	A. de Koter\inst{2,3,4}
	\and
	C. J. Evans\inst{5}
	\and
	C. McEvoy\inst{6}
 	\and
	N. R. Walborn\inst{7}
	\and
	N. Bastian\inst{8}
	\and
	J. M. Bestenlehner\inst{9}
	\and
	G. Gr\"{a}fener\inst{9}
	\and
	A. Herrero\inst{10,11}
	\and
	K. K\"{o}hler\inst{10}
	\and
	J. Ma\'{\i}z Apell\'aniz\inst{13}
	\and
	F. Najarro\inst{14}
	\and
	J. Puls\inst{15}
	\and
	H. Sana\inst{2}
	\and	
	F. R. N. Schneider\inst{12}
	\and
	W. D. Taylor\inst{5}
	\and
	J. Th. van Loon\inst{16}
	\and
	J. S. Vink\inst{9}
          }

   \institute{Department of Physics and Astronomy, University of Sheffield, 
Sheffield S3 7RH, UK\\
              \email{e.doran@sheffield.ac.uk}
	\and
		Astronomical Institute Anton Pannekoek, Amsterdam University, Science Park 904, 1098~XH, Amsterdam, The Netherlands
	\and
		Instituut voor Sterrenkunde, Universiteit Leuven, Celestijnenlaan 200 D, B-3001 Leuven, Belgium
	\and
		Utrecht University, Princetonplein 5, NL-3584 CC Utrecht, The Netherlands
	\and
		UK Astronomy Technology Centre, Royal Observatory Edinburgh, Blackford Hill, Edinburgh, EH9 3HJ, UK
	\and
		Department of Physics \& Astronomy, Queen’s University Belfast, Belfast BT7 1NN, Northern Ireland, UK
	\and
		Space Telescope Science Institute, 3700 San Martin Drive, Baltimore, MD 21218, USA
	\and
		Astrophysics Research Institute, Liverpool John Moores University, Egerton Wharf, Birkenhead, CH41 1LD, UK
	\and
		Armagh Observatory, College Hill, Armagh, BT61 9DG, Northern Ireland, UK
	\and
		Departamento de Astrof\'{\i}sica, Universidad de La Laguna, E-38205 La Laguna, Tenerife, Spain
	\and
		Instituto de Astrof\'isica de Canarias, E-38200 La Laguna, Tenerife, Spain 
	\and
		Argelander-Institut f\"{u}r Astronomie der Universit\"{a}t Bonn, Auf dem H\"{u}gel 71, 53121 Bonn, Germany
	\and
	  	Instituto de Astrof\'{\i}sica de Andaluc\'{\i}a-CSIC, Glorieta de la Astronom\'{\i}a s/n, E-18008 Granada, Spain
	\and
		Centro de Astrobiolog\'{\i}a (CSIC-INTA), Ctra. de Torrej\'on a Ajalvir km-4, E-28850 Torrej\'on de Ardoz, Madrid, Spain
	\and
		Universit\"{a}ts-Sternwarte, Scheinerstrasse 1, 81679 Munchen, Germany
	\and
		Astrophysics Group, School of Physical \& Geographical Sciences, Keele University, Staffordshire, ST5 5BG, UK
	}

   \date{Accepted: 5 August 2013}

 
  \abstract
   {}
   {We compile the first comprehensive census of hot luminous stars in the 30 Doradus (30 Dor) star forming region of the Large Magellanic Cloud. We investigate the stellar content and spectroscopic completeness of early type stars within a 10arcmin (150pc) radius of the central cluster, R136. Estimates were made of the integrated ionising luminosity and stellar wind luminosity. These values were used to re-assess the star formation rate (SFR) of the region and determine the ionising photon escape fraction.}
   {Stars were selected photometrically and combined with the latest spectral classifications. Stellar calibrations and models were applied to obtain physical parameters and wind properties. Their integrated properties were then compared to global observations from ultra-violet (UV) to far-infrared (FIR) imaging as well as the population synthesis code, \emph{Starburst99}.}
   {The census identified 1145 candidate hot luminous stars of which $>700$ were considered genuine early type stars that contribute to feedback. We assess the spectroscopic completeness to reach 85\% in outer regions ($>5$\,pc) but fall to 35\% in the vicinity of R136, giving a total of 500 hot luminous stars with spectroscopy. Only 31 W-R and Of/WN stars were included, but their contribution to the integrated ionising luminosity and wind luminosity was $\sim40$\% and $\sim50$\%, respectively. Similarly, stars with $M_{\rm{init}}>100\,\mathrm{M}_{\odot}$ (mostly H-rich WN stars) also showed high contributions to the global feedback, $\sim25$\% in both cases. Such massive stars are not accounted for by the current \emph{Starburst99} code, which was found to underestimate the integrated ionising luminosity and wind luminosity of R136 by a factor $\sim2$ and $\sim9$, respectively. The census inferred a SFR for 30 Dor of $0.073\pm0.04$\,M$_{\odot}$\,yr$^{-1}$. This was generally higher than that obtained from some popular SFR calibrations but still showed good consistency with the far-UV luminosity tracer and the combined H$\alpha$ and mid-infrared tracer, but only after correcting for H$\alpha$ extinction. The global ionising output was also found to exceed that measured from the associated gas and dust, suggesting that $\sim6^{+55}_{-6}$\% of the ionising photons escape the region.}
   {When studying the most luminous star forming regions, it is essential to include their most massive stars if one is to determine a reliable energy budget. The large ionising outputs of these stars increase the possibility of photon leakage. If 30 Dor is typical of other large star forming regions, estimates of the SFR will be underpredicted if this escape fraction is not accounted for.}


\titlerunning{The VLT-FLAMES Tarantula Survey XI. A census of the hot luminous stars in 30 Doradus} \authorrunning{E. I. Doran et al.}

   \maketitle
%

\section{Introduction}
\label{sec:Introduction}
The Tarantula Nebula (NGC 2070, 30 Doradus - hereafter ``30 Dor'') in the Large Magellanic Cloud (LMC) is optically the brightest H\,{\sc ii} region in the Local Group \citep{Kennicutt:1984}. It has been subject to numerous ground-based photometric and spectroscopic studies, displaying an abundance of hot massive stars \citep{Melnick:1985, SchildTestor:1992, Parker:1993, WalbornBlades:1997}. Hubble Space Telescope observations later revealed their high concentration in the more crowded central cluster, R136 \citep{Hunter:1995, MasseyHunter:1998}, also believed to be the home of stars with masses $>150\,\mathrm{M}_{\odot}$ \citep{Crowther:2010}. Roughly a third of the LMC Wolf-Rayet (W-R) stars compiled in \citet{Breysacher:1999} are contained within the 30 Dor nebular region of N157 \citep{Henize:1956}, with several found within R136 itself. The stellar population of 30 Dor spans multiple ages; from areas of ongoing star formation \citep{Brandner:2001, Walborn:2013} to older ($\sim20-25$\,Myr) supergiants in the Hodge 301 cluster \citep{GrebelChu:2000}.

More recently, the VLT-Flames Tarantula Survey (VFTS, see \citealt{Evans:2011}, hereafter, Paper I) has provided unprecedented multi-epoch spectroscopic coverage of the massive star population. The VFTS greatly extends our sampling of massive stars at different luminosities and temperatures, which are unobtainable from distant galaxies. In particular, it allows several topics to be assessed including: stellar multiplicity \citep{Sana:2013a}; rotational velocity distributions (\citealt{Dufton:2013} \& Ram\'{\i}rez-Agudelo et al. in prep); the dynamics of the region and R136 \citep{Henault-Brunet:2012b}. In this paper we look at 30 Dor in a broader context. Prior to the VFTS, $<20$\% of massive stars within the central $10\,\mathrm{arcmin}\approx150$\,pc (adopting a distance modulus to the LMC of $18.49\pm0.047$\,mag i.e. a distance of $\approx49.97\pm1.11$\,kpc, \citealt{Pietrzynski:2013}) had a known spectral type (SpT). As we will show, this spectroscopic completeness is improved to $\sim85$\% by the VFTS, enabling us to compile the first robust census of the hot luminous stars in 30 Dor.

Its close proximity and low foreground extinction, combined with its rich source of hot luminous stars, make 30 Dor the ideal laboratory for studying starburst environments in the distant universe. In more distant cases, however, stars cannot be individually resolved and starburst regions can only be studied via their integrated properties. This census provides us with a spectral inventory for an archetypal starburst, along with estimates of the stellar feedback. Here, we focus on radiative and mechanical feedback, to which hot luminous stars greatly contribute, especially prior to core-collapse supernova explosions. Their high luminosities and temperatures provide a plethora of extreme UV (EUV) photons ($Q_{\mathrm{0}}$), that ionise the surrounding gas to produce H\,{\sc ii} regions. Meanwhile, their strong stellar winds lead to a high wind luminosity ($L_{\mathrm{SW}}$), capable of sweeping up the ISM into bubbles, even prior to the first supernovae \citep{ChuKennicutt:1994}. R136 itself, appears to show evidence of such effects through an expanding shell \citep[e.g.][]{vanLoon:2013}.

Individual analyses are currently underway of all the VFTS early type stars, allowing determination of their parameters, along with their feedback. Here, we employ various stellar calibrations to estimate the integrated properties for the entire 30 Dor region. This allows for comparisons to the global properties of the nebula. The integrated ionising luminosity is typically obtained by observing the hydrogen recombination lines (such as H$\alpha$) or the radio continuum flux. However, not all ionising photons will necessarily ionise the surrounding gas since they may be absorbed by dust or even escape the region altogether. The star formation rate (SFR) can be determined for such regions using the inferred $Q_{\mathrm{0}}$, but an accurate SFR relies on the knowledge of the fraction of escaping ionising photons.

A similar approach to the present study has been carried out for the Carina Nebula. \citet{Smith:2006} and \citet{SmithBrooks:2007} found the integrated radio continuum flux to provide $\sim75$\% of the $Q_{\mathrm{0}}$ determined from the stellar content, suggesting a significant fraction of the photons were escaping, or were absorbed by dust. Their $Q_{\mathrm{0}}$ determined from the H$\alpha$ luminosity was even smaller, only a third of the stellar output. This was thought to arise from a large non-uniform dust extinction that had been unaccounted for. Lower luminosity H\,{\sc ii} regions have also been studied in the LMC \citep{OeyKennicutt:1997}. Initial comparisons to the H$\alpha$ luminosity suggested 0-51\% of ionising photons escape their regions and could potentially go on to ionise the diffuse warm ionised medium (WIM). However, these H\,{\sc ii} regions were later revisited by \citet{Voges:2008} with updated atmospheric models, indicating a reduced photon escape fraction, with radiation being bound to the regions in all but 20-30\% of cases.

In previous studies, the feedback of up to a few dozen stars are accounted for (70 in the Carina Nebula). 30 Dor is powered by hundreds of massive stars of various spectral types (SpTs) and the census gives an insight into their relative importance. \citet{Parker:1998} already used UV imaging to show how a few of the brightest stars make a significant contribution to the $Q_{\mathrm{0}}$ in the region. \citet{Indebetouw:2009} and \citet{Pellegrini:2010} used infrared (IR) and optical nebular emission lines, respectively, to study the ionised gas itself, with both favouring a photoionisation dominated mechanism in 30 Dor. R136 appeared to play a prominent role as nearly all bright pillars and ionisation fronts were observed to lead back to the central cluster \citep{Pellegrini:2010}. \citet{Indebetouw:2009} also noted regions of more intense and harder radiation which typically coincided with hot isolated W-R stars. \citet{OeyKennicutt:1997} and \citet{Voges:2008} omit W-R stars when studying their H\,{\sc ii} regions, in view of uncertainties in their $Q_{\mathrm{0}}$. However, earlier work by \citet{CrowtherDessart:1998} compiled a list of the hot luminous stars in the inner 10\,pc of 30 Dor and found the W-R stars to play a key role in the overall feedback. We extend their study to a larger radius, employing updated stellar calibrations and models to determine whether the W-R star contribution still remains significant.

Population synthesis codes can predict observable properties of extragalactic starbursts, and 30 Dor equally serves as a test bed for these. Direct comparisons to the stellar content and feedback in similar luminous star forming regions is relatively unstudied. The R136 cluster is believed to be at a pre-supernova age and so has largely preserved its initial mass function (IMF), although mass-loss and binary effects may have affected this to some extent. Its high mass ($M_{\rm{cl}}\lesssim5.5\times10^{4}\,\mathrm{M}_{\odot}$, \citealt{Hunter:1995}) also ensures that its upper mass function (MF) is sufficiently well populated, hence ideal to test synthetic predictions of the stellar feedback. Comparisons to the entire 30 Dor region are less straight forward given its non-coevality, but can still be made by adopting an average age for 30 Dor.

A breakdown of the present paper is as follows. In Section \ref{sec:Photometry}, we present the photometric catalogues used in the census, their magnitude and spatial limits. Section \ref{sec:Candidate selection} sets out the different criteria used for selecting the hot luminous stars from the photometric data. In Section \ref{sec:Spectroscopy}, we match our selected hot luminous stars with any available spectral classifications. We examine the spectroscopic completeness of the census in Section \ref{sec:Completeness}. Section \ref{sec:Stellar Calibrations} discusses the different calibrations assigned to each SpT. Section \ref{sec:Stellar Census} brings together all of the stars in the census, considering their age and mass while Section \ref{sec:Integrated Stellar Feedback} focuses on their integrated feedback and the different contributions from both the central R136 cluster and 30 Dor as a whole. Section \ref{sec:Star Formation Rates} discusses the SFR of 30 Dor and its potential photon escape fraction. Section \ref{sec:Summary} discusses the impact of our results and summarises the main findings of the census.

\section{Photometry}
\label{sec:Photometry}
Our aim was to produce as complete a census as possible of the population of hot luminous stars in the 30 Doradus region. The first step began by obtaining a photometric list of every star in the region, from which our potential hot luminous candidates could be selected. As in Paper I, the overall list had four primary photometric catalogues: Selman, WFI, Parker and CTIO, outlined below.

The `Selman' photometry \citep{Selman:1999}, with Brian Skiff's reworked astrometry\footnote{ftp://cdsarc.u-strasbg.fr/pub/cats/J/A+A/341/98/}, used observations from the Superb-Seeing Imager (SUSI) on the 3.5\,m New Technology Telescope (NTT) at La Silla. Data covered the central 90\,arcsec of 30 Dor in the $UBV$ bands. The completeness limit was $V=19.2$\,mag although sources reached fainter, with typical photometric errors spanning 0.005-0.05\,mag.

The `WFI' photometry was the main source of photometry used for the census, based on the same observations outlined in Section 2.1 of Paper I. $B$- and $V$-band photometry was obtained with the Wide-Field Imager (WFI) at the 2.2\,m Max-Planck-Gesellschaft (MPG)/ESO telescope at La Silla. Data covered $14<V<19$\,mag with photometric errors between 0.002-0.020\,mag although once this was bootstrapped to the \citet{Selman:1999} catalogue, the scatter showed standard deviations of $<0.1$\,mag (see also Section 3.2 of Paper I). WFI sampled the outer sources of 30 Dor, extending at least 12\,arcmin from the centre although the R136 cluster was largely omitted due to saturation.

The `Parker' photometry \citep{Parker:1993}, with Brian Skiff's reworked astrometry, used observations from RCA \#4 CCD on the 0.9\,m telescope at Cerro Tololo Inter-American Observatory (CTIO). The catalogue offered $UBV$ band photometry for a majority of sources and just $BV$ band in other cases. Parker sources predominantly spanned the inner 2\,arcmin with additional coverage of regions north and east of R136. Data reached $B=V=18$\,mag and $U=17$\,mag with average photometric errors from 0.01-0.1\,mag. However, we note that subsequent Hubble Space Telescope/Wide Field Planetary Camera (HST/WFPC2) data showed an incompleteness in the Parker catalogue, revealing some sources to be unaccounted for and others to be spurious (see Footnote 6 of \citealt{Rubio:1998}).

The `CTIO' photometry comes from Y4KCAM camera observations on the 1\,m CTIO telescope, outlined in Section 3.5 of Paper I. It was complete out to a radius of $\approx7.5$\,arcmin and was required to supply photometry for the brighter sources, not covered by the WFI data. The CTIO data reached $V<17.25$\,mag with photometric errors between $\sim0.02-0.1$\,mag. As mentioned in Paper I, the CTIO photometry was not transformed to the exact same system as the WFI photometry but the two remain in reasonable agreement: $\Delta V$ and $\Delta B$ $\leq0.5$\,mag.

\begin{figure*}
\includegraphics[width=\textwidth]{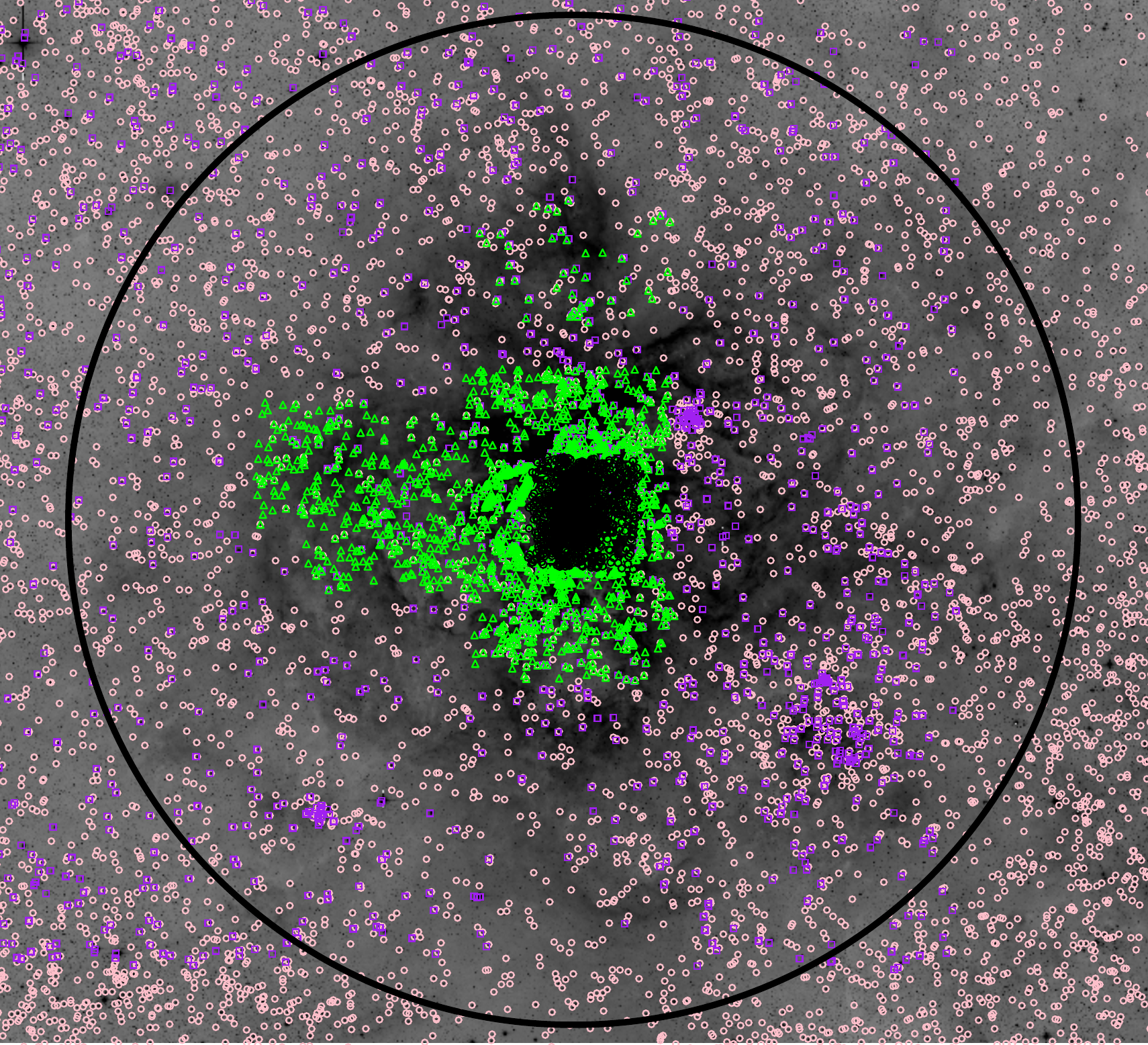}
\caption{The spatial coverage of photometric sources from the four main catalogues used in the census, overlaid on a $V$-band WFI image. Selman (black circles), WFI (pink circles), Parker (green triangles) and CTIO (purple squares). The large black circle marks the $r_{\mathrm{d}}=10$\,arcmin region of the census. North is up and East is to the left.}
\label{fig:Catalogue_spatial_plot}
\end{figure*}

The spatial coverage of these four photometric catalogues is shown in Figure \ref{fig:Catalogue_spatial_plot}. The census itself was chosen to extend out to a radial distance of $r_{\mathrm{d}}=10$\,arcmin (150\,pc) from the centre of R136 (specifically star R136a1, $\alpha=05^{h}38^{m} 42^{s}.39$, $\delta=-69^{\circ}06'02.91''$). The 10\,arcmin radius was selected as it was consistent with the spatial extent of the VFTS. From here onwards, when discussing the census, the term ``30 Dor'' will refer to this $r_{\mathrm{d}}<10$\,arcmin region. However, given the spatial limit of the CTIO photometry ($r_{\mathrm{d}}<7.5$\,arcmin), the brightest targets in the far western and southern regions of 30 Dor had not been covered. To ensure complete coverage, the few remaining brighter objects were taken from the Magellanic Clouds Photometric Survey of \citet{Zaritsky:2004}.

Due to high crowding in the central regions, space-based photometry (and spectroscopy) was favoured for stars within $r_{\mathrm{d}}<20$\,arcsec (5\,pc). From here onwards, when discussing the census, the term ``R136 region'' will refer to this $r_{\mathrm{d}}<20$\,arcsec region. \citet{DeMarchi:2011} provided HST/Wide Field Camera 3 (HST/WFC3) photometry for the R136 region stars in the F336W, F438W and F555W bands. Similarly, HST/Wide Field Planetary Camera (HST/WFPC) photometry (F336W, F439W, F569W bands) from \citet{Walborn:1999} was used for stars within the dense cluster, Brey 73.

As some regions were covered by multiple catalogues, attempts were made to exclude any duplicates. Searches for sources within 0.5\,arcsec of each other were made. Any that were found were subject to the same selection priority as in Paper I, based on the quality of the photometry (Selman was the primary dataset in central regions (excl. the R136 region). If the source lay beyond the Selman region, WFI was used. In the case of brighter sources, Parker was used, otherwise CTIO was used in outer regions of 30 Dor). \citet{DeMarchi:2011} was favoured for all sources in the R136 region. Table \ref{tab:SpT sources} gives a breakdown of the number of candidates, and spectroscopically confirmed, hot luminous stars that were selected from each photometric catalogue.

We assumed that all W-R stars in the region had been previously identified through spectroscopy (although note the recent discovery of the WN\,5h star VFTS 682; \citealt{Bestenlehner:2011}) and while their photometry was not used for their selection (see Section \ref{sec:Candidate selection}), it would still be needed to determine stellar parameters. Issues can arise when using Johnson broad band filters with emission line stars as it can lead to a false reading of the stellar continuum and overestimate their magnitudes. A more reliable measurement is obtained through narrow band filters \citep{Smith:1968}. For all W-R stars, narrow band $b$ \& $v$ magnitudes were either adopted from past work or calculated through spectrophotometry (see Appendix \ref{sec:W-R stars} for more details).

\begin{table}
\begin{center}
\caption{The literature used to select the 1145 photometric candidates and classify the 500 spectroscopically confirmed hot luminous stars in the census. The number in brackets represents those located within the R136 region. Papers which form part of the VFTS series are identified with `(VFTS)'.}
\label{tab:SpT sources}
\begin{tabular} { l r  r } 
\hline\hline
Source of Photometry & Candidates & Confirmed (in R136)\\
\hline
\citet{DeMarchi:2011} & 212 & 70 (70) \\
\citet{Selman:1999} & 200 & 123 (2) \\
WFI (Paper I) & 518 & 196 (0) \\
\citet{Parker:1993} & 71 & 42 (0) \\
CTIO (Paper I) & 126 & 57 (0)\\
\citet{Walborn:1999} & 15 & 10 (0) \\
\citet{Zaritsky:2004} & 3 & 1 (0) \\
\hline
Source of Spectral Type &  & Confirmed (in R136)\\
\hline
\citet{SchildTestor:1992} & & 7 (0) \\
\citet{Parker:1993} & & 3 (0) \\
\citet{WalbornBlades:1997} &  & 8 (0) \\
\citet{MasseyHunter:1998} & & 38 (38) \\
\citet{CrowtherDessart:1998} & & 8 (8)\\
\citet{Bosch:1999} & & 24 (0) \\
\citet{Walborn:1999} & & 9 (0) \\
\citet{Breysacher:1999} & & 3 (0) \\
\citet{Evans:2011} / Paper I & (VFTS) & 16 (1) \\
\cite{Taylor:2011} & (VFTS) & 1 (0) \\
\citet{Dufton:2011} & (VFTS) & 1 (0) \\
\citet{CrowtherWalborn:2011} & & 7 (6) \\
\citet{Henault-Brunet:2012a} & (VFTS) & 15 (14) \\
\citetext{Walborn et al. in prep.} & (VFTS) & 295 (5) \\
\citetext{McEvoy et al. in prep.} & (VFTS) & 35 (0) \\
This work & (VFTS) & 30 (0) \\
\hline
\end{tabular}
\end{center}
\end{table}

\section{Candidate Selection}
\label{sec:Candidate selection}
Of the tens of thousands of photometric sources included in our $r_{\mathrm{d}}<10$\,arcmin spatial cut, we only seek to select the hottest and most luminous stars for the census, to estimate the stellar feedback. The hottest stars will produce the bulk of the ionising photons while the most luminous early type stars will have the strongest stellar winds, hence the largest wind luminosity. Therefore, from all the 30 Dor stars, we wish to account for all the W-R and O-type stars, along with the earliest B-type stars (given their large numbers and high $T_{\mathrm{eff}}$) and B-supergiants (given their strong winds). Various colour and magnitude cuts were applied to the photometric data to extract these stars without prior knowledge of their spectral classification. 

\begin{figure}
\includegraphics[width=0.5\textwidth]{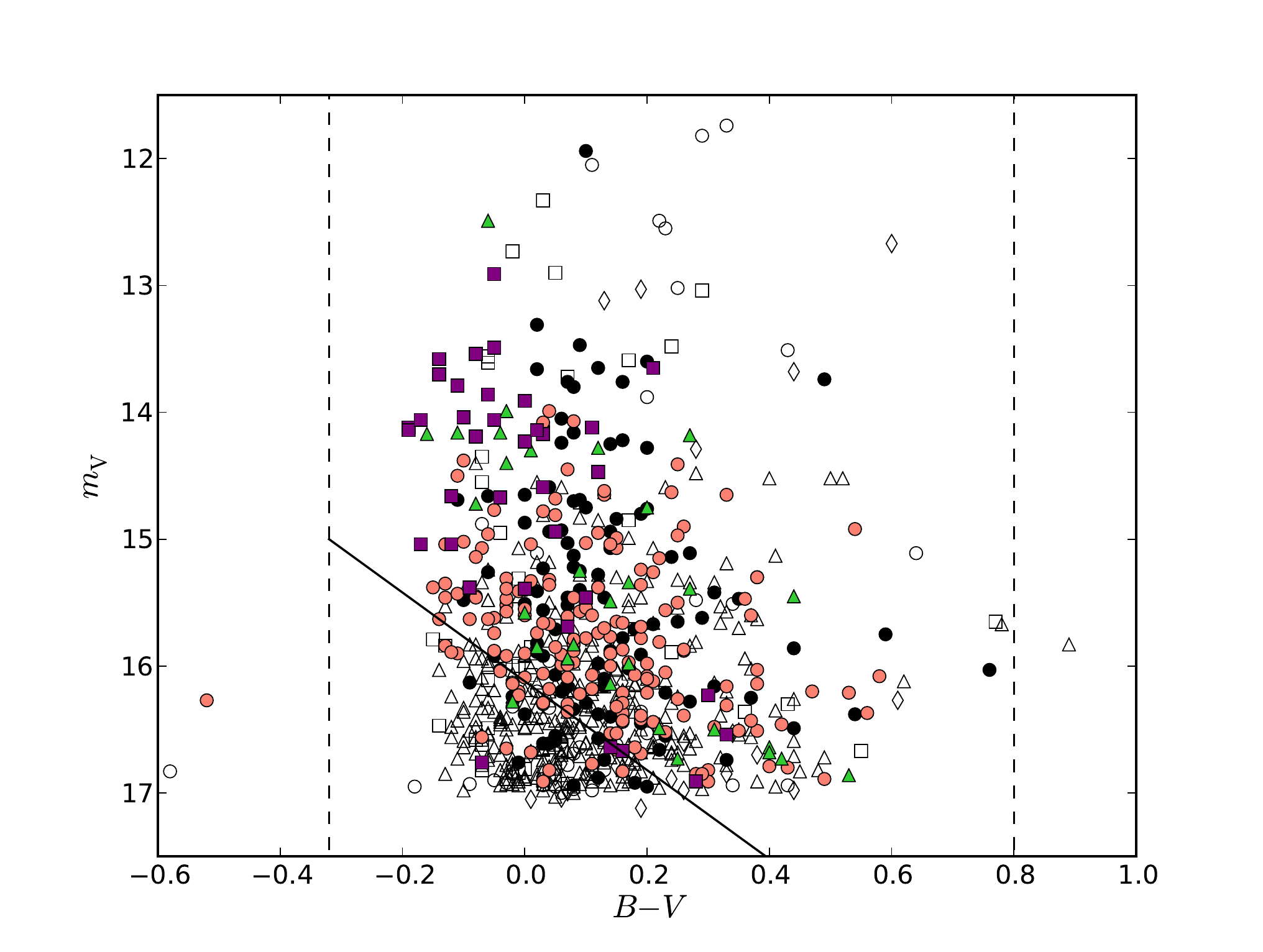}
\caption{A colour magnitude diagram of all the VFTS O-type stars (filled) and B-type stars (open). Photometry is from: Selman (black circles), WFI (pink circles), Parker (green triangles), CTIO (purple squares). Solid and dashed lines mark the $M_{V}$ and $B-V$ colour cuts made on the sample, respectively.}
\label{fig:VFTS_CMD}
\end{figure}

In order to determine the boundaries of these cuts, an initial test was carried out upon the VFTS sample for which we did have a known SpT. Figure \ref{fig:VFTS_CMD} shows a colour-magnitude diagram of all the VFTS O and B-type stars. The left vertical line aims to eliminate all stars with unreliable photometry, given that a typical unreddened O-type star is expected to have an observed colour of $B-V=-0.32$\,mag \citep{Fitzgerald:1970}. Stars that lie to the left of this line would therefore be `too blue'. Similarly, the right vertical line eliminates late-type stars; too cool to contribute to the overall feedback. However, the W-R, O and early B-type stars that we do seek, can still suffer from interstellar reddening which will shift them to the lower right hand side of the diagram. The boundary was therefore set to $B-V=0.8$\,mag to account for as many potentially highly reddened hot stars as possible, whilst keeping the number of unwanted, cooler stars to a minimum.

In an attempt to identify the most luminous stars, a further magnitude cut was made with the diagonal reddening line. This line has a gradient which is equivalent to $R_V=A_{V}/E(B-V)$ \citep{Cardelli:1989}\footnote{Note that the free parameter in the extinction laws of \citet{Cardelli:1989} or \citetext{Ma\'{\i}z Apell\'aniz et al. in prep.} should not strictly be called $R_V$ but $R_{5495}$ instead. The reason is that $R_V \equiv A_V/E(B-V)$ is a function not only of the type of extinction but also of the input spectral energy distribution and the amount of extinction (see Figure 3 of \citealt{MaizApellaniz:2013}). Nevertheless, for hot stars with low extinctions ($A_V < 4$\,mag), the approximation $R_V = R_{5495}$ holds reasonably well and will be used in this paper.}. For candidate selection, we adopt a uniform reddening value of $R_V=3.5$ although a value of $R_V=4.2$ was favoured for stars in the R136 region when determining their stellar parameters (see Appendix \ref{sec:Reddening_appendix} for details). Individual tailored analyses for the extinction of the VFTS O-type stars are currently being carried out \citetext{Ma\'{\i}z Apell\'aniz et al. in prep.}. Comparing these to the extinctions from our uniform $R_V$ value gave a mean difference of $\Delta A_V<0.15$\,mag. We therefore continued with our approach given this relatively small offset and the likely larger uncertainties that would arise in stellar parameters, owing to the use of spectroscopic calibrations.

The magnitude cut begins at $m_V= 15$\,mag and $B-V=-0.32$\,mag. This would be the position of the faintest stars we seek if they suffered from no reddening at all, corresponding to $M_V \approx -3.5$\,mag. A star suffering from interstellar reddening, however, is expected to be shifted along the diagonal line. A value of $R_V = 3.5$ ensured the inclusion of all the most luminous stars. While Figure \ref{fig:VFTS_CMD} shows that some VFTS O-type stars are omitted (those below the line), almost all were found to be O9 or later, and given their position on the diagram, would be subluminous and hence have negligible contributions to the feedback (see Sections \ref{sec:Ionising photon luminosity} \& \ref{sec:Stellar wind luminosity and momentum} for a quantitative discussion). Combining $R_V$ with our limit of $B-V=0.8$\,mag allows us to account for extinctions up to $A_V = 3.9$\,mag.

A further method was also applied to select our hot luminous stars involving the `$Q$-parameter' or reddening free index \citep{Aparicio:1998}. The $Q$-parameter incorporates $U$-band photometry and takes advantage of the fact that different SpTs may have similar $B-V$ colours but have different $U-B$ colours. After adjusting the relationship for 30 Dor using the reddening laws of \citet{Cardelli:1989}, the $Q$-parameter took the form of $Q = (U-B) - 0.67 \times (B-V)$. A limit of $Q<-0.65$ was selected to filter out any mid-late B-type dwarfs and later SpT stars. This removed a further 26 stars from our candidate list and would ideally have been the best selection criterion to use. Unfortunately reliable $U$-band photometry was only available for a small subset of the sources. Furthermore, as some spurious detections had been noted in the \citet{Parker:1993} and CTIO catalogues, a further visual inspection was made of all of their sources that survived the selection cuts, to ensure that they were, indeed, true stellar sources.

An exception was made when selecting W-R and Of/WN stars. Their notably high contribution to the stellar feedback meant that all needed to be accounted for. Some W-R stars were rejected by the selection criteria mentioned, due to the unusual colours that can be brought about by their broad emission lines. For this reason, all of the W-R and Of/WN stars given in Table 2 of Paper I were manually entered into the census along with any further known W-R stars listed by \citet{Breysacher:1999} in our selected region.

A total of 1145 candidate hot luminous stars (see Table \ref{tab:Candidate stars}, Appendix \ref{sec:The Census}) were finally selected from the photometry via the criteria discussed above.

\section{Spectroscopy}
\label{sec:Spectroscopy}
Having selected our candidate hot luminous stars, the aim was now to match as many of these as possible to spectral classifications. The best and most extensive stellar spectroscopy of 30 Dor is offered by the VFTS (Paper I). So far, all of the $\sim360$ VFTS O-type stars have been classified including any binary companions where possible \citetext{Walborn et al. in prep.}. Classification of VFTS B-type stars is currently underway. B-dwarfs and giants included in the census were classified in this work while the identified B-supergiants will become available in McEvoy et al. \citetext{in prep}. Further matches were made to classifications by \citet{Bosch:1999}, \citet{WalbornBlades:1997}, \citet{Parker:1993} and \citet{SchildTestor:1992}.

SpT matching was achieved by setting a proximity distance of $<1$\,arcsec between the position of the photometric candidates and the spectral catalogues. This method relies on accurate and consistent astrometry. The SpTs of \citet{Bosch:1999} followed on from the photometric work of \citet{Selman:1999} and so checks were made to ensure authentic matching. The same was done for SpTs taken from \citet{WalbornBlades:1997} as they provided a \citet{Parker:1993} alias to the stars they classified. For any stars that did become matched with more than one SpT (most likely due to a nearby star in the field of view), the original photometry and that from the spectral catalogue were compared, with the SpT showing the most consistent photometry being selected. About 30 classified stars lacked a luminosity class. These were estimated (only for the purpose of calculating stellar parameters) by taking their derived $M_{V}$ and comparing them to the average $M_{V}$ of stars in the census of that known SpT (see Table \ref{tab:Average absolute magnitudes}). 

\begin{figure*}
\includegraphics[width=\textwidth]{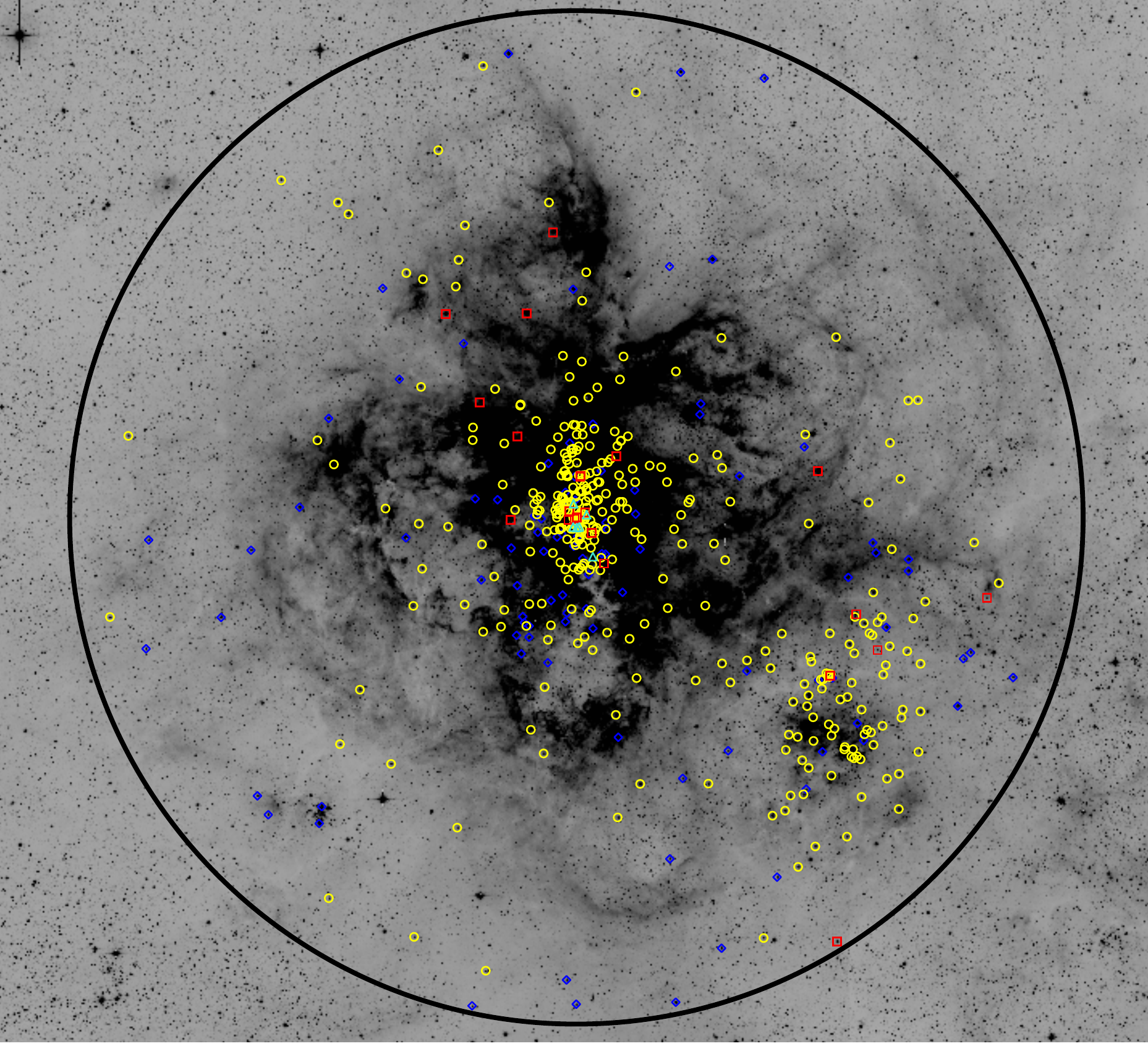}
\caption{All the spectroscopically confirmed hot luminous stars within 30 Dor ($r_{\mathrm{d}}=10$\,arcmin indicated by the large black circle), overlaid on a $V$-band WFI image. W-R stars (red squares), Of/WN stars (cyan triangles), O-type stars (yellow circles) and B-type stars (blue diamonds). North is up and East is to the left.}
\label{fig:10min_Feedback_stars}
\end{figure*}

\begin{figure}
\centering
\includegraphics[width=0.5\textwidth]{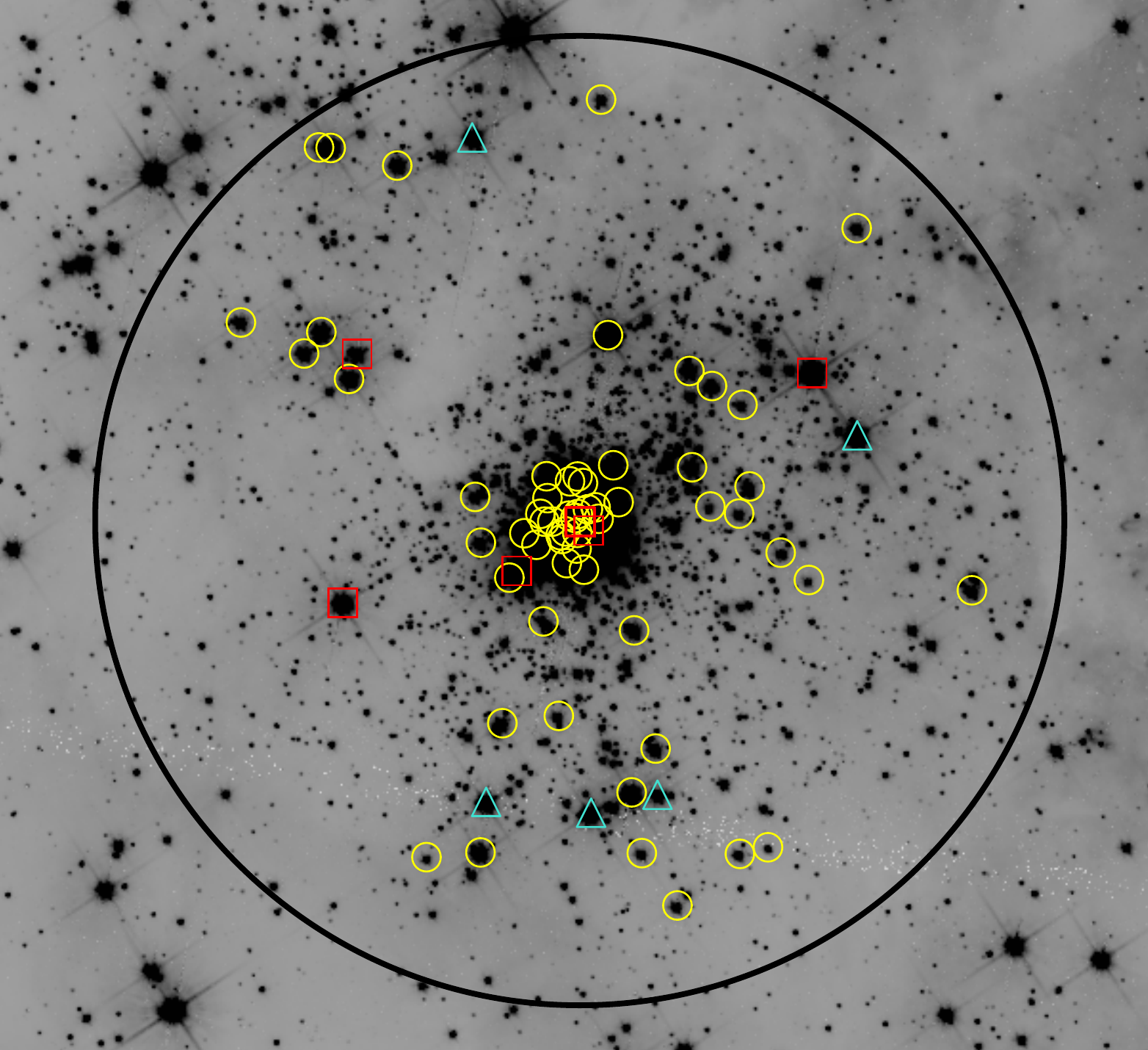}
\caption{All the spectroscopically confirmed hot luminous stars within the R136 region ($r_{\mathrm{d}}=20$\,arcsec indicated by large the black circle), overlaid on a F555W HST/WFC3 image. W-R stars (red squares), Of/WN stars (cyan triangles), O-type stars (yellow circles). North is up and East is to the left.}
\label{fig:0_0p33min_Feedback_stars}
\end{figure}

Within R136, stars were individually matched to the SpTs of \citet{MasseyHunter:1998}, \citet{CrowtherWalborn:2011} and \citet{CrowtherDessart:1998}, along with a few VFTS star classifications \citep{Henault-Brunet:2012a}. For the crowded Brey 73 cluster, we use the same SpTs obtained by \citet{Walborn:1999}.

A total of 31 W-R and Of/WN stars were manually entered into the census. If they were VFTS stars, spectral classifications were taken from Table 2 of Paper I, while non-VFTS stars took their classifications from various literature: \citet{MasseyHunter:1998}, \citet{Breysacher:1999} or \citet{CrowtherWalborn:2011}, and references therein.

Inevitably, duplications occurred, with some stars having been classified in multiple previous studies. As with the photometry, an order of priority was assigned to the literature when selecting a SpT. If available, a VFTS classification (e.g. Paper I, \citealt{Taylor:2011}, \citealt{Dufton:2011}, \citealt{Henault-Brunet:2012a}, Walborn et al. \citetext{in prep.}, McEvoy et al. \citetext{in prep} or this work) was always favoured because of the often high data quality and homogeneity of the approach. Otherwise, classifications by \citet{Bosch:1999} were adopted, followed by \citet{WalbornBlades:1997} and \citet{SchildTestor:1992} and in turn \citet{Parker:1993}. Table \ref{tab:SpT sources} indicates the number of stars classified from each reference. Note that a larger number of stars were matched to spectra as indicated in Table \ref{tab:10min Spectral Breakdown}. Table \ref{tab:SpT sources} only lists the spectroscopically confirmed hot luminous stars for which stellar parameters and feedback were later determined. Figure \ref{fig:10min_Feedback_stars} gives the spatial distribution of the classified stars listed in Table \ref{tab:SpT sources}. Figure \ref{fig:0_0p33min_Feedback_stars} provides a closer look at the central crowded R136 region.

\section{Spectroscopic Completeness}
\label{sec:Completeness}
\begin{figure}
\includegraphics[width=0.5\textwidth]{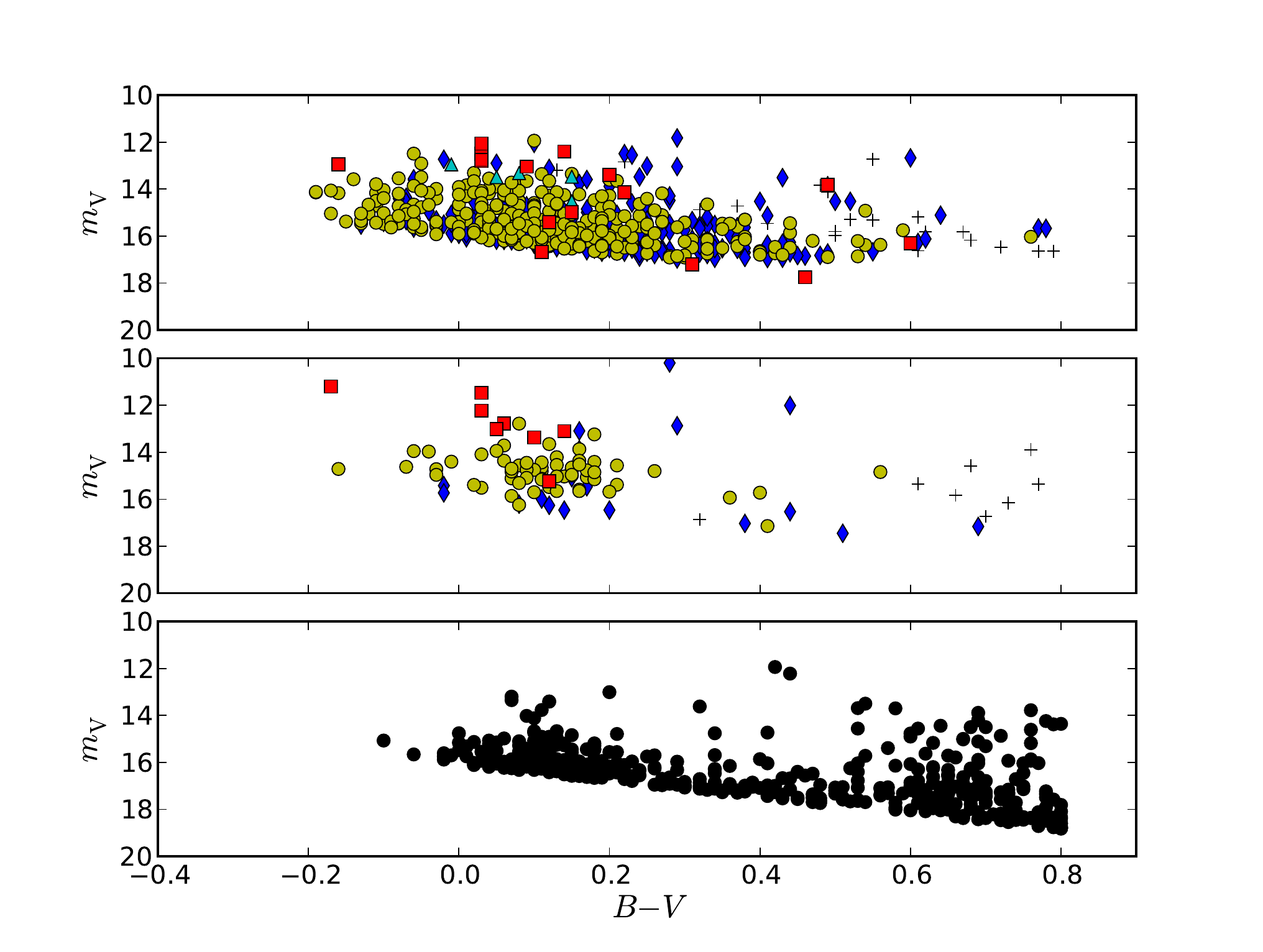}
\caption{ Colour magnitude diagrams of all the stars meeting the selection criteria. Stars with VFTS spectral classification (top), stars with non-VFTS spectral classification (middle) and stars with no spectral classification (bottom). Key for classified stars: W-R stars (red squares), Of/WN stars (cyan triangles), O-type stars (yellow circles), B-type stars (blue diamonds) and stars later than B-type (crosses).}
\label{fig:10_arcmin_spectral_breakdown}
\end{figure}

In this section we aim to estimate the spectroscopic completeness of hot luminous stars in 30 Dor. Figure \ref{fig:10_arcmin_spectral_breakdown} presents a set of colour magnitude diagrams, similar to that in Figure \ref{fig:VFTS_CMD}, now including all photometrically selected stars in the census. Approximately 60\% of stars in the census have available spectroscopy, either from the VFTS or a previous study. The different SpTs of these stars are shown and also listed in Table \ref{tab:10min Spectral Breakdown}.

Overall, 40\% of candidates lacked spectroscopy. However, this fraction rises as one moves to increasingly redder and fainter stars. This is expected since obtaining high quality spectroscopy for fainter stars is much more difficult. As explained in Section \ref{sec:Candidate selection}, more contaminant (later than B-type) stars will be found in the lower right hand side of Figure \ref{fig:10_arcmin_spectral_breakdown}. Ideally, these contaminants would have been removed through the $Q$-parameter cut but the limited availability of $U$-band photometry meant this was rarely possible. The key question is what fraction of the unclassified stars are in fact hot luminous stars, and what would the spectroscopic completeness be if they were taken into account? Figure \ref{fig:10_arcmin_spectral_breakdown} shows that the stars with spectroscopy are governed by the magnitude limit of the VFTS ($m_V<17$\,mag). Therefore, an accurate spectroscopic completeness level can only really be estimated for stars brighter than this limit.

\begin{table}
\begin{center}
\caption{The spectral type distribution of all the stars meeting the photometric selection criteria in Section \ref{sec:Candidate selection}.}
\label{tab:10min Spectral Breakdown}
\begin{tabular} { l  r  r r } 
\hline\hline
ALL STARS & \multicolumn{3}{r}{1145}\\
Stars without spectra & \multicolumn{3}{r}{463}\\
\hline
Stars with spectra & VFTS (\% of Total) & non-VFTS & Total\\
W-R & 17 (68) & 8 & 25\\
Of/WN & 6 (100) & 0 & 6\\
O-type & 322 (84) & 63 & 385\\
B-type  & 219 (92) & 18 & 237\\
later than B-type & 21 (72) & 8 & 29\\
Total & 585 (86) & 97 & 682\\

\hline
\end{tabular}
\end{center}
\end{table}

From this point onwards, we define a separate region, the `MEDUSA region' ($0.33<r_{\mathrm{d}}<10$\,arcmin) along with our previously defined `R136 region' ($r_{\mathrm{d}}<20$\,arcsec). Using a distance modulus of 18.49\,mag these regions span projected radii of about $5<r_{\mathrm{d}}<150$\,pc and $r_{\mathrm{d}}<5$\,pc, respectively. This allowed for a less biased measure of the spectroscopic completeness of the VFTS since its FLAMES/MEDUSA observations (see Paper I) largely avoided the R136 vicinity because of crowdedness.

\subsection{Spectroscopic completeness in the MEDUSA region}
\label{sec:Completeness in the MEDUSA Region}
\begin{figure}
\includegraphics[width=0.5\textwidth]{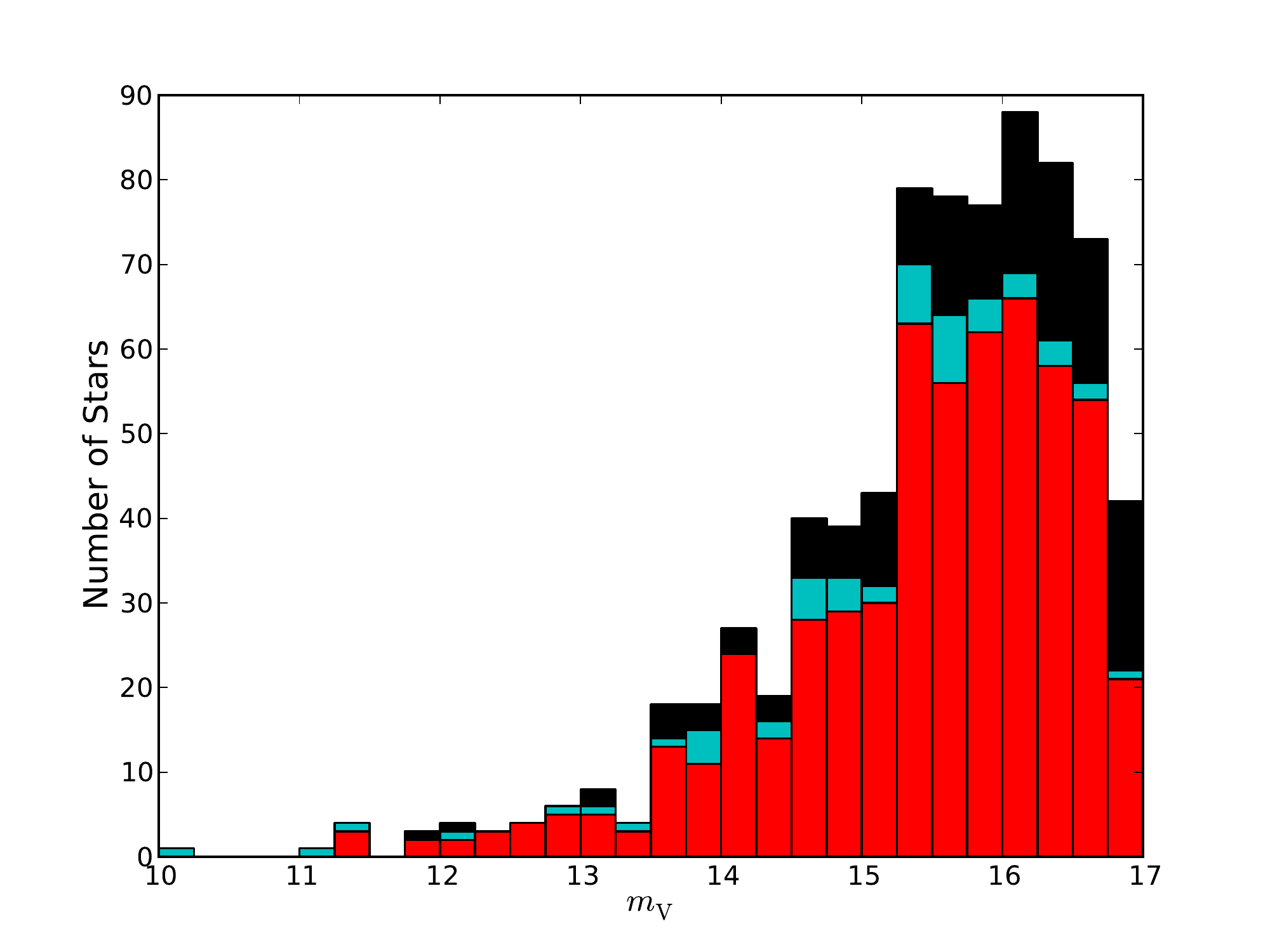}
\caption{The spectroscopic completeness, with respect to $m_{V}$, of all stars in the MEDUSA region meeting the selection criteria and with $m_V<17$\,mag. Stars with VFTS spectroscopy (red), stars with non-VFTS spectroscopy (cyan) and stars with no spectroscopy (black).}
\label{fig:0p33_10min_Vmag_completeness}
\end{figure}

\begin{figure}
\includegraphics[width=0.5\textwidth]{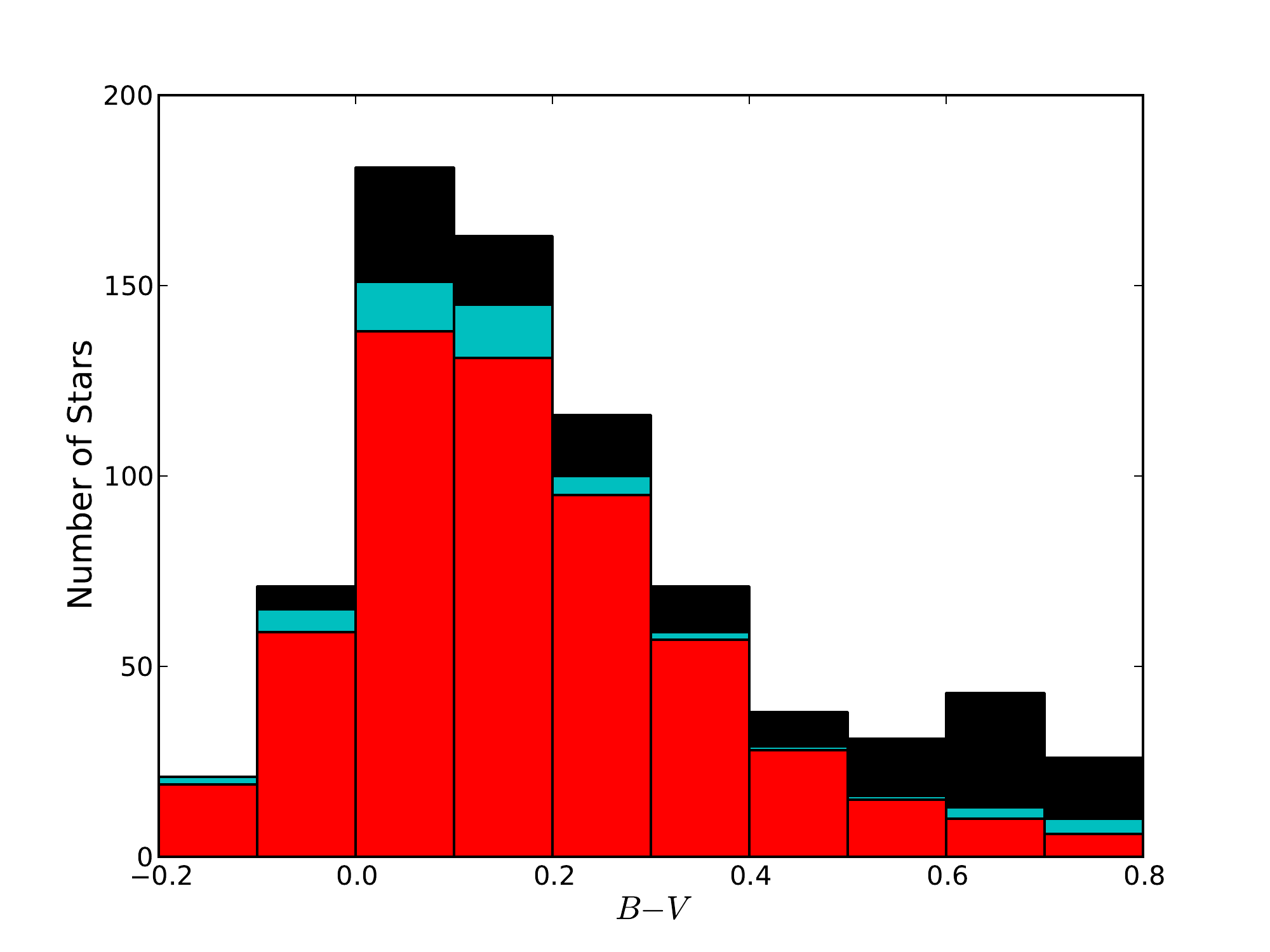}
\caption{The spectroscopic completeness, with respect to $B-V$, of all stars in the MEDUSA region meeting the selection criteria and with $m_V<17$\,mag. Colours are as in Figure \ref{fig:0p33_10min_Vmag_completeness}.}
\label{fig:0p33_10min_colour_completeness_with_contaminants}
\end{figure}

Figures \ref{fig:0p33_10min_Vmag_completeness}
\& \ref{fig:0p33_10min_colour_completeness_with_contaminants} show histograms of the number of photometric candidates for which spectroscopy was available, in relation to magnitude and colour, respectively. Within the MEDUSA region, 80\% of candidates have been spectroscopically observed of which about three quarters were included in the VFTS. Of these, only a subset were spectroscopically confirmed as hot luminous stars. These are the stars given in Table \ref{tab:MEDUSA region spectral breakdown}.

\begin{figure}
\includegraphics[width=0.5\textwidth]{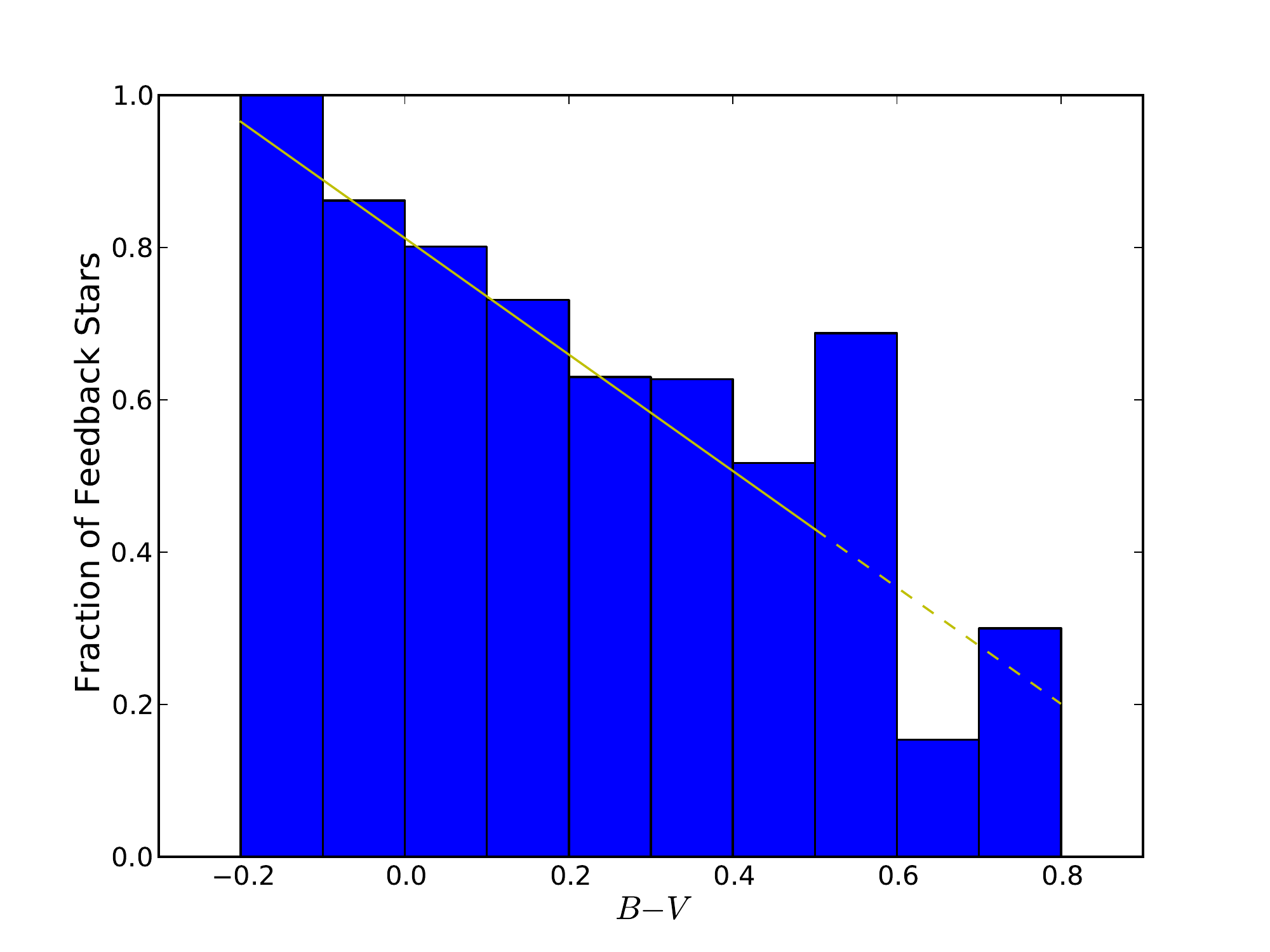}
\caption{The fraction of stars in the MEDUSA region meeting the selection criteria and with both $m_V<17$\,mag and spectra, which were spectroscopically confirmed to be hot luminous stars. A low order polynomial was fitted to the distribution (solid line) and extrapolated redward of $B-V=0.5$\,mag (dashed line).}
\label{fig:0p33_10min_Feedback_star_fraction}
\end{figure}

\begin{figure}
\includegraphics[width=0.445\textwidth]{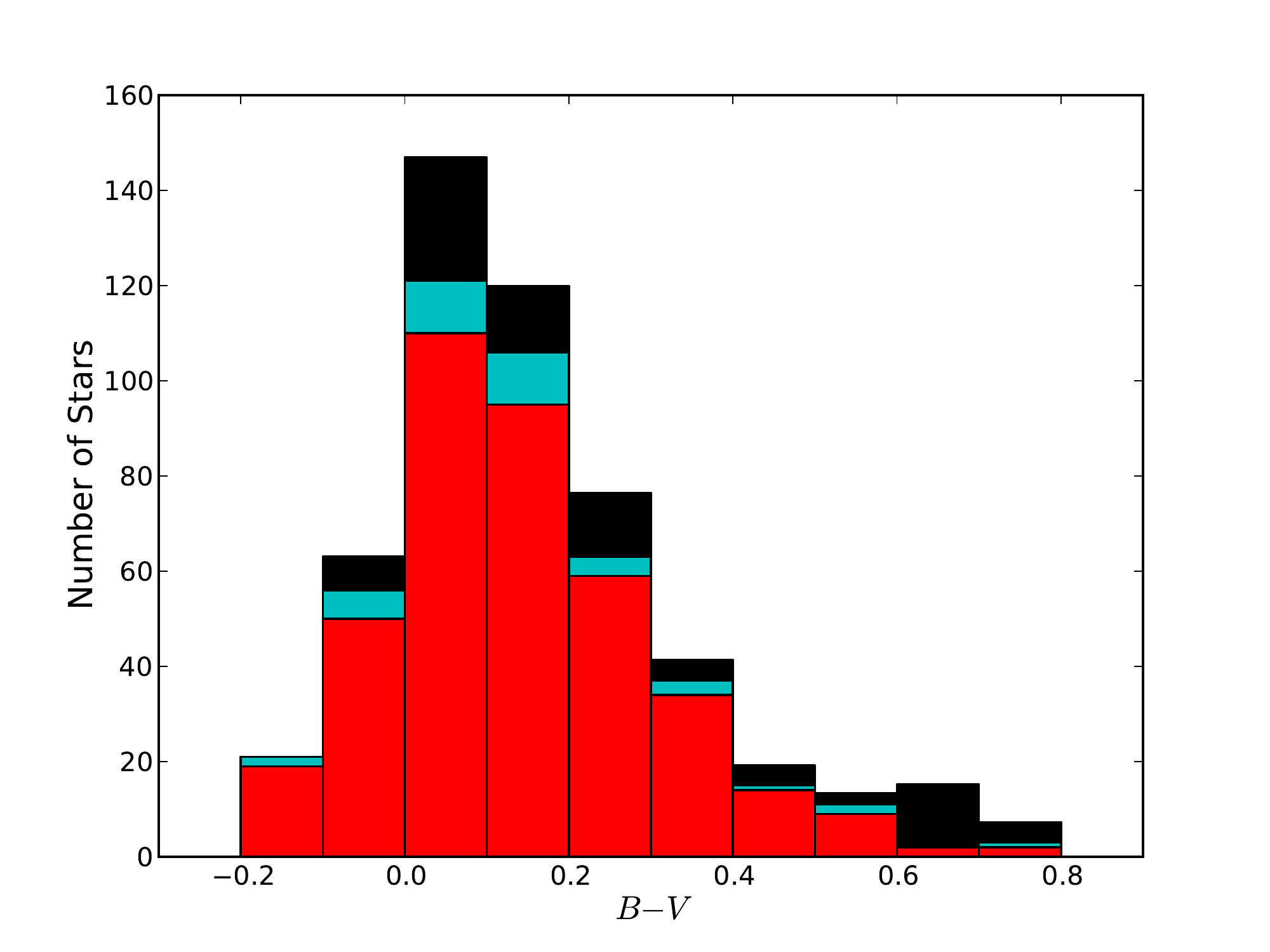}
\caption{The spectroscopic completeness of hot luminous stars with $m_V<17$\,mag in the MEDUSA region, with respect to $B-V$. Colours are as in Figure \ref{fig:0p33_10min_Vmag_completeness}.}
\label{fig:0p33_10min_colour_completeness_without_contaminants}
\end{figure}

In an attempt to obtain an accurate completeness level for the region, Figure \ref{fig:0p33_10min_Feedback_star_fraction} gives the fraction of stars that are spectroscopically confirmed as hot luminous stars as a function of colour. Moving redward of the typical O-type star intrinsic colour $(B-V)_{\mathrm{o}}=-0.32$\,mag, the fraction is seen to fall from unity as more contaminants populate each colour bin. At high values of $B-V$, these fractions become poorly constrained as a result of small number statistics. By accounting for the fraction of contaminants, a new completeness plot can be made as shown in Figure \ref{fig:0p33_10min_colour_completeness_without_contaminants} which, unlike Figure \ref{fig:0p33_10min_colour_completeness_with_contaminants}, has had the predicted contaminant stars removed. Applying this correction, the completeness of the hot luminous stars is estimated at 84\%, of which 76\%  were included in the VFTS.

\begin{table}
\begin{center}
\caption{The spectral type distribution of the spectroscopically confirmed hot luminous stars within the MEDUSA region ($5<r_{\mathrm{d}}<150$\,pc).}
\label{tab:MEDUSA region spectral breakdown}
\begin{tabular} { l | c | c | c | c | c | c | c } 
\hline\hline
W-R and Of/WN stars & \multicolumn{3}{c|}{WR2-WR5} & \multicolumn{3}{c|}{WR6-WR9} & Total\\
\hline
WN or WN/C & \multicolumn{3}{c|}{7} & \multicolumn{3}{c|}{8} & 15 \\
WC & \multicolumn{3}{c|}{3} & \multicolumn{3}{c|}{0} & 3 \\
Of/WN & \multicolumn{3}{c|}{0} & \multicolumn{3}{c|}{1}  & 1 \\
Total & \multicolumn{3}{c|}{10} & \multicolumn{3}{c|}{9} & 19 \\
\hline
O-type stars & \multicolumn{2}{c|}{O2-3.5} & \multicolumn{2}{c|}{O4-6.5} & \multicolumn{2}{c|}{O7-9.7} & Total\\
\hline
V & \multicolumn{2}{c|}{15} & \multicolumn{2}{c|}{60} & \multicolumn{2}{c|}{143} & 218 \\
III & \multicolumn{2}{c|}{5} & \multicolumn{2}{c|}{11} & \multicolumn{2}{c|}{66} & 82\\
I & \multicolumn{2}{c|}{2} & \multicolumn{2}{c|}{3} & \multicolumn{2}{c|}{20} & 25 \\
Total & \multicolumn{2}{c|}{22} & \multicolumn{2}{c|}{74} & \multicolumn{2}{c|}{229} & 325 \\
\hline
B-type stars & \multicolumn{3}{c|}{B0-0.7} & \multicolumn{3}{c|}{B1 and later} & Total\\
\hline
V & \multicolumn{3}{c|}{24} & \multicolumn{3}{c|}{-} & 24 \\
III & \multicolumn{3}{c|}{15} & \multicolumn{3}{c|}{7} & 22 \\
I & \multicolumn{3}{c|}{12} & \multicolumn{3}{c|}{26}  & 38 \\
Total & \multicolumn{3}{c|}{51} & \multicolumn{3}{c|}{33} & 84 \\
\hline
\multicolumn{7}{c}{ GRAND TOTAL} & 428\\
\hline
\end{tabular}
\tablefoot{The numbers given for B-type stars are not complete for 30 Dor. The reader is reminded that late B-type stars were omitted from the census, both during the selection and once SpTs had been matched, due to their negligible contributions to the feedback. These included dwarfs later than B0.5\,V and giants later than B1\,III. All selected supergiants were included. The completeness of the W-R, Of/WN and O-type stars is discussed in Section \ref{sec:Completeness}.
}
\end{center}
\end{table}

\subsection{Spectroscopic completeness in the R136 region}
\label{sec:Completeness of R136 Region}
\begin{figure}
\includegraphics[width=0.5\textwidth]{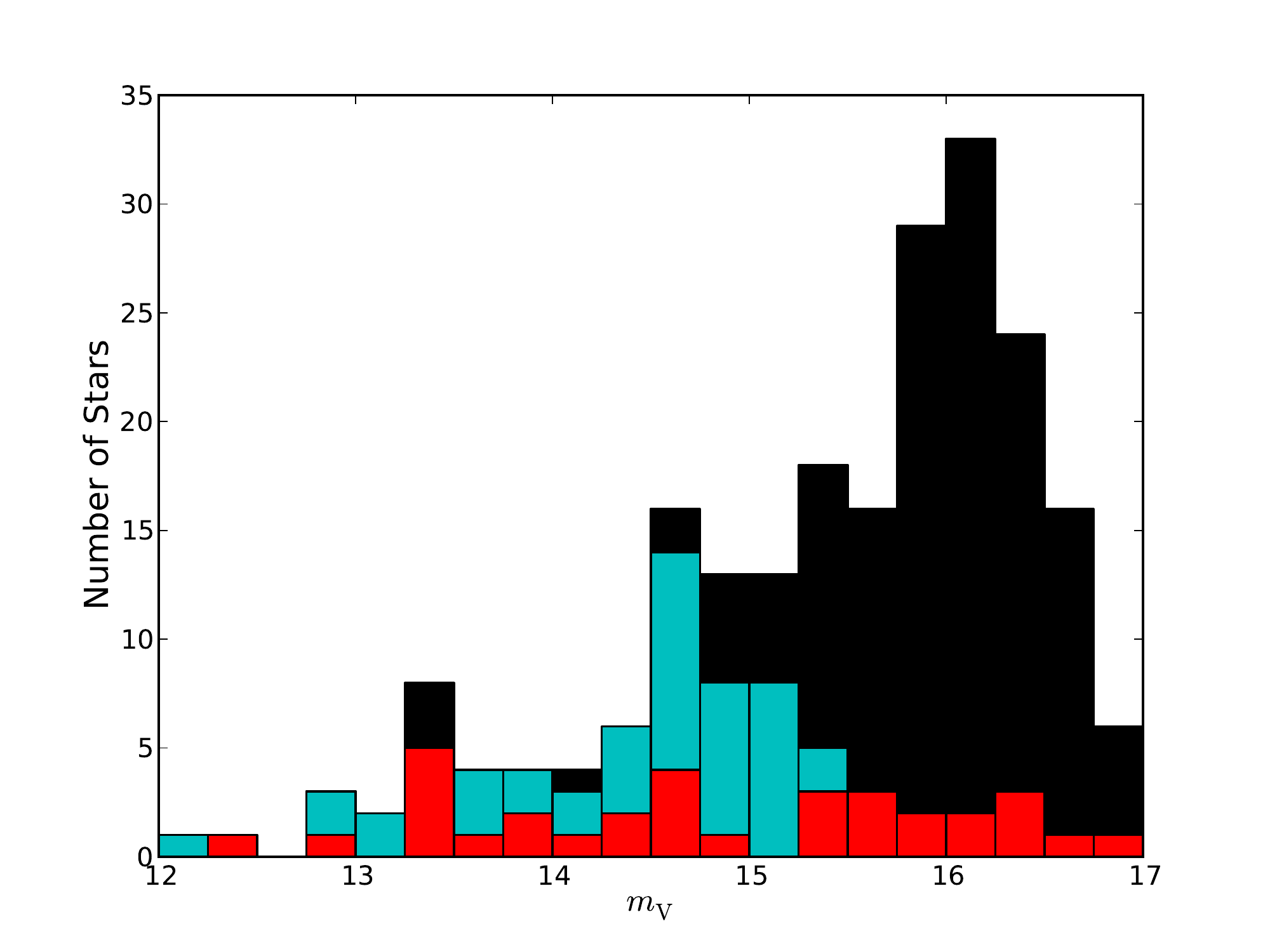}
\caption{The same as Figure \ref{fig:0p33_10min_Vmag_completeness}, only showing completeness in the R136 region. Colours are as in Figure \ref{fig:0p33_10min_Vmag_completeness}.}
\label{fig:0_0p33min_Vmag_completeness}
\end{figure}

\begin{figure}
\includegraphics[width=0.5\textwidth]{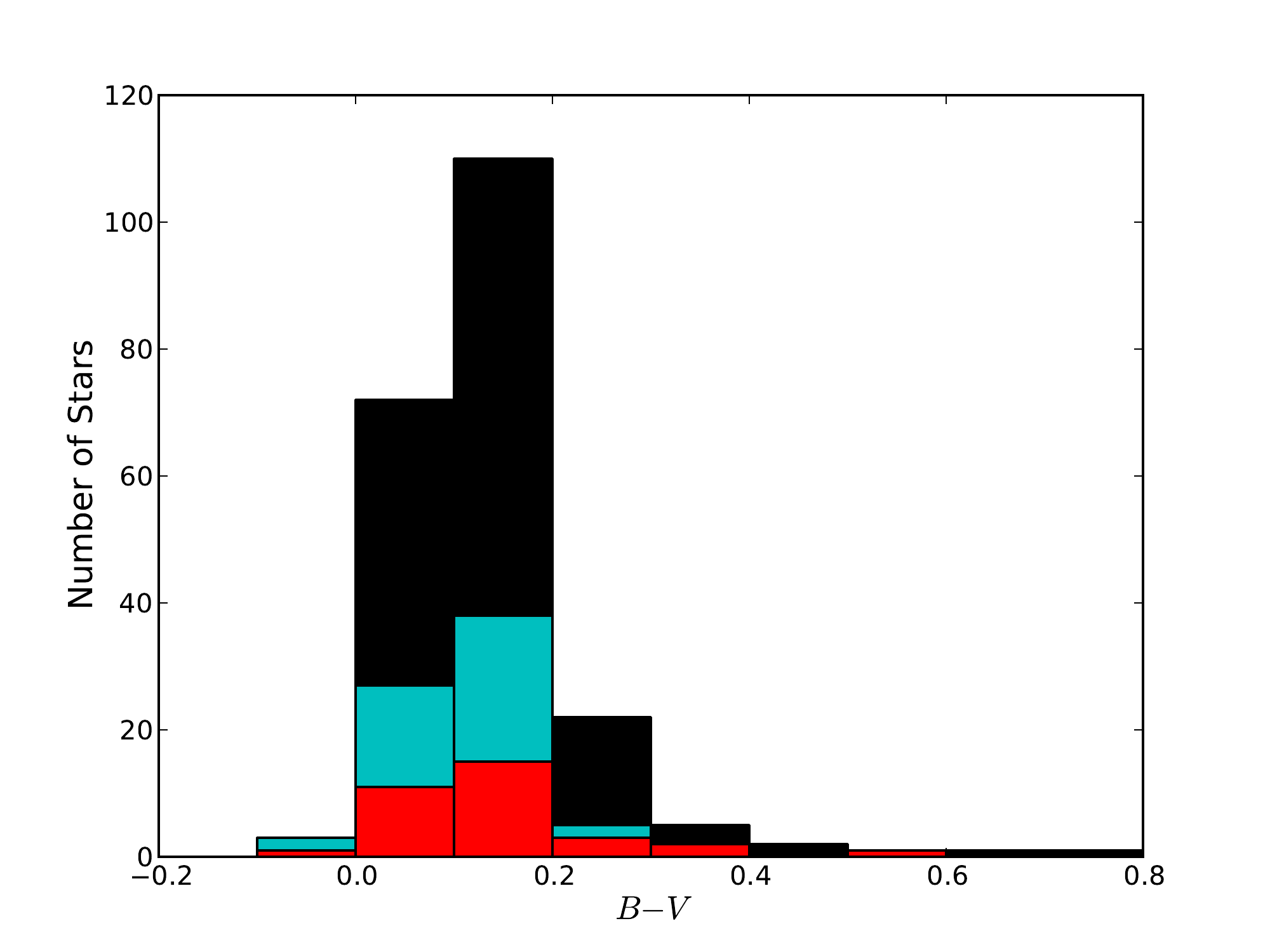}
\caption{The same as Figure \ref{fig:0p33_10min_colour_completeness_with_contaminants}, only showing completeness in the R136 region. Colours are as in Figure \ref{fig:0p33_10min_Vmag_completeness}.}
\label{fig:0_0p33min_colour_completeness_with_contaminants}
\end{figure}

Figure \ref{fig:0_0p33min_Vmag_completeness} shows that only the brightest targets ($\approx35$\%) in the R136 region have archival spectroscopy, primarily obtained with HST/FOS by \citet{MasseyHunter:1998}, with the VFTS covering only a limited number of stars through its FLAMES-ARGUS observations (see Paper I and \citealt{Henault-Brunet:2012a}). A breakdown of their SpTs is listed in Table \ref{tab:R136 Region spectral breakdown}. This will be improved following new HST/STIS spectroscopy of the central parsec (GO 12465/13052, P.I. - P. Crowther). Figure \ref{fig:0_0p33min_colour_completeness_with_contaminants} suggests that these unclassified stars are predominantly early type stars i.e. mid-late O-type stars which eluded spectroscopic confirmation, simply because they are faint. More importantly, completeness is still high at the brighter end, and includes the more luminous W-R, Of/WN and supergiant stars which will contribute most to the feedback of the region. We therefore estimate that 35\% of the hot luminous stars have so far been spectroscopically observed.

\begin{table}
\begin{center}
\caption{The spectral type distribution of the spectroscopically confirmed hot luminous stars within the R136 region ($r_{\mathrm{d}}<5$\,pc).}
\label{tab:R136 Region spectral breakdown}
\begin{tabular} { l | c | c | c | c | c | c | c } 
\hline\hline
W-R and Of/WN stars & \multicolumn{3}{c|}{WR2-WR5} & \multicolumn{3}{c|}{WR6-WR9} & Total\\
\hline
WN or WN/C & \multicolumn{3}{c|}{5} & \multicolumn{3}{c|}{1} & 6 \\
WC & \multicolumn{3}{c|}{1} & \multicolumn{3}{c|}{0} & 1 \\
Of/WN & \multicolumn{3}{c|}{3} & \multicolumn{3}{c|}{2}  & 5 \\
Total & \multicolumn{3}{c|}{9} & \multicolumn{3}{c|}{3} & 12 \\
\hline
O-type stars & \multicolumn{2}{c|}{O2-3.5} & \multicolumn{2}{c|}{O4-6.5} & \multicolumn{2}{c|}{O7-9.7} & Total\\
\hline
V & \multicolumn{2}{c|}{19} & \multicolumn{2}{c|}{11} & \multicolumn{2}{c|}{11} & 41 \\
III & \multicolumn{2}{c|}{9} & \multicolumn{2}{c|}{0} & \multicolumn{2}{c|}{2} & 11\\
I & \multicolumn{2}{c|}{4} & \multicolumn{2}{c|}{4} & \multicolumn{2}{c|}{0} & 8\\
Total & \multicolumn{2}{c|}{32} & \multicolumn{2}{c|}{15} & \multicolumn{2}{c|}{13} & 60 \\
\hline
\multicolumn{7}{c}{GRAND TOTAL} & 72\\
\hline
\end{tabular}
\end{center}
\end{table}

\subsection{Stars unaccounted for}
\label{sec:Stars Unaccounted For}
While we attempt to include all the potential hot luminous stars in 30 Dor, the following factors may have prevented certain candidates from entering our final census:
\begin{enumerate}
\item {\em Crowding} - Our four primary photometry catalogues (Selman, WFI, Parker, CTIO) provided sound coverage of 30 Dor. Potential concern arises in dense regions where candidates may have been lost as a result of less reliable photometry.
\item {\em High extinction} - We calculate an average extinction of $A_V\approx1.5$\,mag for  30 Dor, with our selection criteria allowing for extinctions as high as $A_V=3.92$\,mag. For some stars, the extinction will be higher than this limit. In particular, young embedded stars will have evaded selection through their surrounding gas and dust but could still contribute to the stellar feedback (see Section \ref{sec:Feedback from additional sources}).
\item {\em Spectroscopic Completeness} - Section \ref{sec:Completeness} showed a good spectrosopic completeness for the census but the remaining stars lacking spectroscopy are still important and their contribution is considered in Section \ref{sec:Correcting for unclassified stars}. Furthermore, the magnitude coverage of the spectroscopy was limited to $m_V<17$\,mag. This means that a few faint and highly reddened hot luminous stars could also have avoided classification.
\item {\em Binary systems} - Stars found in binary systems will provide further candidates through their companions as well as changing the photometric properties of each component. Section \ref{sec:Binary systems} addresses the case of double-lined (SB2) binary systems where a knowledge of both stars can provide a combined feedback contribution but single-lined (SB1) binary systems and any further unknown multiple systems, remain uncertain.
\item {\em Unresolved stellar systems} - Spectroscopic binary detection is limited to shorter periods of a few days. While HST imaging is capable of revealing likely composite spectra, many large separation and/or line of sight systems will still appear single and are so far unidentified.
\end{enumerate}

\subsection{Correcting for unclassified stars}
\label{sec:Correcting for unclassified stars}
Section \ref{sec:Completeness in the MEDUSA Region} looked at removing the estimated number of contaminant stars in different colour bins. Even after applying this method to determine the true number of hot luminous stars, some still lack spectroscopy. For the MEDUSA region, a total of 152 of our candidates ($m_V<17$\,mag) lacked spectroscopy, of which it was estimated that only $\approx50$\% would actually be hot luminous stars. Meanwhile, in R136, 141 candidates ($m_V<17$\,mag) lacked spectroscopy but in this case, all were taken to be hot luminous stars capable of contributing to the feedback. This is a reasonable assumption given that all other R136 stars in the census are classified as W-R, Of/WN or O-type stars. Furthermore, as noted earlier, the magnitudes and colours of these unclassified stars suggest them to be mid-late O-type stars.

\citet{Voges:2008} estimated the SpT of unclassified stars in their LMC H\,{\sc ii} regions using a colour-SpT calibration. It was based on $U$- and $B$-band magnitudes as these lay closest to the peaks in the spectral distributions of the OB stars. However, this approach relied upon accurate photometry, and in the case of the $U$-band, this was only available for a select number of our stars. The $M_{V}$ was therefore calculated for each star, assuming a common intrinsic colour of $(B-V)_{\mathrm{o}}=-0.3$\,mag. Its SpT was then estimated based on this $M_{V}$, using the average $M_{V}$ found for each SpT in the rest of the census (see Table \ref{tab:Average absolute magnitudes}). Some crude assumptions had to be made when more than one luminosity class was consistent with the $M_{V}$. In R136, stars were typically assumed to be dwarfs given the young age of the region. All W-R and Of/WN stars were assumed to be already accounted for.

\section{Stellar Calibrations}
\label{sec:Stellar Calibrations}
With photometry and a SpT assigned to as many stars as possible, the aim was now to estimate their feedback. Here we specifically seek the radiative and mechanical stellar feedback. We determined the radius independent Lyman continuum ionising flux per cm$^{-2}$ ($q_{\mathrm{0}}$). This is related to the total number of ionising photons emitted by the star per second ($Q_{\mathrm{0}}$) via $Q_{\mathrm{0}} = 4 \pi R^2 q_{\mathrm{0}}$, where $R$ is the stellar radius at a Rosseland optical depth of two thirds ($R_{2/3}$), or in the case of W-R stars, the radius at an optical depth of $\sim10$ ($R_{*}$). Similarly, if we omit the effects of supernovae, our primary source of mechanical feedback is that produced by the stellar winds. The stellar wind luminosity ($L_{\mathrm{SW}}$) is calculated via $L_{\mathrm{SW}} =\frac{1}{2} \dot{M} \varv_{\infty}^2$, where $\dot{M}$ is the mass-loss rate of the star, $\varv_{\infty}$ is the terminal velocity of the stellar wind and $L_{\mathrm{SW}}$ is given in units of erg\,s$^{-1}$. In addition, we also calculated the modified wind momentum $D_{\mathrm{mom}} = \dot{M} \varv_{\infty} R^{0.5}$.

Detailed atmospheric analyses are currently underway of all the VFTS stars which will eventually be used to assign each stellar parameter. In this study, however, we turn to a variety of calibrations to supply the parameters. Results for W-R and Of/WN stars were based on atmospheric models, given their important role in feedback contribution.

\subsection{O-type stars}
\label{sec:O-type stars}
A total of 385 O-type stars were selected within 30 Dor. The calibrations for the O-type star parameters were based primarily on the models of \citet{Martins:2005}. Their $T_{\mathrm{eff}}$-SpT scale was revised by $+1$\,kK for the LMC environment \citep{Mokiem:2007, RiveroGonzalez:2012a} while the earliest subtypes (O2-O3.5) had $T_{\mathrm{eff}}$ individually selected, based on the works of \citet{RiveroGonzalez:2012b} and \citet{DoranCrowther:2011}. New calibrations for the bolometric correction ($BC$) and $q_{\mathrm{0}}$ were also derived. Tables \ref{tab:O-star Parameters Dwarfs}-\ref{tab:O-star Parameters Supergiants} provide a summary of the O-type star parameters used. Further empirical and theoretical data helped to obtain the stellar wind parameters. Values of $\varv_{\infty}$ were sourced from \citet{Prinja:1990} while mass-loss rates were predicted using the theoretical prescriptions of \citet{Vink:2001}. See Appendix \ref{sec:O-star Calibrations} for more details.

\begin{table}[h]
\begin{center}
\caption{O-type star parameters for luminosity class V.}
\label{tab:O-star Parameters Dwarfs}
\begin{tabular} { l  c  c  c  c  c } 
\hline\hline
SpT & $T_{\mathrm{eff}}$ & $BC_{V}$ & log $q_{\mathrm{0}}$ & $L_{\mathrm{EUV}}$ & $\varv_{\infty}$ \\
     & (kK) &  (mag)    & (ph cm$^{-2}$ s$^{-1}$) & ($L_{\mathrm{Bol}}$) & (km s$^{-1}$) \\
\hline
O2  & 54.0  & $-4.61$  &  24.97 &  0.74 &  3300 \\
O3  & 48.0  & $-4.26$  &  24.70 &  0.61 &  3000 \\
O3.5  & 47.0  & $-4.19$  &  24.65 &  0.58 &  3300 \\
O4  & 43.9  & $-3.99$  &  24.52 &  0.50 &  3000 \\
O5  & 41.9  & $-3.84$  &  24.41 &  0.44 &  2700 \\
O5.5  & 40.9  & $-3.77$  &  24.34 &  0.42 &  2000 \\
O6  & 39.9  & $-3.70$  &  24.27 &  0.39 &  2600 \\
O6.5  & 38.9  & $-3.62$  &  24.20 &  0.36 &  2500 \\
O7  & 37.9  & $-3.55$  &  24.11 &  0.32 &  2300 \\
O7.5  & 36.9  & $-3.47$  &  24.01 &  0.29 &  2000 \\
O8  & 35.9  & $-3.39$  &  23.90 &  0.26 &  1800 \\
O8.5  & 34.9  & $-3.31$  &  23.78 &  0.23 &  2000 \\
O9  & 33.9  & $-3.22$  &  23.64 &  0.19 &  1500 \\
O9.5  & 32.9  & $-3.14$  &  23.49 &  0.16 &  1500 \\
O9.7  & 32.5  & $-3.11$  &  23.43 &  0.14 &  1200 \\
\hline
\end{tabular}
\end{center}
\end{table}

\begin{table}[h]
\begin{center}
\caption{O-type star parameters for luminosity class III.}
\label{tab:O-star Parameters Giants}
\begin{tabular} { l c  c  c  c  c }
\hline\hline
SpT & $T_{\mathrm{eff}}$ & $BC_{V}$ & log $q_{\mathrm{0}}$ & $L_{\mathrm{EUV}}$ & $\varv_{\infty}$ \\
     & (kK) &  (mag)   & (ph cm$^{-2}$ s$^{-1}$) & ($L_{\mathrm{Bol}}$) & (km s$^{-1}$) \\
\hline
O2  & 49.6  & $-4.34$  &  24.78 &  0.65 &  3200 \\
O3  & 47.0  & $-4.20$  &  24.67 &  0.58 &  3200 \\
O3.5  & 44.0  & $-4.01$  &  24.54 &  0.50 &  2600 \\
O4  & 42.4  & $-3.90$  &  24.46 &  0.46 &  2600 \\
O5  & 40.3  & $-3.75$  &  24.34 &  0.40 &  2800 \\
O5.5 & 39.2 & $-3.67$ & 24.27 & 0.37 & 2700 \\
O6  & 38.2  & $-3.58$  &  24.20 &  0.33 &  2600 \\
O6.5  & 37.1  & $-3.50$  &  24.12 &  0.30 &  2600 \\
O7  & 36.1  & $-3.41$  &  24.03 &  0.27 &  2600 \\
O7.5  & 35.0  & $-3.32$  &  23.93 &  0.23 &  2200 \\
O8  & 34.0  & $-3.23$  &  23.82 &  0.20 &  2100 \\
O8.5  & 32.9  & $-3.13$  &  23.70 &  0.16 &  2300 \\
O9  & 31.8  & $-3.03$  &  23.58 &  0.12 &  1900 \\
O9.5  & 30.8  & $-2.93$  &  23.44 &  0.08 &  1500 \\
O9.7  & 30.4  & $-2.89$  &  23.38 &  0.07 &  1200 \\
\hline
\end{tabular}
\end{center}
\end{table}

\begin{table}[h]
\begin{center}
\caption{O-type star parameters for luminosity class I.}
\label{tab:O-star Parameters Supergiants}
\begin{tabular} { l  c  c  c  c  c } 
\hline\hline
SpT & $T_{\mathrm{eff}}$ & $BC_{V}$ & log $q_{\mathrm{0}}$ & $L_{\mathrm{EUV}}$ & $\varv_{\infty}$ \\
     & (kK) &  (mag)  & (ph cm$^{-2}$ s$^{-1}$)  & ($L_{\mathrm{Bol}}$)  & (km s$^{-1}$) \\
\hline
O2  & 46.0 & $-4.10$  &  24.63 &  0.60 &  3000 \\
O3  & 42.0  & $-3.88$  &  24.45 &  0.52 &  3700 \\
O3.5  & 41.1  & $-3.81$  &  24.40 &  0.50 &  2000 \\
O4  & 40.1  & $-3.75$  &  24.35 &  0.48 &  2300 \\
O5  & 38.3  & $-3.61$  &  24.23 &  0.43 &  1900 \\
O5.5 & 37.3 & $-3.53$ & 24.17 & 0.40 & 1900 \\
O6  & 36.4  & $-3.45$  &  24.10 &  0.38 &  2300 \\
O6.5  & 35.4  & $-3.37$  &  24.03 &  0.35 &  2200 \\
O7  & 34.5  & $-3.28$  &  23.95 &  0.32 &  2100 \\
O7.5  & 33.6  & $-3.20$  &  23.86 &  0.29 &  2000 \\
O8  & 32.6  & $-3.10$  &  23.77 &  0.25 &  1500 \\
O8.5  & 31.7  & $-3.01$  &  23.67 &  0.22 &  2000 \\
O9  & 30.7  & $-2.91$  &  23.56 &  0.18 &  2000 \\
O9.5  & 29.8  & $-2.81$  &  23.45 &  0.15 &  1800 \\
O9.7  & 29.4  & $-2.77$  &  23.40 &  0.13 &  1800 \\
\hline
\end{tabular}
\end{center}
\end{table}

\subsection{B-type stars}
\label{sec:B-type stars}
A total of 84 B-type stars were selected within 30 Dor. Most B-type star parameters were estimated using existing calibrations. $T_{\mathrm{eff}}$ was adopted from \citet{Trundle:2007} and $BC_{V}$ calculated from the relations of \citet{Crowther:2006} for supergiants, or \citet{LanzHubeny:2003} in the case of dwarfs and giants. A new calibration was derived for $q_{\mathrm{0}}$ based on the work of \citet{Smith:2002}, \citet{Conti:2008} and \cite{RiveroGonzalez:2012a}. Table \ref{tab:B-star Parameters} gives a summary of the B-type star parameters used. \citet{Prinja:1990} once again supplied $\varv_{\infty}$, along with \citet{KudritzkiPuls:2000}. In the case of early B-dwarfs and giants, where no known $\varv_{\infty}$ are published, we assume a value of $\varv_{\infty}=1000$\,km\,s$^{-1}$ given that wind properties are expected to be similar to their late O-type star counterparts. 

We again employ the \citet{Vink:2001} prescription for B-dwarfs and giants but opted not to use it for B-supergiants following discrepancies with empirical results noted by  \citep{ MarkovaPuls:2008, TrundleLennon:2005}. Instead we turned to using the wind-luminosity relation (WLR, \citealt{Kudritzki:1999}), whereby the modified wind momentum ($D_{\mathrm{mom}} = \dot{M} \varv_{\infty} R^{0.5})$ is shown to relate to the stellar luminosity. See Appendix \ref{sec:B-star Calibrations} for more details.

\begin{table}
\begin{center}
\caption{B-type star parameters for all luminosity classes.}
\label{tab:B-star Parameters}
\begin{tabular} { l  c  c  c  c  c } 
\hline\hline
SpT & $T_{\mathrm{eff}}$ & $BC_{V}$ & $\log q_{\mathrm{0}}$ & $L_{\mathrm{EUV}}$ & $\varv_{\infty}$ \\
     & (kK) & (mag)  & (ph cm$^{-2}$ s$^{-1}$) & ($L_{\mathrm{Bol}}$) & (km s$^{-1}$) \\
\multicolumn{6}{c}{Luminosity Class V}\\
\hline
B0  & 31.4  & $-3.04$  &  23.15 &  0.11 &  1000 \\
B0.2  & 30.3  & $-2.93$  &  22.85 &  0.06 &  1000 \\
B0.5  & 29.1  & $-2.81$  &  22.56 &  0.02 &  1000 \\
\multicolumn{6}{c}{Luminosity Class III}\\
\hline
B0  & 29.1  & $-2.81$  &  22.56 &  0.02 &  1000 \\
B0.2  & 27.9  & $-2.68$  &  22.28 &  $<0.01$ &  1000 \\
B0.5  & 26.7  & $-2.55$  &  22.02 &  $<0.01$ &  1000 \\
B0.7  & 25.4  & $-2.41$  &  21.78 &  $<0.01$ &  1000 \\
B1  & 24.2  & $-2.26$  &  21.56 &  $<0.01$ &  1000 \\
\multicolumn{6}{c}{Luminosity Class I}\\
\hline
B0  & 28.6  & $-2.71$  &  22.86 &  0.10 &  1500 \\
B0.2  & 27.0  & $-2.58$  &  22.51 &  0.03 &  1400 \\
B0.5  & 25.4  & $-2.44$  &  22.17 &  0.03 &  1400 \\
B0.7  & 23.8  & $-2.30$  &  21.84 &  0.01 &  1200 \\
B1  & 22.2  & $-2.14$  &  21.52 &  $<0.01$ &  1100 \\
B1.5 & 20.6 &  $-1.97$ & 21.22  &  $<0.01$ &  800 \\
B2  & 19.0  & $-1.79$  &  20.93 &  $<0.01$ &  800 \\
B2.5 & 17.4  & $-1.60$ &  20.65 &  $<0.01$ &  800 \\
B3  & 15.8  & $-1.38$  &  20.39 &  $<0.01$ &  600 \\
B5  & 14.2  & $-1.14$  &  20.13 &  $<0.01$ &  500 \\
B8  & 12.3  & $-0.83$  &  19.86 &  $<0.01$ &  200 \\
\hline
\end{tabular}
\end{center}
\end{table}

\subsection{Uncertainties in calibrations}
\label{sec:Uncertainties in calibrations}
All of the stellar calibrations are open to various uncertainties. Most important to estimating an accurate ionising luminosity will be our adopted $T_{\mathrm{eff}}$-SpT calibration. \citet{Martins:2005} and \citet{Trundle:2007} would suggest uncertainties on $T_{\mathrm{eff}}$ of about $\pm1-2$\,kK, likely to be higher ($\pm3-4$\,kK) in the case of our earliest O-type stars.

The $BC_{V}$ of the stars are particularly dependent on $T_{\mathrm{eff}}$ \citep{Martins:2005,MartinsPlez:2006}. Given the low extinction of 30 Dor and renewed accuracy in its distance \citep{Pietrzynski:2013}, it is the $BC_{V}$ of the stars which dominates the uncertainty of their bolometric luminosity ($L_{\mathrm{Bol}}$). The uncertainty on $\log L_{\mathrm{Bol}}/{\mathrm{L}}_{\odot}$ is found to be about $\pm0.15-0.25$\,dex. With $T_{\mathrm{eff}}$ similarly affecting the values of $q_{\mathrm{0}}$, our final ionising luminosities, $Q_{\mathrm{0}}$, are expected to be accurate to within $\sim60$\%.

A reliable wind luminosity depends on the accuracy of $\dot{M}$ and $\varv_{\infty}$. The assignment of $\varv_{\infty}$ was primarily based on the observational UV study by \citet{Prinja:1990}. It remains one of the most extensive works yet many SpTs rely on only a couple of measurements. Again, these were carried out on Galactic OB stars, but a metallicity dependence is noted for $\varv_{\infty}$ (\citet{Leitherer:1992} derive $\varv_{\infty} \propto Z^{0.13}$). Given the limited data for some SpT we do not attempt to correct for this dependence. However, in the case of the earliest stars (O2-3.5), values were supplemented by the later work of \citet{PrinjaCrowther:1998}, \citet{Massey:2004} and \citet{DoranCrowther:2011}, who studied further O-type stars, this time in the Magellanic Clouds.

In the case of our mass-loss rates, wind clumping is not directly accounted for in the \citet{Vink:2001} prescription. The effects of clumping have been thought to potentially scale down $\dot{M}$ by a factor of a few. However, \citet{Mokiem:2007b} argued that if a modest clumping correction is applied to empirical mass-loss rates, a better consistency is found with the \citet{Vink:2001} prescription.

The uncertainties on our adopted SpT ($\approx\pm0.5$) should be insignificant when compared to those of our stellar parameters. Any SpT will naturally show a spread in parameters, however, over the entire census, such uncertainties are expected to balance out to provide first order estimates of the feedback. To test this, the properties of a selection of O2-3 type stars in 30 Dor which had been previously modelled \citep{RiveroGonzalez:2012b, Evans:2010, Massey:2005}, were compared to those of our calibrations. Offsets up to a factor of 2 were found but typically our estimates of $Q_{\mathrm{0}}$, $L_{\mathrm{SW}}$ and $D_{\mathrm{mom}}$ were within $\sim40$\% of values obtained from tailored analyses.

\subsection{Wolf-Rayet and Of/WN stars}
\label{sec:Wolf-Rayet and Of/WN Stars}
A total of 25 W-R and 6 Of/WN stars are located within our census region. The significance of such stars to the integrated properties of young star clusters was demonstrated by \citet{CrowtherDessart:1998}, who estimated a contribution of 15\% and 40\% to the ionising and wind luminosities, respectively, out to a radius of 10\,pc from R136a1. However, their results were based upon non-line blanketed W-R models and historical OB star calibrations. 

In view of the importance of W-R stars to our census of 30 Dor, we took the following approach. We analysed a single example of each W-R and Of/WN subtype using the non-local thermodynamic equilibrium (non-LTE) atmospheric code CMFGEN \citep{HillierMiller:1998}, which was used as a template for other stars. CMFGEN solves the radiative transfer equation in the co-moving frame, subject to radiative and statistical equilibrium. Since CMFGEN does not solve the momentum equation, a density/velocity structure is required. For the supersonic part, an exponent of $\beta=1$ is adopted for the velocity law, while the subsonic velocity structure is defined using a plane-parallel TLUSTY model \citep{LanzHubeny:2003}.  

Stellar temperatures, $T_{*}$, correspond to a radius at Rosseland optical depth of 10, and are consequently somewhat higher than effective temperatures, $T_{\mathrm{eff}}$, relating to an optical depth of two thirds. Wind clumping is incorporated using a radially dependent volume filling factor, $f$, with $f_{\infty} = 0.1$ at $\varv_{\infty}$, resulting in a reduction in mass-loss rate by a factor of $f^{-0.5} \sim 3$ with respect to a smooth wind. 

CMFGEN incorporates line blanketing through a super level approximation, in which atomic levels of similar energy are grouped into a single super level that is used to compute the atmospheric structure. For WN and Of/WN subtypes, the model atoms include H\,{\sc i}, He\,{\sc i}-{\sc ii}, C\,{\sc iii}-{\sc iv}, N\,{\sc iii}-{\sc v}, O\,{\sc iii}-{\sc vi}, Si\,{\sc iv}, P\,{\sc iv}-{\sc v}, S\,{\sc iv}-{\sc vi} and Fe\,{\sc iv}-{\sc vii} while WC model atoms comprise He\,{\sc i}-{\sc ii}, C\,{\sc ii}-\,{\sc iv}, O\,{\sc ii}-{\sc vi}, Ne\,{\sc ii}-{\sc vi}, Si\,{\sc iv}, P\,{\sc iv}-{\sc v}, S\,{\sc iv}-{\sc vi}, Ar\,{\sc iii}-{\sc viii} and Fe\,{\sc iv}-{\sc viii}. Other than H, He, CNO, we adopt half-solar abundances \citep{Asplund:2009}.

In some instances, previous CMFGEN tailored analyses of individual stars have been undertaken, which are not repeated here. \citet{Crowther:2010} have studied the bright WN stars within R136, whose properties are adopted here, while we use R136a3 (BAT99-106) as a template for other WN\,5 stars. In addition, \citet{Crowther:2002} undertook a tailored analysis of the WC\,4 star BAT99-90 (Brey 74), which also serves as a template for other WC stars. 

For the remaining template stars, observational datasets are taken from a variety of sources, including IUE, HST (far-ultraviolet spectrophotometry), AAT/RGO, SSO 2.3m/DBS, MSO 1.9m/Coude (optical spectrophotometry) and VLT/UVES, VLT/FLAMES (optical spectroscopy). Inevitably, the stars used as templates suffer from our reliance upon heterogeneous spectroscopic and photometric datasets. Inferred physical and wind properties for these template emission line stars are presented in the Appendix (Table \ref{tab:W-R and Of/WN star parameters}), together with spectroscopic fits (Figures \ref{fig:plot_sk6722}-\ref{fig:plot_bat90}). Narrow-band magnitudes are used in preference to broad-band magnitudes, where available, owing to the non-zero effects of emission lines on the latter. Interstellar extinctions are calculated for an assumed $R_{V} = 4.2$ (R136 region) or 3.5 (elsewhere), with the exception of Sk-67$^{\circ}$ 22 (O2\,If*/WN5, \citealt{CrowtherWalborn:2011}) for which $R_{V}=3.1$ is adopted. Bestenlehner et al. \citetext{in prep.} have undertaken a study of luminous O, Of/WN and WN stars from the VFTS which should be preferred to results from the present study, owing to the use of homogeneous datasets and an extensive grid of CMFGEN models.

For non-template stars, stellar temperatures and wind densities were adopted from the relevant template, with other parameters (mass-loss rates, luminosities) scaled to their respective absolute (narrow-band) magnitudes. In cases where a W-R star was known to be multiple, such as R132 and R140, absolute magnitudes for the W-R component(s) were obtained from the dilution in line equivalent widths with respect to single stars, as illustrated in the Appendix (Table \ref{tab:W-R photometry}). If possible, individual $\varv_{\infty}$ were obtained from literature values \citep[e.g.][]{CrowtherSmith:1997}, or otherwise adopted from the template star. Mass-loss rates were estimated by adopting identical wind densities \citep[i.e. identical transformed radii,][]{Schmutz:1989} to the template stars.

Regarding the reliability of our method, we have compared the parameters inferred for the WN\,5h star VFTS 682 to those from a detailed analysis by \citet{Bestenlehner:2011} and obtain an ionising luminosity that is 50\% lower than the detailed study, with a wind luminosity $\sim60$\% lower. Further comparisons await the results from Bestenlehner et al. (in prep.) for other VFTS stars in common. 

Finally, following the completion of this study, in which R144 (Brey 89, BAT99-118) was selected as the template WN\,6 star, \citet{Sana:2013b} have revealed this to be a WN\,5-6+WN\,6-7 binary system. We are able to estimate the revised parameters for this system using bolometric corrections from our template WN\,5+WN\,7 stars. According to \citet{Sana:2013b}, the mass ratio of the WN\,7 secondary to WN\,5 primary is $\sim1.17$, so for an adopted $L_{\mathrm{Bol}} \propto \mathrm{M}^{1.5}$ for very massive stars \citep{Yusof:2013}, the approximate ratio of their luminosities is $1.17^{1.5} = 1.26$. Based on the systemic absolute magnitude of R144 in Table \ref{tab:W-R photometry}, the inferred absolute magnitudes of the WN\,5 primary and WN\,7 secondary are $-6.4$ and $-7.4$\,mag, respectively, from which bolometric luminosities of $10^{6.2}$ and $10^{6.3}\,\mathrm{L}_{\odot}$ are obtained. Consequently, the systemic bolometric luminosity of R144 may be up to $10^{6.6}\,\mathrm{L}_{\odot}$, i.e. 0.2\,dex higher than that obtained in Table \ref{tab:W-R and Of/WN star parameters}.

\subsection{Binary systems}
\label{sec:Binary systems}
The spectroscopic binary fraction of O-type stars in 30 Dor has been estimated to be $0.51\pm0.04$ \citep{Sana:2013a}, a considerable number of the stars in our census. In the case of these close binary systems, our photometric data will represent the combined light of both components giving an absolute magnitude for the system, $M_{V}^{\mathrm{sys}}$. For now, we have only attempted to correct for a subset of the SB2 systems, where a robust subtype and luminosity class was known for both components. This included 27 of the 48 SB2 systems identified in the census. Average absolute magnitudes from the census (Table \ref{tab:Average absolute magnitudes}) allowed $\Delta M_{V} = M_{V}^{1} - M_{V}^{2}$ to be estimated. Absolute magnitudes for each component (Table \ref{tab:SB2}) were then calculated from $\Delta M_{V}$ and $M_{V}^{\mathrm{sys}}$. Separate stellar parameters were also calculated for each star following the usual calibrations. SB1 and SB2 systems where complete classification of the secondary component was unavailable, were not corrected i.e. all light is assumed to have originated from the primary component.

Correcting W-R stars in multiple systems was even more important, given their feedback contributions. Details of the steps taken are set out in Appendix \ref{sec:W-R Binaries}.

The final stellar parameters derived for all the hot luminous stars with spectroscopy are listed in Table \ref{tab:Feedback from stars}.

\section{Stellar Census}
\label{sec:Stellar Census}
\begin{figure}
\centering
\subfigure[]{\includegraphics[width=0.5\textwidth]{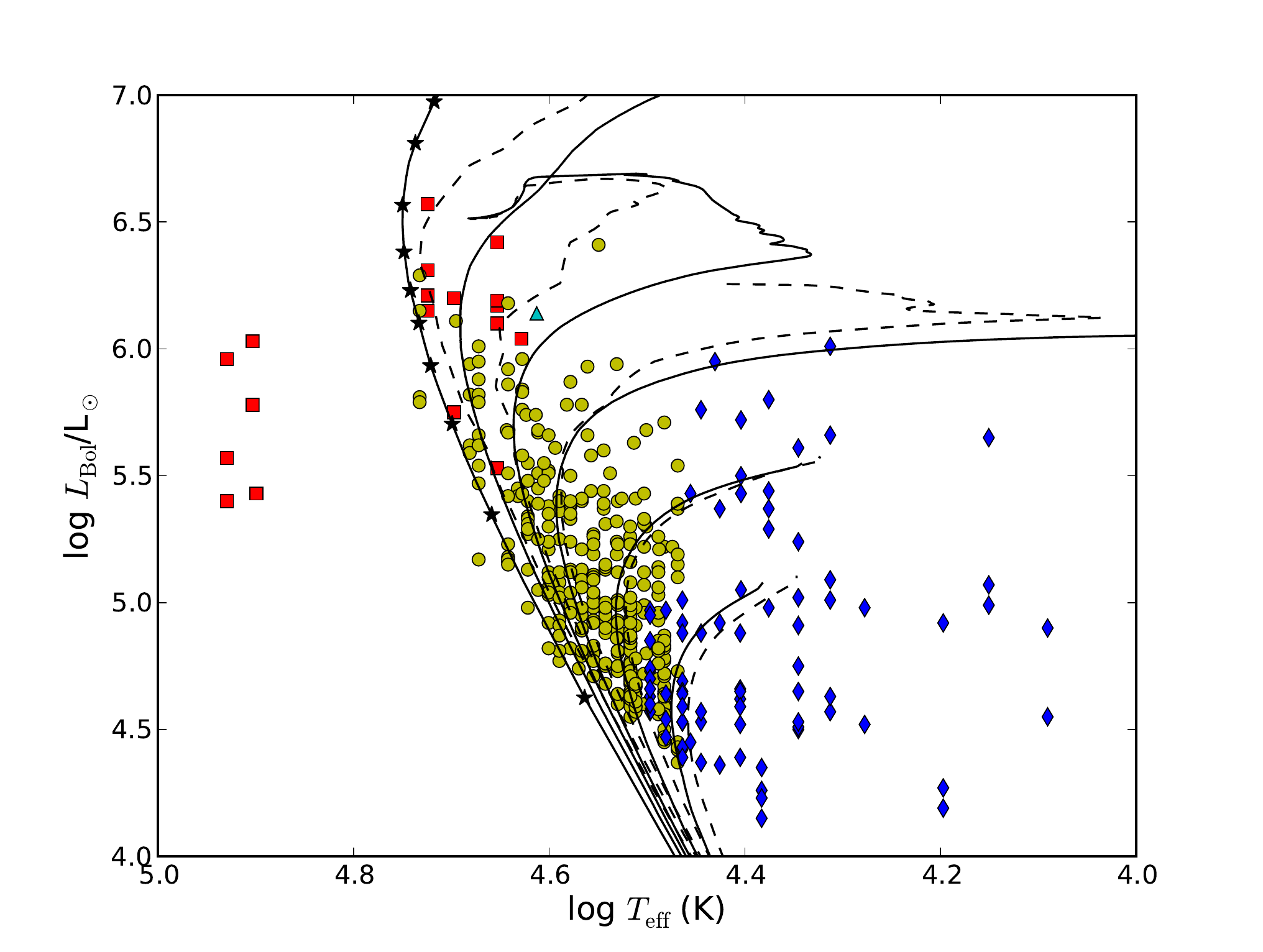}
\label{fig:HRD}}
\subfigure[]{\includegraphics[width=0.5\textwidth]{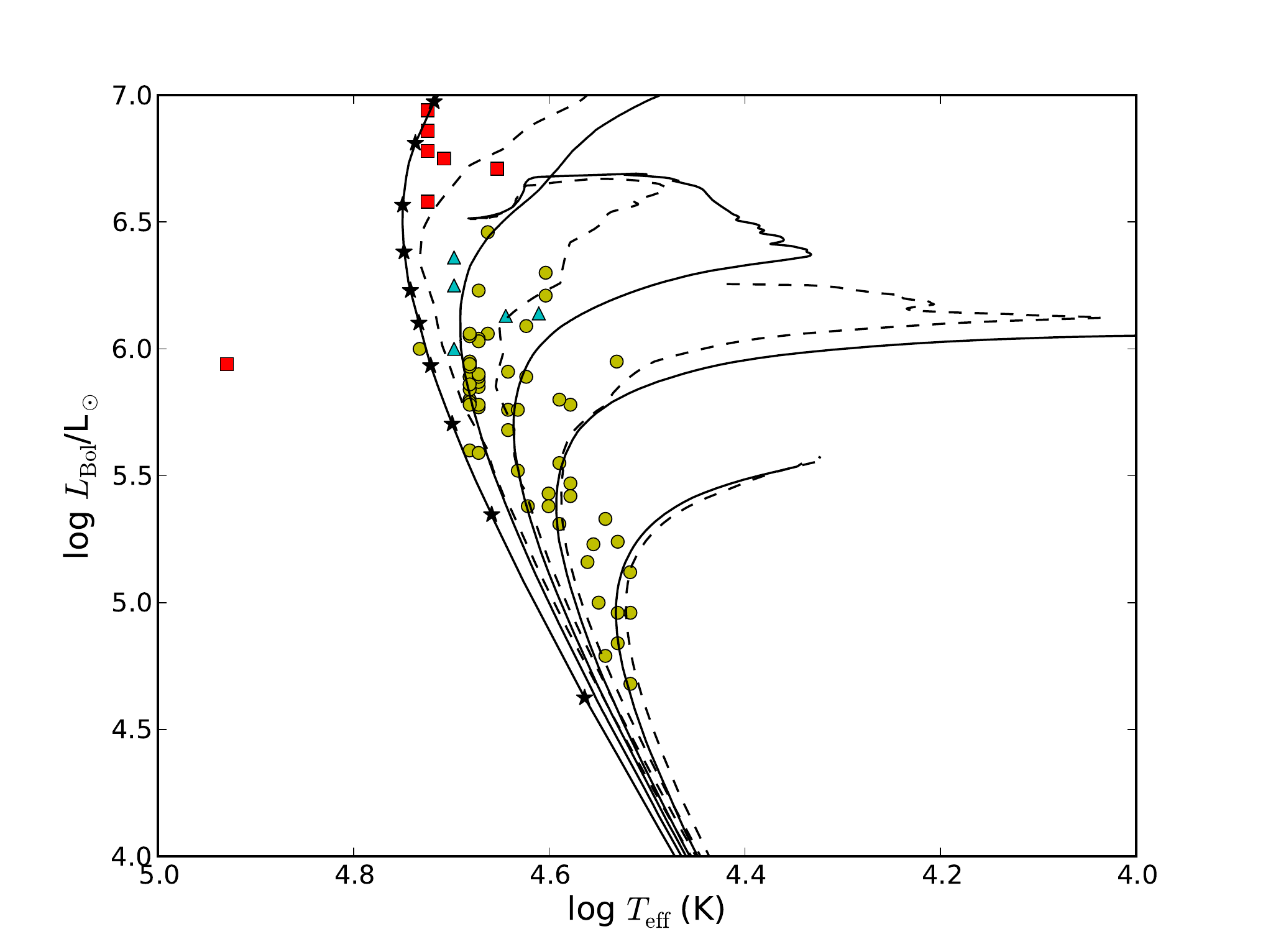}
\label{fig:HRD_R136}}
\caption{ An H-R diagram of all the spectroscopically confirmed hot luminous stars in the MEDUSA region (a) and R136 region (b). W-R stars (red squares), Of/WN stars (cyan triangles), O-type stars (yellow circles) and B-type stars (blue diamonds). Isochrones at 1, 2, 3, \& 5\,Myrs are overlaid in both cases, with 8\,Myr also used for the MEDUSA region. The isochrones are based on rotating ($\varv_{\mathrm{rot}}=300\,\mathrm{km\,s}^{-1}$, dashed line) and non-rotating ($\varv_{\mathrm{rot}}=0\,\mathrm{km\,s}^{-1}$, solid line) evolutionary models of \citet{Brott:2011} and K\"{o}hler et al. \citetext{in prep.}. The ZAMS positions for stars with $M_{\mathrm{init}}$ = 20, 40, 60, 80, 100, 120, 150, 200, 300 \& 400\,M$_{\odot}$ are marked by black stars and joined with a solid line.}
\label{fig:HRDiagram}
\end{figure}

On the basis of our inferred stellar parameters, estimates of age and mass could be made. Figure \ref{fig:HRD} presents an H-R diagram of all the hot luminous stars in the MEDUSA region. As discussed in Section \ref{sec:Uncertainties in calibrations} uncertainties on $T_{\mathrm{eff}}$ are likely to be about $\pm2000$\,K while $\log L_{\mathrm{Bol}}/\mathrm{L}_{\odot}$ is accurate to $\pm0.2$\,dex. The zero-age main sequence (ZAMS) positions are based on the contemporary evolutionary models of \citet{Brott:2011} and K\"{o}hler et al. \citetext{in prep.}. The accompanying isochrones are overlaid for rotating ($\varv_{\mathrm{rot}}=300\,\mathrm{km\,s}^{-1}$) and non-rotating models, spanning ages from 0 to 8\,Myr. The hot stars to the far left of the diagram are the evolved W-R stars. They are not covered by the isochrones as the associated evolutionary tracks only modelled as far as the terminal-age main sequence. However, the isochrones do still reveal a large age spread across more than 8\,Myr. Indeed, \citet{WalbornBlades:1997} found several distinct stellar regions of different ages within 30 Dor, showing a possibility for triggered star formation. 

Identifying individual age groups isn't possible in more distant star forming regions but we attempt it for the massive R136 cluster. Figure \ref{fig:HRD_R136} shows a similar H-R diagram for the R136 region. Its stars are typically more massive, showing a younger age, separate from the rest of 30 Dor. The isochrones suggest the most massive stars to be $\sim1-2$\,Myr old, with an older age being favoured further down the main sequence. This is largely consistent with the work of \citet{MasseyHunter:1998} who found ages $\sim2$\,Myr. \citet{Sabbi:2012} determined a similar 1-2\,Myr age for R136 but also identified an older ($\sim2-5$\,Myr) region extending $\sim5.4$\,pc to the north-east. As our R136 region extends to 5\,pc, there is a high possibility of contamination from older non-coeval stars.

To estimate stellar masses we applied a mass-luminosity relation of $L_{\mathrm{Bol}} \propto M^{1.75}$ for all O-type dwarfs \citep{Yusof:2013}. However, this relation strictly only applies to ZAMS stars within a mass range of $\sim60-150$\,M$_{\odot}$ and as a result would underestimate the masses for lower luminosity dwarfs. In the case of the O-type giants and supergiants, we adopted the $\log\,g$-SpT calibrations of \citet{Martins:2005}. A mass was then calculated using the stellar radius produced from the parameters of our own calibrations. A similar approach was used for the B-type stars, only average $\log\,g$ values were taken for each SpT from the works of \citet{Trundle:2007} and \citet{Hunter:2007}. Rough mass estimates of the WC and WN/C stars were obtained through the calibration of \citet{SchaererMaeder:1992} but in case of the hydrogen burning WN and Of/WN stars, we returned to the H-burning mass-luminosity relation.

\begin{figure}
\includegraphics[width=0.5\textwidth]{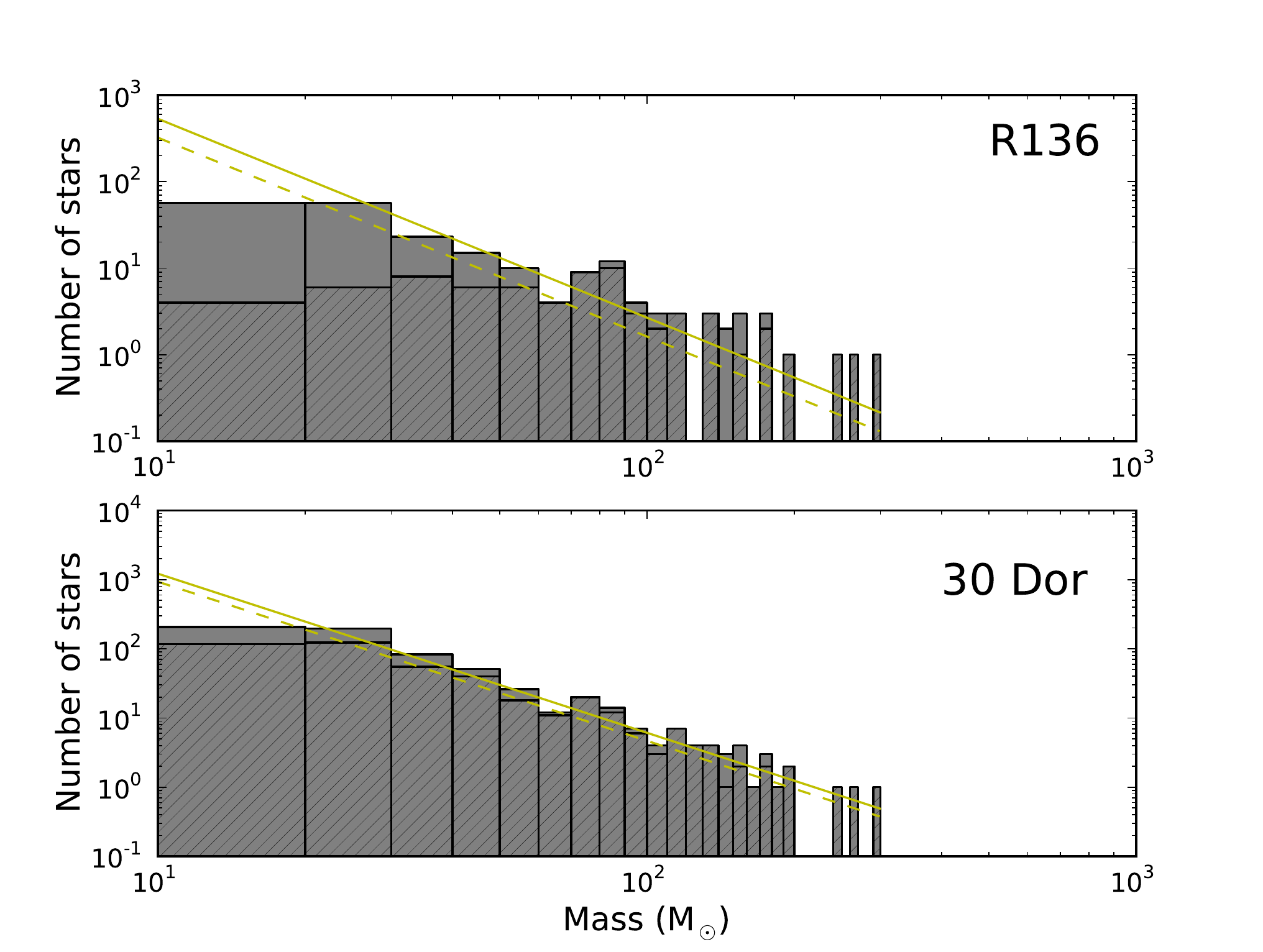}
\caption{The present day mass function of all hot luminous stars classified spectroscopically and photometrically (Sp+Ph) in the R136 region (top) and in 30 Dor (bottom). Hatched regions indicate where the mass function is based solely on stars classified spectroscopically (Sp). A Salpeter $\alpha=-2.3$ slope is overlaid with a yellow line (solid = Sp+Ph, dashed = Sp), for which the integrated mass ($>20$\,M$_{\odot}$) is consistent with that of each region.}
\label{fig:PDMF}
\end{figure}

Our photometry indicates a further 141 hot luminous stars within the R136 region and 222 stars within 30 Dor (Section \ref{sec:Correcting for unclassified stars}). We therefore account for these stars as well when deriving the mass function for each region. In this case, the SpTs for the census were obtained through a combination of spectroscopy and photometry (Sp+Ph), as opposed to relying solely on stars that were spectroscopically classified (Sp). Our inferred mass functions for 30 Dor and the R136 region are plotted in Figure \ref{fig:PDMF}. There is consistency with the \citet{Salpeter:1955} $\alpha=-2.3$ slope but this notably deviates at $\sim20$\,M$_{\odot}$, due to incomplete photometry and spectroscopy. Assuming the Sp+Ph stellar census to be complete above 20\,M$_{\odot}$ and adopting a \citet{Kroupa:2001} initial mass function (IMF), the total stellar mass for each region was obtained: $M_{\mathrm{30 Dor}}\sim1.1\times10^{5}$\,M$_{\odot}$ and $M_{\mathrm{R136}}\sim5.0\times10^{4}$\,M$_{\odot}$, the latter being within $\sim10$\% of the R136 mass estimated by \citet{Hunter:1995} when a similar Kroupa IMF is adopted.

\section{Integrated Stellar Feedback}
\label{sec:Integrated Stellar Feedback}
For an estimate of the total stellar feedback of 30 Dor, the contributions of individual stars were combined to produce the cumulative plots presented in Figures \ref{fig:Ionising_Flux} - \ref{fig:Kinetic_Energy}. While the R136 region showed a lower completeness compared to 30 Dor, it still contained a high proportion of the brightest W-R, Of/WN and O-type stars, and is therefore considered separately. In addition, feedback estimates are also made for the stars lacking spectroscopy (Section \ref{sec:Correcting for unclassified stars}). The integrated feedback at different projected radial distances can be found in Table \ref{tab:Breakdown of feedback at different radii} along with a breakdown of the contributions from the different SpTs.

\begin{table*}
\begin{center}
\caption{The cumulative properties of hot luminous stars in 30 Dor with increasing projected radius, $r_{\mathrm{d}}$, given in (pc), from R136a1. The top half of the table gives the more robust values, based solely on the number of stars (N) classified spectroscopically (Sp). The bottom half combines both spectroscopically and photometrically classified stars (Sp+Ph). Note the contribution from the W-R and Of/WN stars remains unchanged as none of these stars were taken to lack spectroscopy. $\log L_{\mathrm{Bol}}$ is given in $(\mathrm{L}_{\odot})$, $Q_{\mathrm{0}}$ in $(\times10^{49}\,\mathrm{ph}\,\mathrm{s}^{-1})$, $L_{\mathrm{SW}}$ in $(\times10^{37}\,\mathrm{erg}\,\mathrm{s}
^{-1})$ and $D_{\mathrm{mom}}$ in $(\times10^{36}\,\mathrm{cgs\,units})$.}
\label{tab:Breakdown of feedback at different radii}
\begin{tabular} {l | @{\hspace{2mm}} c @{\hspace{2mm}} r @{\hspace{2mm}} r @{\hspace{2mm}} r @{\hspace{2mm}} r | @{\hspace{2mm}} c @{\hspace{2mm}} r  @{\hspace{2mm}} r @{\hspace{2mm}} r @{\hspace{2mm}} r | @{\hspace{2mm}} c @{\hspace{2mm}} r  @{\hspace{2mm}} r @{\hspace{2mm}} r @{\hspace{2mm}} r} 
\hline\hline
Region & \multicolumn{5}{c}{OB Stars} & \multicolumn{5}{c}{W-R and Of/WN Stars} & \multicolumn{5}{c}{ Grand Total} \\

$<r_{\mathrm{d}}$ & N & $\log L_{\mathrm{Bol}}^{\mathrm{Sp}}$ & $Q_{\mathrm{0}}^{\mathrm{Sp}}$  & $L_{\mathrm{SW}}^{\mathrm{Sp}}$ & $D_{\mathrm{mom}}^{\mathrm{Sp}}$ & N & $\log L_{\mathrm{Bol}}^{\mathrm{Sp}}$ & $Q_{\mathrm{0}}^{\mathrm{Sp}}$  & $L_{\mathrm{SW}}^{\mathrm{Sp}}$ & $D_{\mathrm{mom}}^{\mathrm{Sp}}$ & N & $\log L_{\mathrm{Bol}}^{\mathrm{Sp}}$ & $Q_{\mathrm{0}}^{\mathrm{Sp}}$ & $L_{\mathrm{SW}}^{\mathrm{Sp}}$
& $D_{\mathrm{mom}}^{\mathrm{Sp}}$ \\

\hline

5 & 60 & 7.61 & 242 & 37.6 &  2.72 & 12 & 7.66  & 334 & 63.9 & 7.69 & 72 & 7.93 & 576 & 101 &  10.4 \\
10 & 120 & 7.77 & 319 & 48.6 & 3.86   & 13 & 7.67  & 343 & 65.8 & 8.03 & 133 & 8.02 & 662 & 114 & 11.9 \\
20 & 191 & 7.88 & 397 & 59.0 &  4.83   & 19 & 7.74 & 398 & 85.3 &  9.98    & 210 & 8.12 & 795 & 144  & 14.8 \\
150 & 469 & 8.07  & 554 & 75.6 & 6.22  & 31 & 7.84 & 503 & 122 & 13.7   & 500 & 8.27 & 1056 & 197 & 19.9 \\

$<r_{\mathrm{d}}$ & N & $\log L_{\mathrm{Bol}}^{\mathrm{Sp+Ph}}$ &  $Q_{\mathrm{0}}^{\mathrm{Sp+Ph}}$  & $L_{\mathrm{SW}}^{\mathrm{Sp+Ph}}$ & $D_{\mathrm{mom}}^{\mathrm{Sp+Ph}}$ & N & $\log L_{\mathrm{Bol}}^{\mathrm{Sp+Ph}}$ & $Q_{\mathrm{0}}^{\mathrm{Sp+Ph}}$  & $L_{\mathrm{SW}}^{\mathrm{Sp+Ph}}$ & $D_{\mathrm{mom}}^{\mathrm{Sp+Ph}}$ & N & $\log L_{\mathrm{Bol}}^{\mathrm{Sp+Ph}}$ & $Q_{\mathrm{0}}^{\mathrm{Sp+Ph}}$ & $L_{\mathrm{SW}}^{\mathrm{Sp+Ph}}$ &  $D_{\mathrm{mom}}^{\mathrm{Sp+Ph}}$ \\

\hline

5 & 201 & 7.85 & 388 & 60.3 & 4.54   & 12 & 7.66  & 334 & 63.9 & 7.69   & 213 & 8.06 & 723 & 124 &  12.2 \\
10 & 284 & 7.96 & 474 & 71.9 & 5.74   & 13 & 7.67  & 343 & 65.8 & 8.03   & 297 & 8.14 & 786 & 138 & 13.8 \\
20 & 374 & 8.04 & 555 & 82.4 &  6.72   & 19 & 7.74 & 398 & 85.3 &  9.98   & 393 & 8.22 & 952 & 168  & 16.7  \\
150 & 691 & 8.20 & 734 & 102 &  8.49   & 31 & 7.84 & 503 & 122 & 13.7   & 722 & 8.36 & 1237 & 224 & 22.2  \\
\hline
\end{tabular}
\end{center}
\end{table*}

\subsection{Ionising photon luminosity}
\label{sec:Ionising photon luminosity}
\begin{figure}
\centering
\subfigure[]{\includegraphics[width=0.5\textwidth]{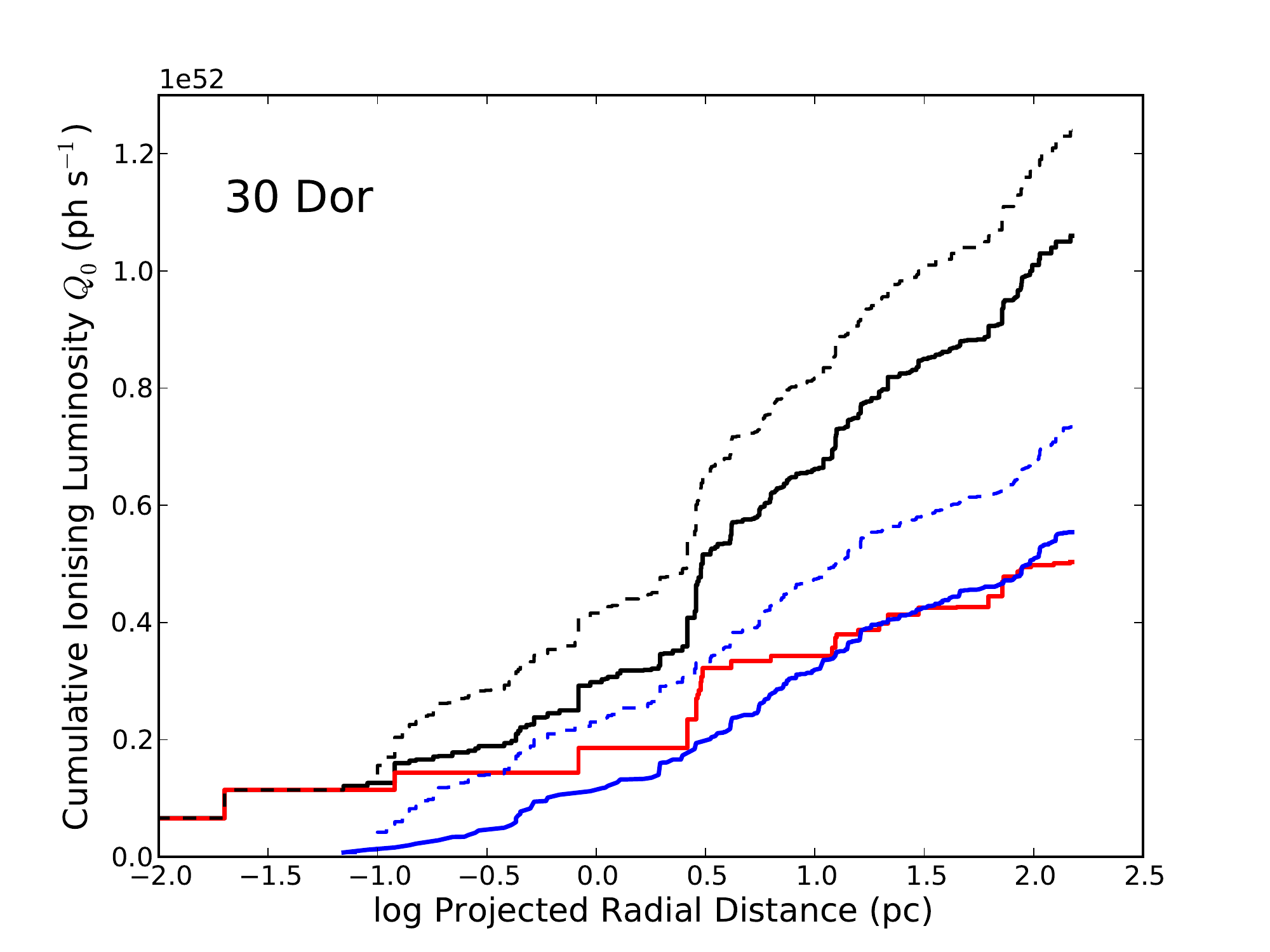}
\label{fig:10min_Ionising_Flux}}
\subfigure[]{\includegraphics[width=0.5\textwidth]{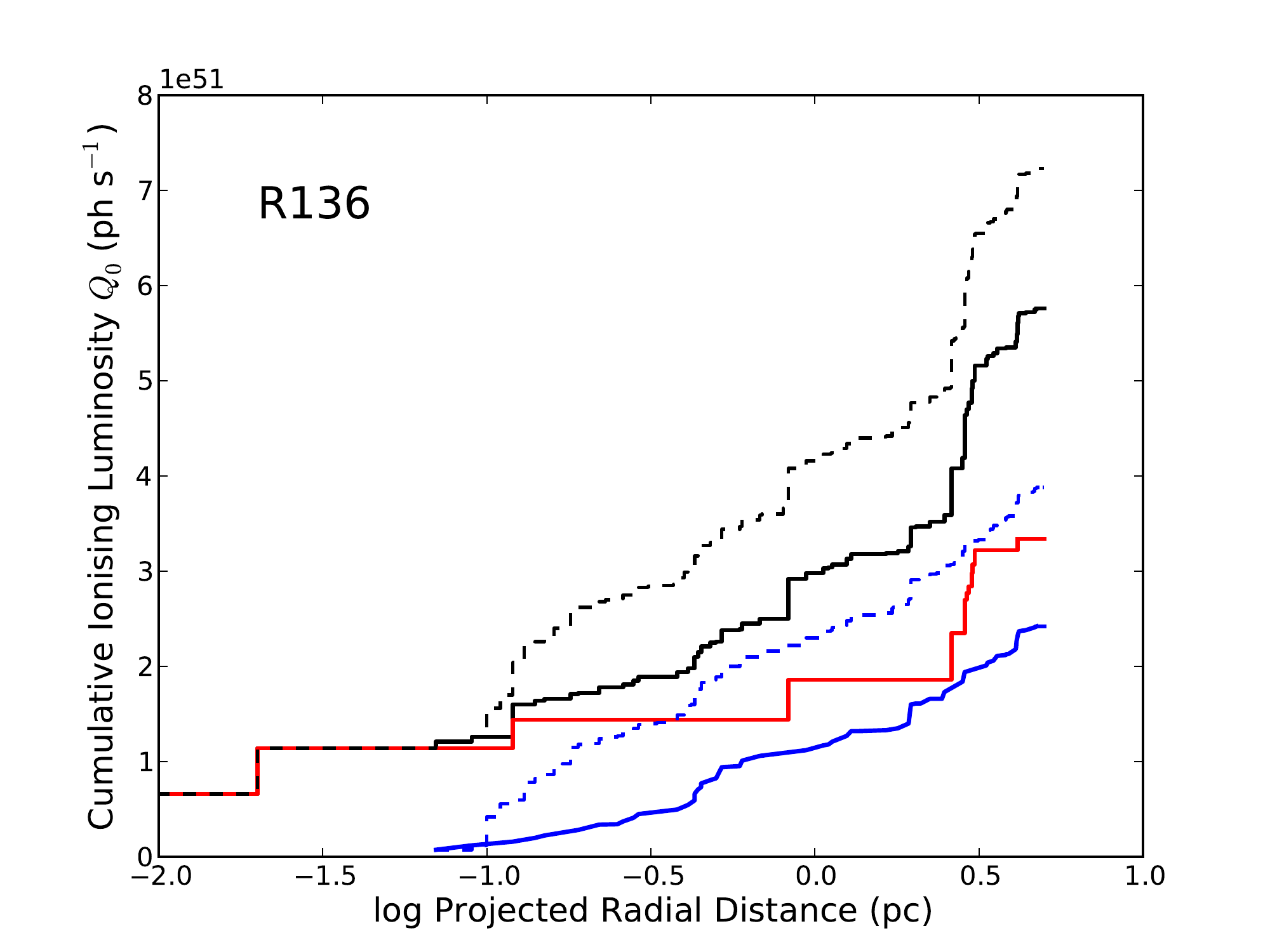}
\label{fig:0_0p33min_Ionising_Flux}}
\caption{Cumulative ionising luminosities of the hot luminous stars in 30 Dor (a) and the R136 region (b). Feedback from all stars (black), W-R and Of/WN stars (red) and OB stars (blue). Solid lines include the feedback from stars which were spectroscopically classified as hot and luminous (Sp) while dashed lines include feedback from photometrically classified stars (Sp+Ph).}
\label{fig:Ionising_Flux}
\end{figure}

Figure \ref{fig:10min_Ionising_Flux} shows the total ionising photon luminosity from stars with spectroscopy (solid line) to be $Q^{\mathrm{Sp}}_{\mathrm{0}}$ $\approx 1.0\times10^{52}$ ph s$^{-1}$. We see a sharp increase within the inner 20\,pc (80\,arcsec) of 30 Dor from which $\sim75$\% of the ionising luminosity is produced. This relates to the large number of hot massive stars in the vicinity of R136. The increases reflect the contributions from W-R and Of/WN stars. In spite of only 31 such stars being within 30 Dor, they contribute an equivalent output to the 469 OB stars present. Taking a closer look at the feedback from the R136 region (see Figure \ref{fig:0_0p33min_Ionising_Flux}), we see that it is analogous to 30 Dor as a whole. The ionising luminosity is shared roughly equally between the W-R \& Of/WN stars and OB stars. An exception to this is seen at the very core of the cluster where the four WN\,5h stars \citep{Crowther:2010} dominate the luminosity of the cluster.

The dashed lines represent the total ionising luminosity when accounting for stars which had their SpTs estimated from photometric data ($Q^{\mathrm{Sp+Ph}}_{\mathrm{0}}$). The contribution from these stars increases the total ionising luminosity by about 25\% and 15\%, in the R136 region and 30 Dor, respectively. While these values are less robust, it shows that the remaining hot luminous stars in 30 Dor without spectroscopy (predominantly in R136) should not be neglected. However, it also indicates that the stars with known SpT likely dominate the total output so that any uncertainties due to stars lacking spectroscopy should not be too severe.

\citet{CrowtherDessart:1998} made earlier estimates of the ionising luminosity of 30 Dor using this same method of summing the contribution of the individual stars. Their study extended out to $r_{\mathrm{d}}=10$\,pc and found $Q_{\mathrm{0}}=4.2\times10^{51}\,\mathrm{ph\,s}^{-1}$. They also made SpT assumptions for some stars and so we compare this to our $Q_{\mathrm{0}}^{\mathrm{Sp+Ph}}$ estimate and find our revised value to be almost 90\% larger (see Table \ref{tab:Breakdown of feedback at different radii}). The offset will largely be due to the higher $T_{\mathrm{eff}}$ calibrations and updated W-R star models used, along with the new photometric and spectroscopic coverage of the stars within the region.

Table \ref{tab:Top 10} lists the ten stars with the highest ionising luminosities. These are all found to be W-R or Of/WN stars, except for Mk 42 (O2\,If*), and are mainly located within the R136 region. We see that these 10 stars alone produce 28\% of $Q^{\mathrm{Sp+Ph}}_{\mathrm{0}}$. It is accepted that some less luminous stars will have been missed from the census but their impact on the integrated luminosity will be negligible in comparison to these W-R stars. For example, Figure \ref{fig:VFTS_CMD} showed 38 subluminous VFTS O-type stars being removed from the census by our selection criteria. Their combined ionising luminosity is estimated to be $<0.5$\% of $Q^{\mathrm{Sp}}_{\mathrm{0}}$, which is below the output from B-type stars in the census.

\subsection{Stellar wind luminosity and momentum}
\label{sec:Stellar wind luminosity and momentum}
\begin{figure}
\centering
\subfigure[]{\includegraphics[width=0.5\textwidth]{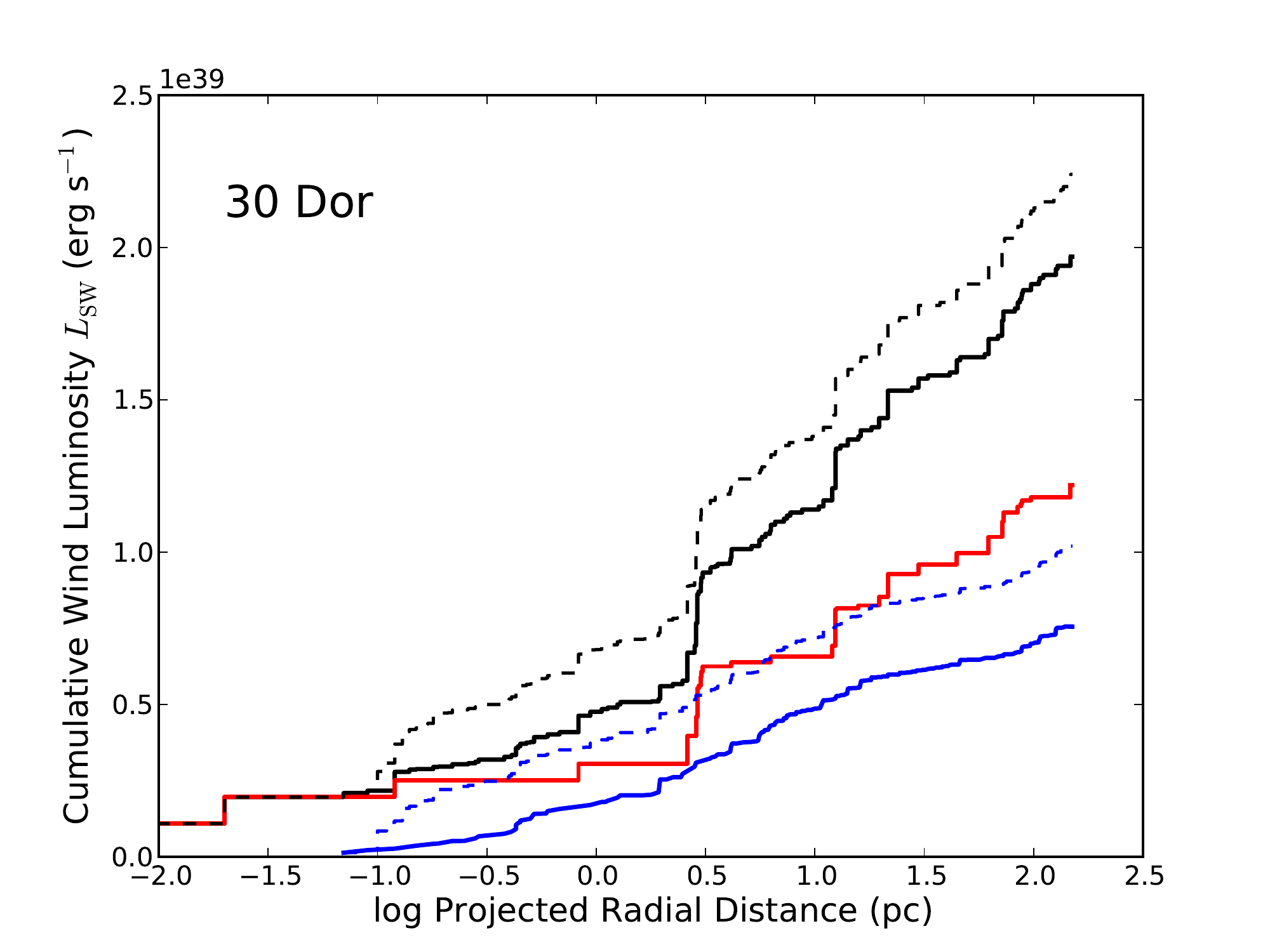}
\label{fig:10min_Kinetic_Energy}}
\subfigure[]{\includegraphics[width=0.5\textwidth]{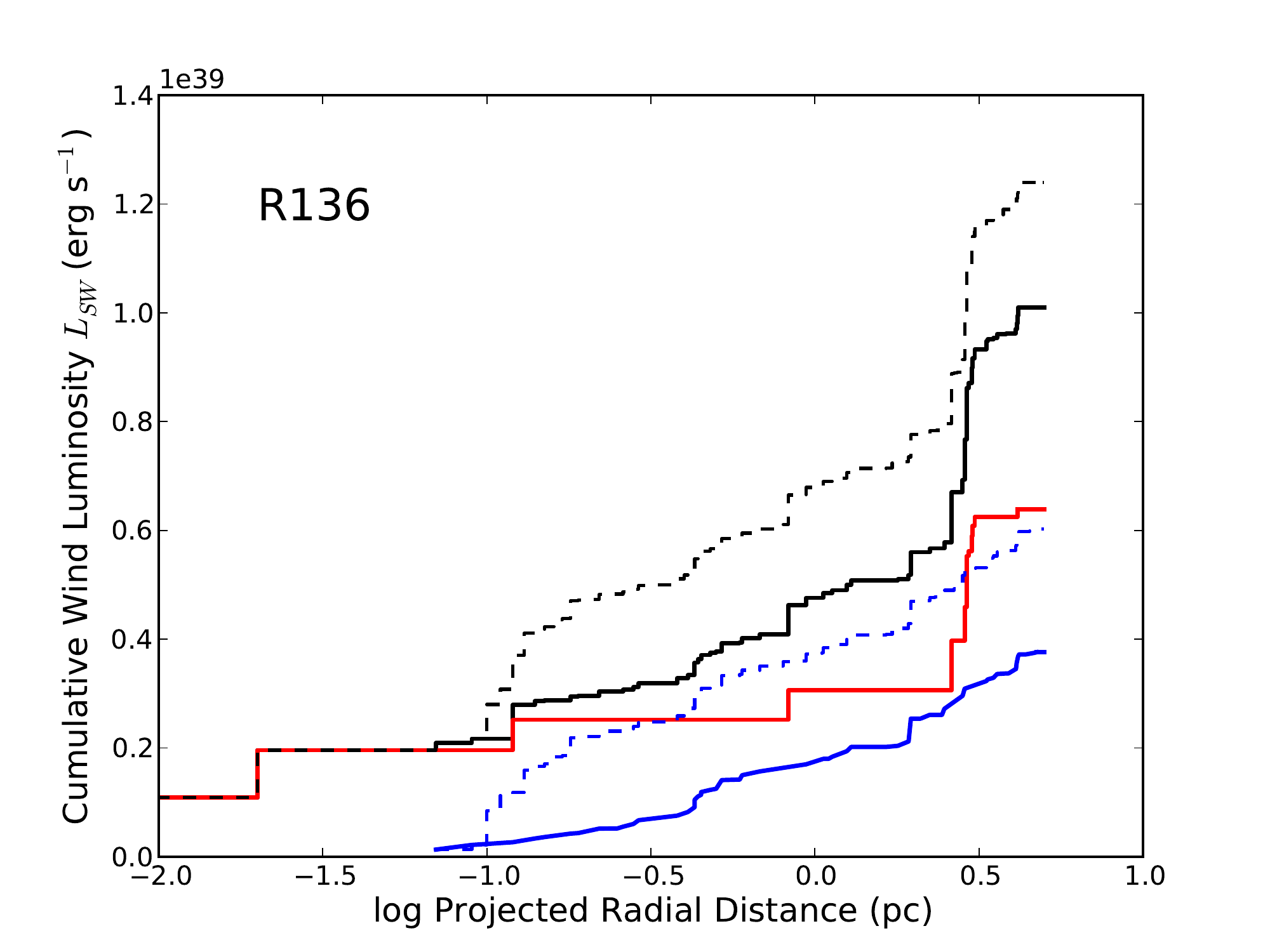}
\label{fig:0_0p33min_Kinetic_Energy}}
\caption{Cumulative stellar wind luminosities of the hot luminous stars in 30 Dor (a) and the R136 region (b). Lines are as in Figure \ref{fig:Ionising_Flux}.}
\label{fig:Kinetic_Energy}
\end{figure}

The total stellar wind luminosity of 30 Dor was estimated at $L_{\mathrm{SW}}^{\mathrm{Sp}}$ $\approx 1.97\times10^{39}$ erg s$^{-1}$. Figure \ref{fig:10min_Kinetic_Energy} shows the radial cumulative wind luminosity to behave similarly to the ionising luminosity: rapidly increasing within the inner 20\,pc from which $\sim75$\% of the total wind luminosity is produced. W-R and Of/WN stars dominate the mechanical output, providing about two thirds of the total. Their higher contribution arises from their high mass-loss rates and fast winds. This behaviour is repeated in the R136 region (see Figure \ref{fig:0_0p33min_Kinetic_Energy}), where the four WN\,5h stars again provide significant mechanical feedback to the cluster. The dashed lines again represent our combined spectroscopically and photometrically classified sample. $L_{\mathrm{SW}}^{\mathrm{Sp+Ph}}$ values suggest wind luminosities in the R136 region and within 30 Dor could increase roughly a further 25\% and 15\%, respectively.

Again, we can compare to the work of \citet{CrowtherDessart:1998} who found $L_{\mathrm{SW}}=1.61\times10^{39}\,\mathrm{erg\,s}^{-1}$ out to $r_{\mathrm{d}}=10$\,pc, i.e. about 15\% higher than our $L_{\mathrm{SW}}^{\mathrm{Sp+Ph}}$ estimate (see Table \ref{tab:Breakdown of feedback at different radii}). The offset results again from the updated stellar calibrations and mass-loss rates that are corrected for wind clumping.

Table \ref{tab:Top 10} also lists the ten stars with the highest wind luminosities. These are all W-R stars, most of which overlap with the dominant ionising stars and are also located within the R136 region. The importance of the W-R stars is echoed here as these 10 stars alone contribute 35\% of $L_{\mathrm{SW}}^{\mathrm{Sp+Ph}}$. Once again, contributions of subluminous stars become negligible, such as the VFTS O-type stars excluded from our census whose combined wind luminosity is $<0.05$\% of $L_{\mathrm{SW}}^{\mathrm{Sp}}$ and just a few \% of the B-type star output.

The integrated modified wind momenta, $D_{\mathrm{mom}}$, behave very similarly to $L_{\mathrm{SW}}$. Table \ref{tab:Breakdown of feedback at different radii} shows how the W-R and Of/WN stars are again the key contributors while the contribution from OB stars is slightly reduced, likely due to the lower dependence that their high $\varv_{\infty}$ play on $D_{\mathrm{mom}}$. 

We note that the contribution of the W-R stars to the stellar bolometric luminosity is different to the stellar feedback. OB stars contribute $\sim70$\% of the bolometric luminosity of 30 Dor (Table \ref{tab:Breakdown of feedback at different radii}). W-R and Of/WN stars, while individually luminous compared to many of the OB stars, contribute $\sim30$\% of the integrated luminosity of the region.

\subsection{Feedback from additional sources}
\label{sec:Feedback from additional sources}
While the census focusses on selecting hot luminous stars, our estimates are potentially influenced by additional sources of feedback. Embedded stars, shock emission and sources outside the census could all contribute. Nevertheless, here we argue that they will only have a limited impact on our integrated values and that the W-R stars still have the greatest influence.

First we address the impact of stars beyond our selected $r_{\mathrm{d}}\leq150$\,pc region of 30 Dor. Figures \ref{fig:10min_Ionising_Flux} \& \ref{fig:10min_Kinetic_Energy} clearly show the cumulative effect of hot luminous stars to drop off at large distances from R136. There are no significant young clusters close to our census border to provide extra feedback. Indeed, even if we were to extend our census to $r_{\mathrm{d}}\leq225$\,pc (more than doubling the projected area on the sky), we would only expect to enclose a further $\sim300$ candidate hot luminous stars (25\% of the census). Spectral coverage is poor beyond $r_{\mathrm{d}}\sim150$\,pc but only one W-R star is known, BAT99-85/VFTS 002, WC\,4+(O6-6.5\,III).

The isochrones in Figure \ref{fig:HRD} show the age of massive stars within 30 Dor to extend beyond 8\,Myr, with ages of 20-25\,Myr derived for the Hodge 301 cluster \citep{GrebelChu:2000}. This clearly extends well beyond the epoch of the first SNe that would have occurred in the region and is demonstrated by the N157B supernova remnant \citep{Chu:1992}. A complex structure of gas filaments and bubbles is generated by stellar winds and SNe, many of which coincide with bright X-ray sources \citep{Townsley:2006}. \citet{ChuKennicutt:1994}. identify a number of gas shells expanding at velocities up to 200\,km\,s$^{-1}$. Despite the evidence for shocks, the ionising output from hot stars seems capable of keeping the shells ionised \citep{ChuKennicutt:1994}. 

A further lack of highly ionised species, such as [O \,{\sc iv}] and [Ne\,{\sc v}] leaves photoionisation as the dominant process in the energy budget \citep{Indebetouw:2009, Pellegrini:2010}. As neither optical nor IR nebular emission lines show strong correlation between high excitation regions and listed embedded stars, the ionisation structure of 30 Dor primarily arises from the optically known hot stars.

When considering mechanical feedback, however, SNe do become significant. While some of the smaller shells are thought to be wind-blown bubbles, the larger and faster moving shells are only likely to be carved by the energy input of SNe \citep{ChuKennicutt:1994}. An exception would be the shell surrounding R136. Taking our derived wind luminosity for the R136 region with a pre-SNe age of $\sim2$\,Myr, the kinetic energy generated by the stellar winds could amount to a few $10^{52}\,\mathrm{erg}$, sufficiently high to drive the bubble \citep{ChuKennicutt:1994}. However, determining a combined stellar/SNe energy budget for 30 Dor, awaits a better understanding of its star formation history (SFH).

\subsection{R136 comparison with population synthesis models}
\label{sec:Comparison with population synthesis models}

Population synthesis codes seek to mimic the combined properties of a stellar system of a given age and mass. With only a few initial parameters being required, expected observable properties of extragalactic star forming regions can be predicted. However, there are limited targets available to test the reliability of such models. R136 is one of the few resolved massive star clusters which allows an empirical study of the stars to be compared to synthetic predictions. In particular, its high mass ($M_{\rm{cl}}=5-10\times10^{4}$\,M$_{\odot}$, \citealt{Hunter:1995, Andersen:2009}) and young age ($1-2$\,Myr, Section \ref{sec:Stellar Census}), means that the upper MF is well populated and stochastic effects are minimal \citep{Cervino:2002}.

A synthetic model of an R136-like cluster was generated by the population synthesis code, \emph{Starburst99} \citep{Leitherer:1999}. An instantaneous burst model with total mass $10^6$\,M$_{\odot}$ and metallicity $Z=0.4\,Z_{\odot}$\footnote{This is the most appropriate LMC metallicity provided by \emph{Starburst99}, and while different to $Z=0.5\,Z_{\odot}$ adopted for our stellar wind calibrations, this difference should have minor effects when comparing results.} was adopted, which was scaled to $M_{\rm{cl}}=5.5\times10^{4}$\,M$_{\odot}$\footnote{This mass was favoured as it was derived from lower mass stars $2.8-15$\,M$_{\odot}$ \citep{Hunter:1995} and then scaled to a Kroupa IMF.}, in order to mimic R136. A \citet{Kroupa:2001} IMF with $M_{\rm{up}}=100$\,M$_{\odot}$ was selected along with the ``high mass-loss rate'' Geneva evolutionary tracks of \citet{Meynet:1994}. The theoretical wind model was chosen, meaning that the wind luminosity produced was based on mass-loss rates and terminal wind velocities from Equations 1 \& 2 of \citet{Leitherer:1992}, respectively. Ionising luminosities used the calibration from \citet{Smith:2002}.

\begin{figure}
\centering
\subfigure[]{\includegraphics[width=0.5\textwidth]{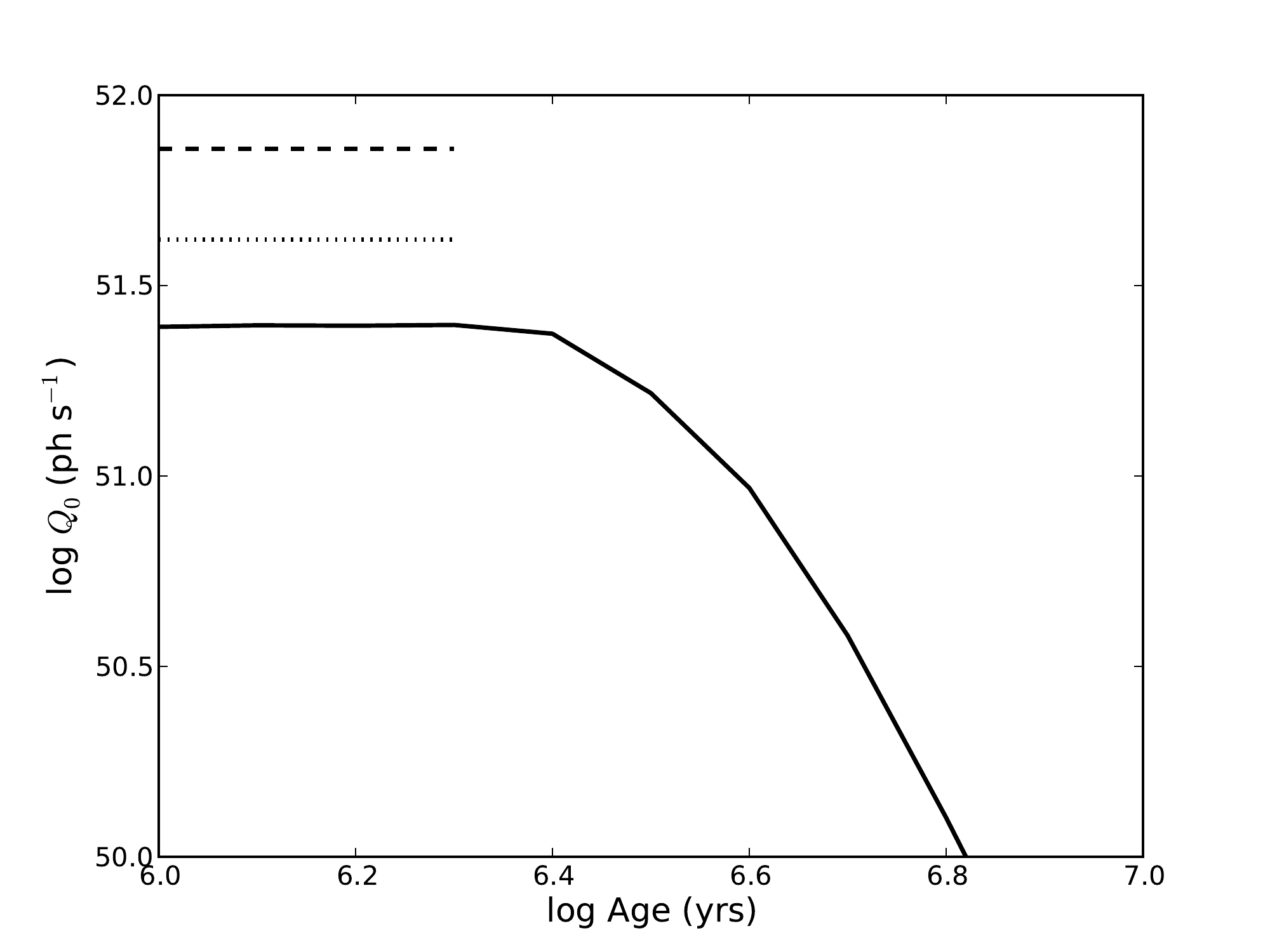}
\label{fig:Q0_S99_vs_Empirical}}
\subfigure[]{\includegraphics[width=0.5\textwidth]{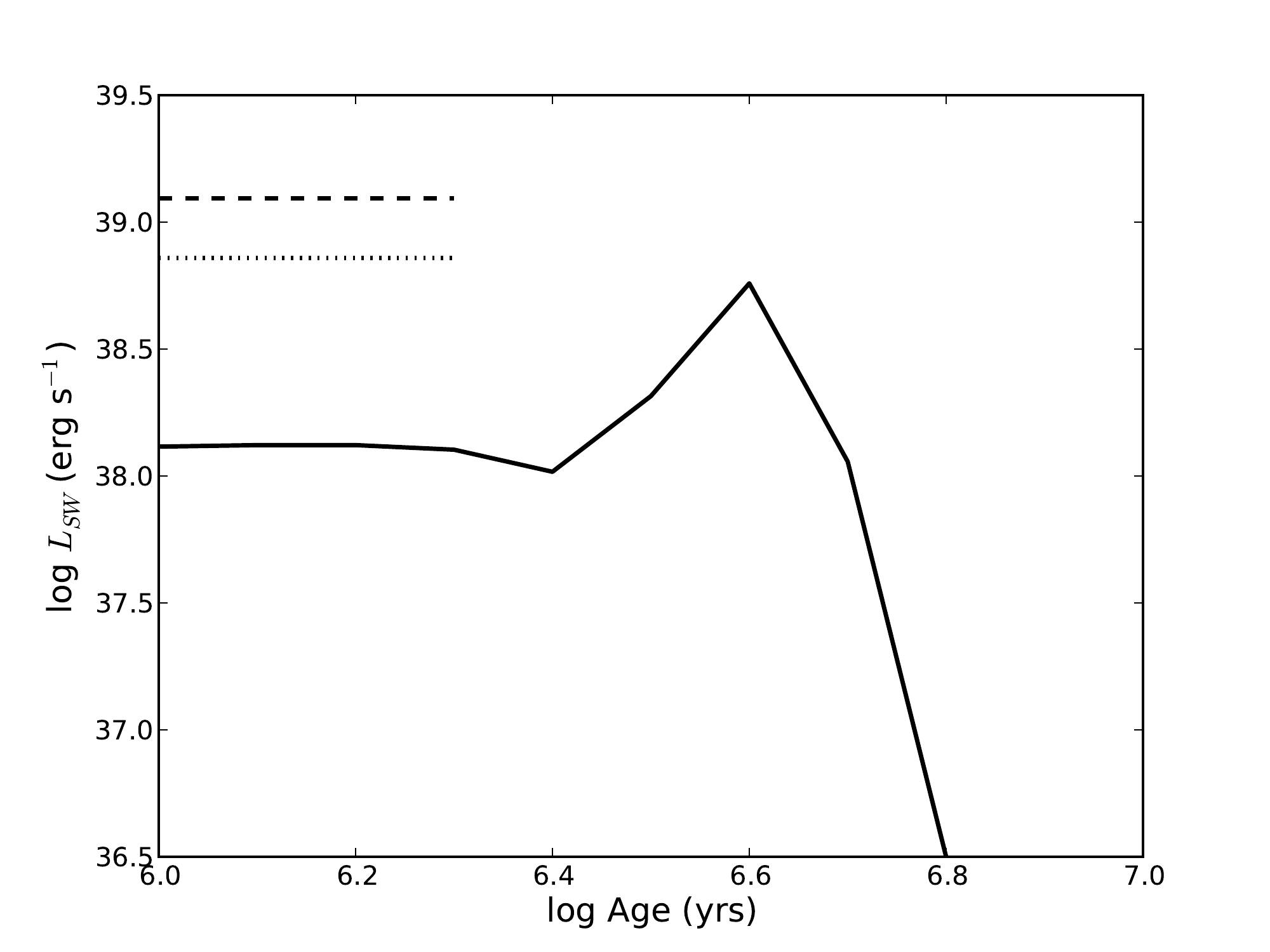}
\label{fig:KE_S99_vs_Empirical}}
\caption{The integrated ionising luminosity (a) and stellar wind luminosity (b) of the R136 region. Predictions from \emph{Starburst99} are plotted with a solid line. Values from our census, with and without $M_{\mathrm{init}}>100$\,M$_{\odot}$ stars, are plotted in dashed and dotted lines, respectively.}
\label{fig:S99_vs_Empirical}
\end{figure}

Our empirical feedback for R136 is compared with \emph{Starburst99} in Figure \ref{fig:S99_vs_Empirical}. Here we compare the combined spectroscopic and photometric values ($Q_{\mathrm{0}}^{\mathrm{Sp+Ph}}$ \& $L_{\mathrm{SW}}^{\mathrm{Sp+Ph}}$). The \emph{Starburst99} predictions only represent contributions from stars (i.e. excluding supernovae). Their changing values, over time, reflect the different SpTs that contribute to the feedback. Comparisons to \emph{Starburst99} are made between 1-2\,Myr, given the age obtained by \citet{MasseyHunter:1998} for the most luminous members of R136 (see also Figure \ref{fig:HRD_R136}). We see that in both cases, our empirical results exceed the predictions of \emph{Starburst99}. The predicted ionising luminosity underestimates the empirical results by a factor of two, while the wind luminosity is underestimated by a factor of nine.

The standard feedback recipe for \emph{Starburst99} is somewhat dated and may help explain this disagreement. In the case of the ionising luminosity, we have seen how W-R stars contribute significantly to the output of R136. Recent spectral modelling of the four core WN\,5h stars by \citet{Crowther:2010} has estimated all to have initial masses $M_{\rm{init}}>150$\,M$_{\odot}$. However, the IMF used for \emph{Starburst99} was limited to $M_{\mathrm{up}}=100$\,M$_{\odot}$ and the incorporated evolutionary tracks only extended up to $120$\,M$_{\odot}$. The \emph{Starburst99} output would therefore have excluded stars with $M_{\rm{init}}>100$\,M$_{\odot}$. 

In an attempt to identify other stars with $M_{\rm{init}}>100$\,M$_{\odot}$, we examined the latest Bonn evolutionary tracks \citetext{K\"{o}hler et al. in prep.}, which extend to $500$\,M$_{\odot}$. For LMC metallicity and at an age of $\leq2$\,Myr, the luminosity of a $100$\,M$_{\odot}$ star was expected to be $\log(L_{\mathrm{Bol}}/\mathrm{L}_{\odot})\gtrsim6.3$. Table \ref{tab:Top 10} lists ten census stars to which this applies, of which eight are located within the R136 region. The dotted lines in Figure \ref{fig:S99_vs_Empirical}, show our empirical results if such stars are excluded. A better agreement is obtained with \emph{Starburst99}, with differences in $Q_{0}$ reduced to $\sim0.2$\,dex. We find minimal differences between our $q_{\mathrm{0}}$ calibration and the one used by \emph{Starburst99} \citep{Smith:2002}, so this remaining offset most likely arises from the higher $T_{\mathrm{eff}}$ that we assign to our SpTs.

When comparing the wind luminosities, excluding stars with $M_{\rm{init}}>100$\,M$_{\odot}$ has less of an impact and \emph{Starburst99} is still seen to underestimate our empirical results by $\sim0.7$\,dex. In this case, the disagreement largely arises from the different mass-loss prescriptions. \emph{Starburst99} predictions follow \citet{Leitherer:1992} and are found to differ from the \citet{Vink:2001} mass-loss rates by a factor of $2-10$, depending on SpT. Any differences in $\varv_{\infty}$ should be minimal, however, (see Figure 2 in \citealt{Leitherer:1992}).

We note that neither rotation or binarity are accounted for in evolutionary models used in \emph{Starburst99}. However, more recent versions \citep{Vazquez:2007, Levesque:2012} do attempt to address the effects of rotation on stellar outputs, as well as adopting the \citet{Vink:2001} mass-loss prescription. They typically found that rotating models prolonged the ionising output of stars, particularly those in lower mass bins. Rotation led to increased temperatures and luminosities and hence a greater and harder ionising luminosity, especially $>3$\,Myr in the presence of W-R stars. Rotational mixing would also increase mass-loss rates and so the wind luminosity is potentially higher as well. The situation is more complex when incorporating massive binaries, but \citet{EldridgeStanway:2009} indicate an increase in ionising luminosity at later ages. Despite these updates, the ability to model the most massive stars still appears crucial for accurate predictions of the feedback. Evolutionary models for very massive stars are in the process of being included in synthetic codes \citetext{K\"{o}hler et al. in prep.}.

\begin{table}
\begin{center}
\caption{The census stars with the highest ionising luminosities, wind luminosities and masses and their combined contribution to 30 Dor. Entries given in bold are based on tailored W-R analyses (see Table \ref{tab:W-R and Of/WN star parameters}). Other entries rely either on the less reliable W-R star template method or the calibrations discussed in Section \ref{sec:Stellar Calibrations}. For more reliable outputs for the Of/WN and WN stars, we direct the reader to Bestenlehner et al. \citetext{in prep.}.}
\label{tab:Top 10}
\begin{tabular} {l @{\hspace{2mm}} l @{\hspace{2mm}} l @{\hspace{2mm}} r  @{\hspace{2mm}} r @{\hspace{2mm}} r }
\hline\hline
ID & Star &  SpT & $r_{\mathrm{d}}$ & $Q_{\mathrm{0}}$ & $L_{\mathrm{SW}}$\\
\# &  &  & (pc) & ($10^{49}$\,ph\,s$^{-1})$ & ($10^{37}$erg s$^{-1}$)\\
\multicolumn{6}{c}{ 10 STARS WITH HIGHEST IONISING LUMINOSITY}\\
\hline
{\bf 630} & {\bf R136a1} &  {\bf WN\,5h}  & {\bf 0.00} & {\bf 66.0} & {\bf 10.9}  \\
\emph{770} & \emph{Mk34} & \emph{WN\,5h} & \emph{2.61} & \emph{49.2}  & \emph{9.2}  \\
{\bf 633} & {\bf R136a2} &  {\bf WN\,5h} &  {\bf 0.02} & {\bf 48.0} &  {\bf 8.7}  \\
{\bf 706} & {\bf R136c} &  {\bf WN\,5h} & {\bf 0.83} & {\bf 42.0} &  {\bf 5.4}  \\
\emph{493} & \emph{R134} & \emph{WN\,6(h)} & \emph{2.86} & \emph{35.3} & \emph{6.2} \\
{\bf 613} & {\bf R136a3} &  {\bf WN\,5h} & {\bf 0.12} & {\bf 30.0} &  {\bf 5.6}  \\
\emph{1001} & \emph{BAT99-122} & \emph{WN\,5h} & \emph{71.71} & \emph{25.0} & \emph{5.5} \\
\emph{580} & \emph{Mk 42} & \emph{O2\,If*} & \emph{1.96} & \emph{18.8} & \emph{4.0} \\
{\bf 916} & {\bf BAT99-118} &  {\bf WN\,6h} & {\bf 62.27} & {\bf 18.2} & {\bf 5.3} \\
\emph{375} & \emph{BAT99-95} &  \emph{WN\,7h} & \emph{21.56} &  \emph{15.8}  & \emph{7.6} \\
& \cline{4-5}
\multicolumn{4}{r}{Total (\% of $Q_{\mathrm{0}}^{\mathrm{Sp+Ph}}$ or $L_{\mathrm{SW}}^{\mathrm{Sp+Ph}}$):} & 348.3 (28\%) & 68.4 (29\%) \\
\multicolumn{6}{c}{ 10 STARS WITH HIGHEST WIND LUMINOSITY}\\
\hline
{\bf 630} & {\bf R136a1} &  {\bf WN\,5h}  & {\bf 0.00} & {\bf 66.0} & {\bf 10.9}  \\
\emph{543} & \emph{R140a1} & \emph{WC\,4} & \emph{12.43} & \emph{6.6}  &  \emph{9.7} \\
\emph{762} & \emph{Mk33Sb} &  \emph{WC\,5} & \emph{2.90} & \emph{6.3}  &  \emph{9.4}  \\
\emph{770} & \emph{Mk34} & \emph{WN\,5h} & \emph{2.61} & \emph{49.2}  & \emph{9.2}  \\
{\bf 633} & {\bf R136a2} &  {\bf WN\,5h} &  {\bf 0.02} & {\bf 48.0} &  {\bf 8.7}  \\
\emph{375} & \emph{BAT99-95} &  \emph{WN\,7h} & \emph{21.6} &  \emph{15.8}  & \emph{7.6} \\
\emph{493} & \emph{R134} & \emph{WN\,6(h)} & \emph{2.86} & \emph{35.3} & \emph{6.2} \\
{\bf 613} & {\bf R136a3} &  {\bf WN\,5h} & {\bf 0.12} & {\bf 30.0} &  {\bf 5.6}  \\
\emph{1001} & \emph{BAT99-122} & \emph{WN\,5h} & \emph{71.7} & \emph{25.0} & \emph{5.5} \\
{\bf 706} & {\bf R136c} &  {\bf WN\,5h} & {\bf 0.83} & {\bf 42.0} &  {\bf 5.4}  \\
\cline{4-5}
\multicolumn{4}{r}{Total (\% of $Q_{\mathrm{0}}^{\mathrm{Sp+Ph}}$ or $L_{\mathrm{SW}}^{\mathrm{Sp+Ph}}$):} & 324.2 (28\%) & 78.2 (35\%) \\
\multicolumn{6}{c}{ STARS WITH $M_{\rm{init}}>$100\,M$_{\odot}$}\\
\hline
{\bf 630} & {\bf R136a1} &  {\bf WN\,5h}  & {\bf 0.00} & {\bf 66.0} & {\bf 10.9}  \\
\emph{770} & \emph{Mk34} & \emph{WN\,5h} & \emph{2.61} & \emph{49.2}  & \emph{9.2}  \\
{\bf 633} & {\bf R136a2} &  {\bf WN\,5h} &  {\bf 0.02} & {\bf 48.0} &  {\bf 8.7}  \\
{\bf 706} & {\bf R136c} &  {\bf WN\,5h} & {\bf 0.83} & {\bf 42.0} &  {\bf 5.4}  \\
\emph{493} & \emph{R134} & \emph{WN\,6(h)} & \emph{2.86} & \emph{35.3} & \emph{6.2} \\
{\bf 613} & {\bf R136a3} &  {\bf WN\,5h} & {\bf 0.12} & {\bf 30.0} &  {\bf 5.6}  \\
\emph{1001} & \emph{BAT99-122} & \emph{WN\,5h} & \emph{71.71} & \emph{25.0} & \emph{5.5} \\
\emph{580} & \emph{Mk 42} & \emph{O2\,If*} & \emph{1.96} & \emph{18.8} & \emph{4.0} \\
{\bf 916} & {\bf BAT99-118} &  {\bf WN\,6h} & {\bf 62.27} & {\bf 18.2} & {\bf 5.3} \\
\emph{375} & \emph{Mk35} & \emph{O2\,If*/WN\,5} & \emph{3.07} & \emph{15.6}  & \emph{1.7}  \\
\cline{4-5}
\multicolumn{4}{r}{Total (\% of $Q_{\mathrm{0}}^{\mathrm{Sp+Ph}}$ or $L_{\mathrm{SW}}^{\mathrm{Sp+Ph}}$):} & 348.1 (28\%) & 62.5 (28\%) \\
\hline
\end{tabular}
\end{center}
\end{table}

\section{Star Formation Rates}
\label{sec:Star Formation Rates}
There are various methods to estimate the SFR of a region, from UV through to IR. Each have advantages and disadvantages, as they trace the fates of the ionising photons from the young stellar population \citep{Kennicutt:1998}. Our census of 30 Dor now provides us with the total ionising luminosity produced by its stars. The UV stellar continuum is often used to directly probe this emission while gaseous nebular recombination lines (e.g. H$\alpha$) and the far-infrared (FIR) dust continuum are more widely used. In this section we compare the findings of these different indicators for both \citet{Salpeter:1955} and \citet{Kroupa:2001} IMFs. Comparing their estimates to those of our census also allows us to determine the possible fraction of ionising photons which may be escaping the region, $f_{\mathrm{esc}}$.

\citet{Kennicutt:1998} provide a series of widely used calibrations to estimate the SFR from various wavelengths. They are derived from population synthesis models which predict the respective luminosities ($Q_{\mathrm{0}}$, H$\alpha$ etc.) for a given set of parameters, including the IMF, over a period of constant star formation. They were revisited in a review by \citet{KennicuttEvans:2012} and references therein, where changes arose from updates to the IMF and stellar population models. However, these calibrations are more applicable to galaxies and starburst regions, for which star formation has occurred for at least 100\,Myr. In the case of 30 Dor, findings from \citet{WalbornBlades:1997} and our census suggest stellar ages $>10$\,Myr but information on the region's SFH is limited. Nevertheless, \citet{DeMarchi:2011} had used their HST/WFC3 photometry to focus on pre main-sequence stars, albeit within a smaller central region of 30 Dor compared to our census. They estimated a relatively constant SFH for 30 Dor for the past $\sim10$\,Myr.

Based on these findings, we opted to derive our own SFR calibrations using \emph{Starburst99}. For the case of 30 Dor, our synthetic model was set to run for 10\,Myr with a continuous $\mathrm{SFR} = 1\,\mathrm{M}_{\odot}\,\mathrm{yr}^{-1}$. All other model parameters were identical to our R136 burst model, i.e. a Kroupa IMF,  ``high mass-loss rate'' Geneva evolutionary tracks and $Z=0.4\,Z_{\odot}$. The calibrations given in Equations \ref{eq:LyC Stars} - \ref{eq:FIR} were then derived from the respective luminosities, as predicted by \emph{Starburst99}, at an age of 10\,Myr. However, we showed earlier, the exclusion of $M_{\rm{init}}>100$\,M$_{\odot}$ stars can lead to integrated stellar luminosities being underestimated. On this occassion, we do not attempt to adjust the coefficients in Equations \ref{eq:LyC Stars} - \ref{eq:FIR}, noting that the most massive stars could lead to SFR discrepancies. Nevertheless, they should still be more reliable than those calibrated for galaxies.

\subsection{Lyman continuum from census}
\label{sec:Lyman continuum from stars}
30 Dor offers us the rare opportunity to measure the integrated Lyman continuum (LyC) ionising luminosity directly from its stars. As only massive ($>10$\,M$_{\odot}$) and young ($<20$\,Myr) stars significantly contribute to this quantity, it provides a nearly instantaneous measure of the SFR \citep{Kennicutt:1998}. We have already estimated this integrated value as $Q_{\mathrm{0}}^{\mathrm{Sp+Ph}}$. The resulting SFR for 30 Dor is then:
\begin{equation}
\label{eq:LyC Stars}
\mathrm{SFR}\,(\mathrm{M}_{\odot}\,\mathrm{yr}^{-1} ) = 5.8\times10^{-54} Q_{\mathrm{0}}\,(\mathrm{ph\,s}^{-1}).
\end{equation}
When accounting for photometrically classified stars, the census gives a SFR of 0.105 or 0.073 $\mathrm{M}_{\odot}\,\mathrm{yr}^{-1}$ for a Salpeter or Kroupa IMF, respectively (see Table \ref{tab:SFR Comparison}). Having accounted for stars individually, this is one of the most direct methods of quantifying their ionising luminosity. We now discuss the alternative methods available and compare their results in Table \ref{tab:SFR Comparison}.

\subsection{Far-UV continuum}
\label{sec:FUV continuum}
Probing the young massive stars can also be achieved in the far-UV (FUV), since they will dominate the integrated UV spectrum of a star forming region. However, it is particularly sensitive to extinction (as the cross-section of dust peaks in the UV). The FUV continuum flux of 30 Dor was obtained from Ultraviolet Imaging Telescope (UIT) images using its B5 (1615\,\AA) filter \citep{Parker:1998}. An aperture, consistent to our census region was used and corrected for a uniform extinction using the \citet{Fitzpatrick:1986} law, adopting an average $R_V\approx3.5$ and $E(B-V)\approx0.4$. Our SFR calibration for 30 Dor from the FUV is:
\begin{equation}
\label{eq:FUV}
\mathrm{SFR}\,(\mathrm{M}_{\odot}\,\mathrm{yr}^{-1} ) = 1.2\times10^{-40} L_{1615}\,\mathrm{(erg\,s^{-1}\,\AA^{-1})},
\end{equation}
where $L_{1615}$ is the continuum luminosity at 1615\,\AA. This gave a SFR in very good agreement with the census. As a direct tracer of the hot luminous stars, the consistency of the FUV continuum luminosity would be expected. The sensitivity of this diagnostic to extinction shows that the  adoption of a uniform \citet{Fitzpatrick:1986} law is reasonable for 30 Dor.

\subsection{Lyman continuum from H$\alpha$}
\label{sec:Lyman continuum from Halpha}
For most giant H\,{\sc ii} regions, the ionising stars are not individually observable and hydrogen recombination lines, primarily H$\alpha$, serve as the main indicator for a young massive population. \citet{Kennicutt:1995} measured the H$\alpha$ flux from 30 Dor through a series of increasing circular apertures, centred on R136, from which we selected a 10\,arcmin radius, consistent with our census region. Observations from \citet{Pellegrini:2010} gave an integrated nebular flux ratio of $F(\mathrm{H}\alpha)/F(\mathrm{H}\beta) =4.37$.  Assuming the intrinsic ratio to be 2.86 for an electron temperature of $T_{\mathrm{e}}=10^{4}$\,K and electron density of $n_{\mathrm{e}}=100\,\mathrm{cm}^{-3}$, we find $A_{\mathrm{H\alpha}}=1.20$\,mag when applying $R_{V}=3.5$ as before. The SFR for 30 Dor based on H$\alpha$ is then:
\begin{equation}
\label{eq:LyC Halpha}
\mathrm{SFR}\,(\mathrm{M}_{\odot}\,\mathrm{yr}^{-1} ) = 4.2\times10^{-42} L_{\mathrm{H}\alpha}\,(\mathrm{erg\,s}^{-1})
\end{equation}
where $L_{\mathrm{H}\alpha}$ is the H$\alpha$ luminosity. The SFR from this method is $\sim30$\% lower than estimates from the census. As with the FUV, this method can be sensitive to extinction as well as the IMF. Furthermore, absorption of ionising photons by dust and their potential leakage make this SFR a lower limit.

\subsection{Far-infrared continuum}
\label{sec:Far-IR continuum}
Stellar UV photons may be absorbed by dust and re-emitted at FIR wavelengths. Measuring the FIR luminosity ($L_{\mathrm{FIR}}$) allows us to account for this absorption and in cases of high dust opacity, it provides a SFR tracer too. \citet{Skibba:2012} recently produced dust luminosity ($L_{\mathrm{dust}}$) surface density maps of the Magellanic Clouds\footnote{The $L_{\mathrm{dust}}$ values from \citet{Skibba:2012} incorporated 24\,$\mu$m MIPS images that were saturated at the very core of 30 Dor. These few pixels did not list values. To correct for this, we substituted in the values of neighbouring pixels to obtain the final integrated $L_{\mathrm{FIR}}$ given in Table \ref{tab:SFR Comparison}.}. These integrated observations from the \emph{Spitzer} SAGE \citep{Meixner:2006} and \emph{Herschel Space Observatory} HERITAGE \citetext{Meixner et al., ApJ, submitted} surveys. $L_{\mathrm{dust}}$ covered 5.8-500\,$\mu$m, and while limits are not completely consistent with the FIR provided by \emph{Starburst99}, the FIR peak ($\sim100\mu$m) is well covered and the overall difference should be small. Our equivalent SFR calibration for 30 Dor in the FIR is:
\begin{equation}
\label{eq:FIR}
\mathrm{SFR}\,(\mathrm{M}_{\odot}\,\mathrm{yr}^{-1} ) = 4.2\times10^{-44} L_{\mathrm{FIR}}\,(\mathrm{erg\,s}^{-1}),
\end{equation}
where $L_{\mathrm{FIR}}$ refers to the far infrared luminosity. The coefficient is based on the total bolometric luminosity of the stars as predicted by \emph{Starburst99}, assuming that all of the stellar luminosity is absorbed and re-radiated by the dust. The FIR continuum implies a SFR significantly lower (factor of $\sim10$) than our census. This would indicate that only a small proportion of the ionising photons contribute to heating the dust with the remainder either ionising gas or escaping the region. It should also be noted that older stellar populations $(>10$\,Myr) are still capable of heating the dust and contributing to the FIR emission, even if they no longer produce ionising photons. Such stars are not related to the recent star formation.

\begin{table*}
\begin{center}
\caption{Comparison of SFR tracers for 30 Dor.}
\label{tab:SFR Comparison}
\begin{tabular} { l  l  r r  r }
\hline\hline
\multirow{2}{*}{SFR Tracer }& & & \multicolumn{2}{c}{SFR [M$_{\odot}$\,y$r^{-1}$]} \\
& & & Salpeter IMF & Kroupa IMF \\
\hline
\multirow{2}{*}{Lyman Continuum (Stars)} & $Q_{\mathrm{0}}^{\mathrm{Sp}}$ [ph\,s$^{-1}$] & $10.56\times10^{51}$ & $0.088\pm0.053$ & $0.061\pm0.037$\tablefoottext{3} \\
& $Q_{\mathrm{0}}^{\mathrm{Sp+Ph}}$ [ph\,s$^{-1}$] & $12.37\times10^{51}$ & $0.103\pm0.062$ & $0.073\pm0.044$\tablefoottext{3} \\
\hline
\multirow{3}{*}{FUV Continuum (1615\AA)} & $F_{1615}$ [erg\,cm$^{-2}$\,s$^{-1}$\,\AA] & $4.88\times10^{-11}$ & & \\
& $I_{1615}$\tablefoottext{1} [erg\,cm$^{-2}$\,s$^{-1}$\,\AA] & $2.01\times10^{-9}$ & & \\
& $L_{1615}$ [erg\,s$^{-1}$\,A] & $6.02\times10^{38}$ & $0.102\pm0.033$ & $0.071\pm0.023$\tablefoottext{4} \\
\hline
\multirow{3}{*}{Lyman Continuum (H$\alpha$)} & $F_{\mathrm{H}\alpha}$ [erg\,cm$^{-2}$\,s$^{-1}$] &  $1.37\times10^{-8}$ & & \\
& $I_{\mathrm{H}\alpha(\mathrm{corr})}$\tablefoottext{2} [erg\,cm$^{-2}$\,s$^{-1}$] &  $4.12\times10^{-8}$ & & \\
& $L_{\mathrm{H}\alpha(\mathrm{corr})}$ [erg\,s$^{-1}$] & $1.23\times10^{40}$ & $0.075\pm0.018$ & $0.052\pm0.012$\tablefoottext{5} \\
\hline
\multirow{4}{*}{Combined H$\alpha$ and MIR (22$\mu$m)} & $F_{22}$ [Jy] & 2800 & & \\
& $\nu L_{22}$ [erg\,s$^{-1}$] & $1.13\times10^{41}$ & & \\
& $L_{\mathrm{H}\alpha(\mathrm{obs})}$ [erg\,s$^{-1}$] & $4.09\times10^{39}$ & $0.050\pm0.003$ & $0.035\pm0.002$\tablefoottext{6} \\
& $L_{\mathrm{H}\alpha(\mathrm{corr})}$ [erg\,s$^{-1}$] & $1.23\times10^{40}$ & $0.100\pm0.018$ & $0.070\pm0.012$\tablefoottext{6} \\
\hline
FIR Continuum & $L_{\mathrm{FIR}}$ [erg\,s$^{-1}$] & $1.58\times10^{41}$ & $0.010\pm0.001$ & $0.007\pm0.001$\tablefoottext{7} \\
\hline
\end{tabular}
\tablefoot{
\tablefoottext{1}{$A_{1600}=4.0$\,mag from \citet{Fitzpatrick:1986} law}.
\tablefoottext{2}{$A_{\mathrm{H\alpha}}=1.20$\,mag from integrated F(H$\alpha$)/F(H$\beta$) ratio}.
\tablefoottext{3}{From Equation \ref{eq:LyC Stars}.}
\tablefoottext{4}{From Equation \ref{eq:LyC Halpha}.}
\tablefoottext{5}{From Equation \ref{eq:FUV}.}
\tablefoottext{6}{From Equation \ref{eq:LyC + MIR}.}
\tablefoottext{7}{From Equation \ref{eq:FIR}.}
Salpeter IMF SFRs are obtained by scaling Kroupa IMF SFRs by a factor of 1.44.
}
\end{center}
\end{table*}

\subsection{Combined H$\alpha$ and mid-IR}
\label{Halpha and MIR}
By combining a dust-emission-based tracer with one tracing ionised gas, it should be possible to account for all the ionising photons, except for those escaping the region. \citet{Calzetti:2007} and \citet{Kennicutt:2009} provide such a SFR diagnostic by combining the observed H$\alpha$ and 24\,$\mu$m luminosities. The observed H$\alpha$ luminosity was calculated from \citet{Kennicutt:1995} as before, only not corrected for extinction. \emph{Spitzer} 24\,$\mu$m observations were unfortunately saturated at the core of 30 Dor. We therefore turned to the similar 22\,$\mu$m filter aboard the Wide-field Infrared Survey Explorer (WISE), and measured a luminosity from the archival image coadd\_id=0837m697\_ab41. While the SFR calibration of \citet{Calzetti:2007} is suited for H\,{\sc ii} regions (their Equation 7), for consistency we replace their first coefficient with the one derived in our Equation \ref{eq:LyC Halpha}. In addition, to account for the narrower bandwidth and shorter wavelength of the WISE filter, their second coefficient was adjusted to give:
\begin{eqnarray}
\label{eq:LyC + MIR}
\mathrm{SFR}\,(\mathrm{M}_{\odot}\,\mathrm{yr}^{-1} ) &=& 4.2\times10^{-42}[ L_{\mathrm{H}\alpha(\mathrm{obs})} \,(\mathrm{erg\,s}^{-1}) \nonumber \\
&& + 0.039 \times \nu L_{22}\,(\mathrm{erg\,s}^{-1})],
\end{eqnarray}
where $L_{22}$ is the observed luminosity density at $22\,\mu$m and $\nu$ is the central frequency of the $22\,\mu$m filter. This method also estimated a SFR lower than our empirical census, by a factor of $\sim2$. However, we note that by substituting the extinction corrected H$\alpha$ luminosity ($L_{\mathrm{H}\alpha(\mathrm{corr})}$) into Equation \ref{eq:LyC + MIR}, results are much more consistent. This approach would therefore be recommended for star forming regions lacking UV imaging.

\subsection{Photon escape fraction}
\label{sec:Photon Escape Fraction}
The leakage of ionising photons from 30 Dor might be expected given the open structure and filaments of the region. We can only consider the $f_{\mathrm{esc}}$ for 30 Dor by dealing in numbers of photons. We have already determined the number of ionising photons emitted by its stars, $Q_{\mathrm{0}}^{\mathrm{Sp+Ph}}$. The number of photons which ionise the gas ($Q_{\mathrm{0}}^{\mathrm{H}\alpha}$) can be calculated from Equations \ref{eq:LyC Stars} and \ref{eq:LyC Halpha}, and is shown to be directly proportional to $L(\mathrm{H}\alpha)$.

The number of photons absorbed by dust ($Q_{\mathrm{0}}^{\mathrm{Dust}}$) is more complicated since both EUV and non-EUV photons contribute to heating the dust. We therefore use the $L_{\mathrm{EUV}}/L_{\mathrm{Bol}}$ fraction which was estimated for each SpT  (see Tables \ref{tab:O-star Parameters Dwarfs} - \ref{tab:B-star Parameters}) using the models of \citet{Martins:2005} and \citet{DoranCrowther:2011}. We obtain an average $L_{\mathrm{EUV}}/L_{\mathrm{Bol}} \approx 0.45$ for 30 Dor as a whole, representing a typical O5\,V star. If we were to apply this same fraction to the dust then $L^{\mathrm{Dust}}_{\mathrm{EUV}}\sim0.45\times L_{\mathrm{FIR}}$, where $L^{\mathrm{Dust}}_{\mathrm{EUV}}$ is the dust luminosity heated by EUV photons. Of course, we have already determined that a fraction of the EUV photons leaving our stars must ionise the gas, while variations in the dust opacity could result in further photons escaping the region. An additional test was therefore made with the grain models of \citet{WeingartnerDraine:2001} to calculate the fraction of ionising luminosity of a typical O5\,V star, that is absorbed by dust. These estimates suggested that $L^{\mathrm{Dust}}_{\mathrm{EUV}}\approx0.55\times L_{\mathrm{EUV}}$
for a variety of $R_{V}$ values \citetext{M. Min, priv. comm.}. Therefore, in terms of the number of ionising EUV photons: 
\begin{eqnarray}
\label{eq:Dust Q0}
Q_{\mathrm{0}}^{\mathrm{Dust}} \sim 0.55\times L_{\mathrm{FIR}}/L_{\mathrm{EUV}} \times Q_{\mathrm{0}}^{\mathrm{Sp+Ph}}.
\end{eqnarray}
Table \ref{tab:Photon Escape Fraction} gives a breakdown of the ionising photons. The ratio of $(Q_{\mathrm{0}}^{\mathrm{H\alpha}}+Q_{\mathrm{0}}^{\mathrm{Dust}})/Q_{\mathrm{0}}^{\mathrm{Sp+Ph}} \sim0.94$, therefore $f_{\mathrm{esc}}\sim0.06$. This would suggest 30 Dor to be `density bounded' with just under 6\% of its ionising photons escaping the region. Given the uncertainties on our values, $f_{\mathrm{esc}}$ potentially ranges from $0-0.61$. This could equally see 30 Dor as a `radiation bounded' region. However, we recall that $Q_{\mathrm{0}}^{\mathrm{Sp+Ph}}$ is still likely an underestimate due to unaccounted contributors (see Section \ref{sec:Stars Unaccounted For}). $Q_{\mathrm{0}}^{\mathrm{Dust}}$ is also likely to be an upper limit. This arises from the omission of luminous cool supergiants from the census due to their minimal ionising luminosity. However, these older stellar populations would still have contributed to $L_{\mathrm{Bol}}$, hence lowering the integrated $L_{\mathrm{EUV}}/L_{\mathrm{Bol}}$ fraction.

\citet{SmithBrooks:2007} followed the same approach as our work on the Carina Nebula and found a quarter and one third of the ionising photons to be unaccounted for when comparing to radio and H$\alpha$ fluxes, respectively. Similar work by \citet{Voges:2008} showed 20-30\% of their less luminous LMC H\,{\sc ii} regions to be in a similar density bounded state. A slightly different approach was made by \citet{Pellegrini:2011b} who implemented nebular line ratios to detect ionisation fronts around H\,{\sc ii} regions in the Magellanic Clouds. They too, concluded the regions to be density bounded with average $f_{\mathrm{esc}}>0.4$.

\begin{table}[h]
\begin{center}
\caption{Comparison of $Q_{\mathrm{0}}$ tracers for 30 Dor.}
\label{tab:Photon Escape Fraction}
\begin{tabular} { l  l @{\hspace{3mm}} r @{\hspace{3mm}} r } 
\hline\hline
$Q_{\mathrm{0}}$ Tracer & & & $Q_{\mathrm{0}}$ [$10^{51}$\,ph\,s$^{-1}$] \\
\hline
LyC (Stars$^{\mathrm{Sp}}$) & $L_{\mathrm{EUV}}$ [erg\,s$^{-1}$] & $3.49\times10^{41}$ & $10.6\pm6.3$\tablefoottext{1} \\
LyC (Stars$^{\mathrm{Sp+Ph}}$) & $L_{\mathrm{EUV}}$ [erg\,s$^{-1}$] & $4.01\times10^{41}$ & $12.4\pm7.4$\tablefoottext{1} \\
\\
LyC (H$\alpha$) & $L_{\mathrm{H}\alpha}$ [erg\,s$^{-1}$] & $1.23\times10^{40}$ & $9.0\pm1.8$\tablefoottext{2} \\
FIR Continuum & $L_{\mathrm{FIR}}$ [erg\,s$^{-1}$] & $1.58\times10^{41}$ & $2.6\pm2.0$\tablefoottext{3} \\ 
\hline
\end{tabular}
\tablefoot{
\tablefoottext{1}{From census}.
\tablefoottext{2}{From equating Equations \ref{eq:LyC Stars} \& \ref{eq:LyC Halpha}}.
\tablefoottext{3}{From Equation \ref{eq:Dust Q0}}.
}
\end{center}
\end{table}

\section{Discussion and Summary}
\label{sec:Summary}

The global properties of 30 Dor can be obtained directly from its individually resolved stars or through observations of the entire region. For extragalactic star forming regions, only the latter is a possibility. As one of the most massive and closest star forming regions, 30 Dor therefore serves as the ideal template for more distant targets.

The global ionising luminosity and wind luminosity of R136 indicate significant contributions from the most massive stars ($M_{\rm{init}}>100\,M_{\odot}$) in the region, many of which are WN stars. Accurate modelling of such stars is therefore essential if we are to correctly synthesise the energy budget of a region, which is not currently the case for population synthesis codes. The most massive stars are now being included in stellar evolutionary grids K\"{o}hler et al. \citetext{in prep.} with their ionising and mechanical feedback now beginning to be simulated \citep{Dale:2013, RogersPittard:2013}.

Through these higher ionising luminosities, the $f_{\mathrm{esc}}$ for previously studied regions may be subject to revision. In the case of 30 Dor, we estimate that the percentage of ionising photons which ionise the gas is just over 70\%. Estimating the fraction which heats the surrounding dust is more complicated but a fraction of $\sim20$\% is obtained from FIR observations. This would mean $\sim5-10$\% escape 30 Dor although large uncertainties potentially allow all ionising photons to be retained, or in an extreme case, up to 61\% to escape the region. These limits primarily arise from the temperature calibrations applied to our stars. The uncertainty in $T_{\mathrm{eff}}$, especially for earlier O-type stars, restricts the accuracy of $Q_{\mathrm{0}}^{\mathrm{Sp+Ph}}$ and in turn LyC (Stars$^{\mathrm{Sp+Ph}}$). As a substantial fraction of these stars will be individually analysed as part of the VFTS project \citetext{Ram\'{\i}rez-Agudelo et al. in prep, Sab\'{\i}n-Sanjulia\'{\i}n et al. in prep., McEvoy et al. in prep.}, these parameters will soon be more reliably constrained. Accuracy will also benefit from improved knowledge of the extinction law in 30 Dor (\citealt{Tatton:2013}, Ma\'{\i}z Apell\'aniz et al. in prep.). Nevertheless, should such photon leakage be genuine, young hot luminous stars provide an essential ionising source to the WIM. This is a potential reason for the discrepancies measured between H$\alpha$ and FUV continuum tracers in nearby galaxies \citep{Lee:2009,Relano:2012} and could help to partially explain the offsets we see in our own SFR results.

Similarly, at young ages, the higher stellar wind luminosities could be comparable to supernovae, offering additional methods for driving superbubbles and galactic winds, as well as triggering star formation. We only considered the stellar winds when considereing the mechanical feedback from R136 and comparing to \emph{Starburst99}. However, the age spread of 30 Dor undoubtedly means SNe will have contributed to its global feedback. \citet{Voss:2012} made similar population synthesis comparisons in the Carina region, for which \citet{SmithBrooks:2007} had created an equivalent census. When accounting for SNe, their results showed a consistent energy budget to that of the \citet{SmithBrooks:2007} census.

When considering short-lived ionising stars, the age of the population is important to ensure that star formation has occurred for a sufficient timescale. The ionising stars of 30 Dor are distributed over ages beyond 8\,Myr but are largely concentrated between $2-5$\,Myr. Is this consistent with the representative age of the entire population in extragalactic cases? We have estimated the representative age for 30 Dor using the H$\alpha$ nebular observations of \citet{Pellegrini:2010}. The integrated H$\alpha$+[N\,{\sc ii}] emission line flux was measured and corrected for [N\,{\sc ii}] contamination\footnote{assuming $F([\mathrm{N}\,${\sc ii}]$\lambda6548,\lambda6583)/F(\mathrm{H}\alpha) = 0.15$} resulting in $W_{\lambda}(\mathrm{H}\alpha)=1100\,\AA$ (E. Pellegrini priv. comm.). $W_{\lambda}(\mathrm{H}\alpha)$ will decline with age, since hot stars gradually end their lives and the hydrogen gas can no longer be ionised \citep{SchaererVacca:1998}. At LMC metallicity, $W_{\lambda}(\mathrm{H}\alpha)=1100\,\AA$ corresponds to an age of $\sim3.5$\,Myr, consistent with the average stellar age.

The standard extragalactic SFR tracers underestimated the result from our direct census by up to a factor of two, with the exception of the FIR continuum which underestimated by a factor of 10. Drawing conclusions from this single result could be premature given the uncertainties on our census calibrations. However, \citet{ChomiukPovich:2011} suggest similar discrepancies when comparing the SFR tracers available for well resolved regions and unresolved extragalactic regions. They too, recognise the importance of the most massive stars to the ionising output but also stress how the uncertainty of the adopted IMF could lead to inconsistencies. The SFR calibrations in Equations \ref{eq:LyC Stars} - \ref{eq:LyC + MIR} were derived for compatibility with 30 Dor, particularly allowing for the region's relatively young SFH in comparison to galaxies, but there are further factors to be considered. For example, \citet{Leitherer:2008} offer alternative calibrations accounting for the effects of stellar rotation which produce even lower SFRs. 
 
So when attempting to determine the SFR of a young unresolved star forming region, similar to 30 Dor, what might be the favoured SFR tracer? Returning to Table \ref{tab:SFR Comparison}, the very good agreement from the FUV continuum tracer might be preferred but this approach still relies heavily on an accurate extinction law and observations in the FUV are not always readily available. Both the H$\alpha$ and FIR continuum tracers fail to account for ionising photons absorbed by dust or ionising gas, respectively. The combined H$\alpha$ and MIR tracer therefore might be expected to the most accurate although we see it underestimate by a factor of 2. However, if instead we were to use the extinction corrected H$\alpha$ luminosity ($L_{\mathrm{H}\alpha(\mathrm{corr})}$) in Equation \ref{eq:LyC + MIR}, we find much better agreement with the census. In regions of relatively low amounts of dust, as suggested by $L_{\mathrm{FIR}}$, the accuracy of this approach therefore depends more on the reliability of $L(\mathrm{H}\alpha)$ and $A_{\mathrm{H}\alpha}$ although it should be noted that any escaping photons would still not be accounted for.

In summary, we have compiled a census of the hot luminous stars within the inner 10\,arcmin of 30 Dor, based on their $UBV$ band photometry. These stars were matched to as many spectral classifications as possible and their stellar parameters determined via calibrations and models. The integrated mechanical and radiative feedback of 30 Dor and its central cluster R136 were estimated, with comparisons made to the nebular properties and population synthesis models, respectively. Our main findings are as follows:

\begin{enumerate}
\item A total of 1145 candidate hot luminous stars were photometrically selected in the census of which 722 were believed to be relevant in the context of feedback. Following recent observations of the VFTS, spectroscopy was available to confirm 500 of these, including: 25 W-R stars, 6 Of/WN stars, 385 O-type stars and 84 B-type stars. 
\item We estimate the spectral completeness of hot luminous stars in 30 Dor to be $\sim85$\% down to $m_V=17$\,mag. This completeness falls to $\sim35$\% in the central R136 cluster, in view of the crowding in the region although this will improve greatly with recent HST/STIS observations.\ 
\item This allowed estimates of the stellar mass of both R136, $M_{\mathrm{R136}}\sim5.0\times10^{4}$\,M$_{\odot}$ (consistent with \citealt{Hunter:1995}) and 30 Dor itself, $M_{\mathrm{30 Dor}}\sim1.1\times10^{5}$\,M$_{\odot}$.

\item The total ionising and stellar wind luminosity of the hot luminous stars in 30 Dor was estimated to be $12.4\times10^{51}$\,ph\,s$^{-1}$ and $2.24\times10^{39}$\,erg\,s$^{-1}$, respectively. $10^{8.36}$\,L$_{\odot}$ was derived for total bolometric luminosity. While these values incorporate stars which currently do not have spectroscopy, they are still likely to be lower limits given the other sources that are unaccounted for in the census.

\item The W-R and Of/WN stars are crucial to this feedback in 30 Dor. Just 31 of these stars were found to have a comparable ionising luminosity, and even greater wind luminosity, than the 469 OB stars in the region. This behaviour is replicated within R136, which alone was seen to contribute over 50\% of the overall ionising luminosity and wind luminosity of 30 Dor. The contribution of R136 is so dominant because it hosts one third of 30 Dor's W-R and Of/WN stars, along with a majority of its most massive O-type stars.

\item Stars with the highest initial masses, exceeding $100\,\mathrm{M}_{\odot}$ were also predominantly W-R stars. Comparisons to the population synthesis code \emph{Starburst99} indicate that inaccurate modelling of such massive stars could result in significant underestimates in the feedback of star forming regions.\

\item The integrated ionising luminosity was used to derive a SFR for 30 Dor, estimated at $0.073\pm0.04$\,M$_{\odot}$\,yr$^{-1}$ adopting a Kroupa IMF. Standard SFR tracers largely underestimated this value but the FUV continuum showed good agreement as did a modified H$\alpha$ + MIR tracer. Comparisons to the global observations allowed a photon escape fraction of $f_{\mathrm{esc}}\sim0.06^{+0.55}_{-0.06}$ to be determined for 30 Dor. This could help explain the discrepancies in the SFRs and provide a potential ionising source for the WIM.\
\end{enumerate}

\begin{acknowledgements}
We are grateful to an anonymous referee for their suggestions in improving the paper. We are grateful to Guido De Marchi for providing the HST/WFC3 photometry of R136. We thank Ramin Skibba for passing on his dust maps of the LMC and Eric Pellegrini for his H$\alpha$ images of 30 Dor. We also thank Michiel Min for computing the expected EUV dust absorption. EID would like to thank the Science and Technology Facilities Council for its studentship. J.M.A. and F.N. acknowledge support from [a] the Spanish Government Ministerio de Econom\'{\i}a y Competitividad through grants AYA2010-15081, AYA2010-17631 and AYA2010-21697-C05-01 and [b] the Consejer\'{\i}a de Educaci\'on of the Junta de Andaluc\'{\i}a through grant P08-TIC-4075. N.R.W. acknowledges support from STScI, which is operated by AURA, Inc., under NASA contract NAS5-26555.
\end{acknowledgements}

\bibliographystyle{aa} 
\bibliography{VFTS_Feedback} 

\appendix

\section{Stellar Calibrations}

\subsection{O-type star calibrations}
\label{sec:O-star Calibrations}
\subsubsection{O-type star temperatures}
\citet{Martins:2005} (hereafter MSH05) provided a $T_{\mathrm{eff}}$-SpT scale for Galactic O-type stars. Following \citet{Mokiem:2007},  a small upward adjustment to the scale was necessary before applying it to our stars in 30 Dor. Furthermore, as MSH05 relied on He\,{\sc i}-{\sc ii} lines, its accuracy could only be maintained as early as O3 type stars, meaning that the scale needed to be extended for earlier type stars in the census.

We turned to the recent work of \citet{RiveroGonzalez:2012a, RiveroGonzalez:2012b} in order to determine a refined calibration for the $T_{\mathrm{eff}}$ of LMC O-type stars. Figure 12 from \citet{RiveroGonzalez:2012b} compares the $T_{\mathrm{eff}}$ of their models against the calibrations of MSH05, showing a typical offset of $\approx1$\,kK for dwarfs, O4 type and later. Only limited models are available of LMC giants and even fewer for supergiants. However, given their minimal offsets from the MSH05 calibration, it was assigned to all our giants and supergiants, O4 type and later. In the case of O2-O3.5 type stars, where possible, $T_{\mathrm{eff}}$ was based on averages of the \citet{RiveroGonzalez:2012b}, \citet{DoranCrowther:2011} and \citet{Massey:2004, Massey:2005} models, with a double weighting going toward \citet{RiveroGonzalez:2012b}. Extrapolations were made from the MSH05 calibrations in the case of unmodelled SpTs.

\subsubsection{O-type star intrinsic colours and bolometric borrections}
Intrinsic colours were adopted from Table 2.1 of \citet{Conti:2008}. In the case of earlier giants and supergiants, where colours were lacking, a $(B-V)_{\mathrm{o}}$-SpT relation was assumed consistent with the models of \citet{MartinsPlez:2006}. 

MSH05 provide a T$_{\mathrm{eff}}-BC_{V}$ calibration for O-type stars, (their equation 4). A renewed calibration was derived from our new temperatures, which extended to SpT$<$O4 by combining it with the stellar parameters from \citet{RiveroGonzalez:2012b} and \citet{DoranCrowther:2011}. Only very minimal variations in $BC_{V}$ were shown between different luminosity classes.

\subsubsection{O-type star ionising luminosities}
MSH05 also derived a $q_{0}$($T_{\mathrm{eff}}$, $\log g$) function, but with the new $T_{\mathrm{eff}}$ that we assigned, their same $q_{\mathrm{0}}$ values could not be directly adopted. Instead, as was done with the $BC_{V}$, we combined these values with data from \citet{RiveroGonzalez:2012a, RiveroGonzalez:2012b} and \citet{DoranCrowther:2011}, along with \citet{Mokiem:2004} and \citet{Smith:2002}, in order to sample the higher $T_{\mathrm{eff}}$ range and determined new separate relations for the dwarf, giant and supergiant luminosity classes.

\subsubsection{O-type star wind velocities and mass-loss rates}
An extensive observational study in the UV of the terminal velocities of Galactic OB stars was made by \citet{Prinja:1990}. The $\varv_{\infty}$ assigned to our different SpTs were primarily based on the mean values given in their Table 3 with slight updates made to incorporate the later work of \citet{PrinjaCrowther:1998}, \citet{Massey:2004} and \citet{DoranCrowther:2011}, who studied further O-type stars, this time in the Magellanic Clouds. Measurements did not extend to late SpTs and so a value for $\varv_{\infty}$ was estimated in these cases. Given that the velocities of Prinja et al. (1990) are for Galactic stars, the metallicity dependence of $\varv_{\infty}$ is noted, Leitherer et al. (1992) deriving a dependence of $\varv_{\infty} \propto Z^{0.13}$. With $L_{\mathrm{SW}} \propto \varv_{\infty}^{2}$, the effect could be substantial. However, $\dot{M}$ is the more dominant term in this case so that any metallicity dependence of $\varv_{\infty}$ should only change the $L_{\mathrm{SW}}$ by up to $\sim10$\%. 

For O-type star mass-loss rates we applied the theoretical prescriptions of \citet{Vink:2001}, entering a LMC metallicity of $Z=0.5\,Z_{\odot}$, throughout. These predicted mass-loss rates do not directly account for wind clumping but systematic offsets found by \citet{Mokiem:2007b} when comparing to empirically determined mass-loss rates, could suggest them to be consistent with a clumped wind, nonetheless.

\begin{table}
\begin{center}
\caption{The average absolute magnitudes of OB stars in the census. The number of stars (N) the average $M_{V}$ is based on, and their spread ($\sigma$), are both given. Note the number of stars used to determine these magnitudes varies from those in Table \ref{tab:10min Spectral Breakdown} as some stars were omitted due to spurious values.}
\label{tab:Average absolute magnitudes}
\centering
\begin{tabular} { l   r @{\hspace{2mm}} c @{\hspace{2mm}} c  r @{\hspace{2mm}} c @{\hspace{3mm}} c  r @{\hspace{3mm}} c @{\hspace{3mm}} c }
\hline\hline
SpT & \multicolumn{9}{c}{Luminosity Class} \\
&  \multicolumn{3}{c}{V}  & \multicolumn{3}{c}{III}  & \multicolumn{3}{c}{I} \\
& $M_{V}$ & N & $\sigma$ & $M_{V}$ & N & $\sigma$ & $M_{V}$ & N & $\sigma$ \\
\hline
O2     & $-5.6$ & 5 & 0.5 & \ldelim \{ {3}{2mm}   \multirow{3}{*}{$-5.9$} & \multirow{3}{*}{14} & \multirow{3}{*}{0.3} & \ldelim \{ {3}{2mm} \multirow{3}{*}{$-6.3$} & \multirow{3}{*}{6} & \multirow{3}{*}{0.6} \\
O3     &  $-5.5$ & 22 & 0.4  &    &  \\
O3.5	&   $-5.0$ & 8 & 0.5  &  		 &  \\
O4	&   $-4.8$ & 13 & 0.5 &  $-5.7$ & 5 & 0.5 &  \ldelim \{ {7}{2mm} \multirow{7}{*}{$-6.2$} & \multirow{7}{*}{5} & \multirow{7}{*}{0.6} \\
O5	& \ldelim \{ {2}{2mm} \multirow{2}{*}{$-4.7$} & \multirow{2}{*}{14} & \multirow{2}{*}{0.4} &  \ldelim \{ {3}{2mm}  \multirow{3}{*}{$-5.6$} & \multirow{3}{*}{4}& \multirow{3}{*}{0.4} & \\
O5.5	&   	        &  		 &  \\
O6	&   $-4.7$ & 17 & 0.5 &       \\
O6.5	&   $-4.4$  & 18 & 0.6 & \ldelim \{ {3}{2mm} \multirow{3}{*}{$-5.5$}	& \multirow{3}{*}{7} & \multirow{3}{*}{0.2}	 &  \\
O7    &   $-4.7$  & 18 & 0.5  &                &  \\
O7.5	&   $-4.7$  & 13 & 0.5 &  		 &  \\
O8	&   $-4.4$  & 20 & 0.4 & \ldelim \{ {2}{2mm}  \multirow{2}{*}{$-4.9$} & \multirow{2}{*}{11} & \multirow{2}{*}{0.4} &  \ldelim \{ {5}{2mm}  \multirow{5}{*}{$-5.7$}  & \multirow{5}{*}{12} & \multirow{5}{*}{0.4} \\
O8.5	  & $-4.4$  & 18 & 0.5 &  		  &  \\
O9	  & $-4.3$  & 25 & 0.5 &   $-4.9$ & 8 & 0.5 &  \\
O9.5	  & $-4.0$  & 43 & 0.4 &   $-4.6$ & 14 & 0.5 &  \\
O9.7	  & $-3.9$  & 12 & 0.3 &   $-4.0$ & 22 & 0.4 &  \\
B0	  & $-3.8$  & 11 & 0.3 & & - & & & - & \\
\hline
\end{tabular}
\end{center}
\end{table}

\subsection{B-type star calibrations}
\label{sec:B-star Calibrations}
\subsubsection{B-type star temperatures}
\citet{Trundle:2007} and \citet{Hunter:2007} carried out atmospheric modelling on over 100 B-type stars as part of the previous VLT-FLAMES Survey of Massive Stars, many of which were in the LMC. Our B-type star temperature calibration adopts the same scale given in Table 10 of \citet{Trundle:2007}. Whilst some SpTs needed to be interpolated, we recall that the study only needed to consider the earliest B-type dwarfs and giants and B-type supergiants.

\subsubsection{B-type star intrinsic colours and bolometric corrections}
Intrinsic colours for B-type stars were adopted from Table 2.1 of \citet{Conti:2008}. \citet{Crowther:2006} produced a T$_{\mathrm{eff}}-BC_{V}$ calibration based on their B-supergiant studies along with the work of \citet{Trundle:2004} and \citet{TrundleLennon:2005}. This was applied to all the B-supergiants in the sample. In the case of the earliest B-dwarfs and giants, the relation from \citet{LanzHubeny:2003} (their equation 2) was used although as with the O-type stars, $BC_{V}$ shows only small changes with luminosity class, with $T_{\mathrm{eff}}$ being the dominant factor.

\subsubsection{B-type star ionising luminosities}
As with the O-type stars, a $q_{0}(T_{\mathrm{eff}})$ relation was determined for the cooler B-type stars. Based on the data of \citet{Mokiem:2007}, \citet{Conti:2008} and \citet{Smith:2002}, separate relations for dwarf, giant and supergiant luminosity classes were made as before.

\subsubsection{B-star Wind Velocities and Mass-loss Rates}
\citet{Prinja:1990} was once again used to assign $\varv_{\infty}$ to the B-supergiants with \citet{KudritzkiPuls:2000} offering values to the later SpTs. The early B-dwarfs and giants in the census were assumed to have similar $\varv_{\infty}$ to their late O-type star counterparts, while their $\dot{M}$ was supplied by the \citet{Vink:2001} prescription. An exception was made for B-supergiants, however, following a discrepancy between the prescription and empirical results \citep{MarkovaPuls:2008, Crowther:2006, TrundleLennon:2005}. In these cases we employed the Wind-Luminosity Relation (WLR) of \citet{Kudritzki:1999}, whereby the modified wind momentum ($D_{\mathrm{mom}} = \dot{M} \varv_{\infty} R^{0.5})$ relates to the stellar luminosity. We adopt B-supergiant parameters from past studies \citep{MarkovaPuls:2008, Lefever:2007, Crowther:2006, Kudritzki:1999} to plot $\log D_{\mathrm{mom}}$ vs $\log L$. The sample was split into early and mid/late type stars (as in \citet{Crowther:2006}, their figure 7) before a least squares fit was made and used to estimate $\dot{M}$. Despite discrepancies with the \citet{Vink:2001} predictions, given the relatively small number of B-supergiants in the census, similar integrated properties are expected for 30 Dor, with either mass-loss rate approach.

\begin{table}
\begin{center}
\caption{A breakdown of the SB2 systems that were corrected for in the census. $\Delta M_{V} = M_{V}^{1} - M_{V}^{2}$. The sum of the ionising luminosity and wind luminosity from the individual components was calculated and used in our final integrated values of 30 Dor.}
\label{tab:SB2}
\centering
\begin{tabular} { l @{\hspace{1mm}} r @{\hspace{1mm}} r @{\hspace{1.5mm}} r @{\hspace{1mm}} r @{\hspace{1.5mm}} l @{\hspace{1.5mm}} l @{\hspace{1mm}} r @{\hspace{1mm}} r @{\hspace{1mm}} r}
\hline\hline
ID & $V$ & $B-V$ & $A_{V}$ & $M_{V}^{\mathrm{sys}}$ & SpT$^{1}$ & SpT$^{2}$ & $\Delta M_{V}$ & $M_{V}^{1}$ & $M_{V}^{2}$\\
\# & (mag) & (mag) & (mag) & (mag) & & & (mag) & (mag) & (mag) \\
\hline
  56 & 16.91 & $0.28$ & 2.1 & $-3.6$ & O9V & O9.5V & $-0.3$ & $-3.0$ & $-2.7$  \\
  63 & 15.56 & $0.23$ & 1.9 & $-4.8$ & O8.5V & O9.5V & $-0.4$ & $-4.2$ & $-3.8$  \\
  64 & 16.23 & $0.30$ & 2.2 & $-4.4$ & O6.5V & O6.5V & $-0.0$ & $-3.7$ & $-3.7$  \\
  73 & 15.15 & $0.22$ & 1.9 & $-5.2$ & O8.5III & O9.7V & $-1.0$ & $-4.8$ & $-3.8$  \\
 126 & 16.44 & $0.21$ & 1.8 & $-3.8$ & O9.7V & B0V & $-0.1$ & $-3.1$ & $-3.0$  \\
 191 & 15.50 & $0.25$ & 2.0 & $-5.0$ & O8V & B0V & $-0.6$ & $-4.5$ & $-3.9$  \\
 194 & 14.78 & $0.03$ & 1.2 & $-4.9$ & O6V & O9.5V & $-0.8$ & $-4.5$ & $-3.7$  \\
 206 & 15.81 & $0.22$ & 1.9 & $-4.5$ & O9V & B0V & $-0.5$ & $-4.0$ & $-3.5$  \\
 235 & 13.79 & $-0.11$ & 0.7 & $-5.4$ & O4V & O5V & $-0.1$ & $-4.7$ & $-4.6$  \\
 333 & 14.38 & $-0.10$ & 0.8 & $-4.9$ & O4.5V & O5.5V & $-0.2$ & $-4.2$ & $-4.0$  \\
 422 & 15.71 & $0.18$ & 1.8 & $-4.5$ & O7V & O7V & $-0.0$ & $-3.8$ & $-3.8$  \\
 424 & 14.75 & $0.10$ & 1.5 & $-5.2$ & O3.5V & O5.5V & $-0.3$ & $-4.6$ & $-4.3$  \\
 439 & 15.86 & $0.44$ & 2.7 & $-5.3$ & O7.5V & O7.5V & $-0.0$ & $-4.5$ & $-4.5$  \\
 512 & 14.16 & $-0.11$ & 0.7 & $-5.0$ & O8V & O9.5V & $-0.4$ & $-4.5$ & $-4.1$  \\
 525 & 14.19 & $-0.08$ & 0.8 & $-5.1$ & O6.5V & O6.5V & $-0.0$ & $-4.4$ & $-4.4$  \\
 576 & 14.72 & $-0.03$ & 1.2 & $-5.0$ & O3V & O4V & $-0.7$ & $-4.5$ & $-3.8$  \\
 620 & 11.94 & $0.10$ & 1.4 & $-7.9$ & O6.5I & O6.5I & $-0.0$ & $-7.2$ & $-7.2$  \\
 635 & 14.50 & $0.08$ & 1.7 & $-5.7$ & O4V & O6V & $-0.1$ & $-5.0$ & $-4.9$  \\
 722 & 13.99 & $-0.03$ & 0.9 & $-5.4$ & O7.5I & O9I & $-0.5$ & $-4.8$ & $-4.3$  \\
 735 & 15.41 & $0.03$ & 1.2 & $-4.3$ & O9V & O9.7V & $-0.4$ & $-3.7$ & $-3.3$  \\
 782 & 15.91 & $0.19$ & 1.7 & $-4.3$ & O9.7III & B0V & $-0.2$ & $-3.6$ & $-3.4$  \\
 812 & 14.93 & $0.06$ & 1.3 & $-4.9$ & O8V & O9.7V & $-0.5$ & $-4.3$ & $-3.8$  \\
 885 & 16.03 & $0.38$ & 2.4 & $-4.9$ & O5V & O8V & $-0.3$ & $-4.3$ & $-4.0$  \\
 906 & 15.13 & $0.08$ & 1.4 & $-4.8$ & O6.5V & O6.5V & $-0.0$ & $-4.0$ & $-4.0$  \\
 944 & 16.91 & $0.30$ & 2.2 & $-3.8$ & O7V & O8V & $-0.3$ & $-3.1$ & $-2.8$  \\
1024 & 16.89 & $0.49$ & 2.8 & $-4.4$ & O7.5V & O8.5V & $-0.3$ & $-3.8$ & $-3.5$  \\
1058 & 14.06 & $-0.17$ & 0.5 & $-5.0$ & O5.5V & O7V & $-0.0$ & $-4.2$ & $-4.2$  \\
\hline
\end{tabular}
\end{center}
\tablefoot{A total of 48 SB2 systems were identified in the census. However, in order to determine accurate feedback values of both components, only the 27 SB2 systems, for which both a subtype and luminosity class was known for the secondary component, are listed in this table.
}
\end{table}

\section{W-R and Of/WN Stars}
\label{sec:W-R stars}
All of the W-R stars relied on narrow band photometry for accurate determination of their stellar properties. In many cases, this could be adopted from the works of \citet{CrowtherHadfield:2006}, \citet{SchmutzVacca:1991}, and \citet{Torres-DodgenMassey:1988}. For the remainder, spectrophotometry of various data sets was required such as \citet{CrowtherSmith:1997} and the HST/FOS observations of \citet{deKoter:1997} and \citet{MasseyHunter:1998} (see Table \ref{tab:W-R photometry}).

Intrinsic colours of all WN stars used the $(b-v)_{0}$-$W_{\lambda}(4686)$ relation from \citet{CrowtherHadfield:2006}, where $W_{\lambda}(4686)$ was measured from the available spectroscopy or adopted from \citet{Schnurr:2008}. In the case of WC stars, $(b-v)_{0}$ was adopted from \citet{CrowtherHadfield:2006} if available, otherwise an average $(b-v)_{0}=-0.28$ was applied to single WC stars and $(b-v)_{0}=-0.32$ to WC binaries where an O-type star dominated the light.

\begin{figure}
\includegraphics[height=0.45\textheight]{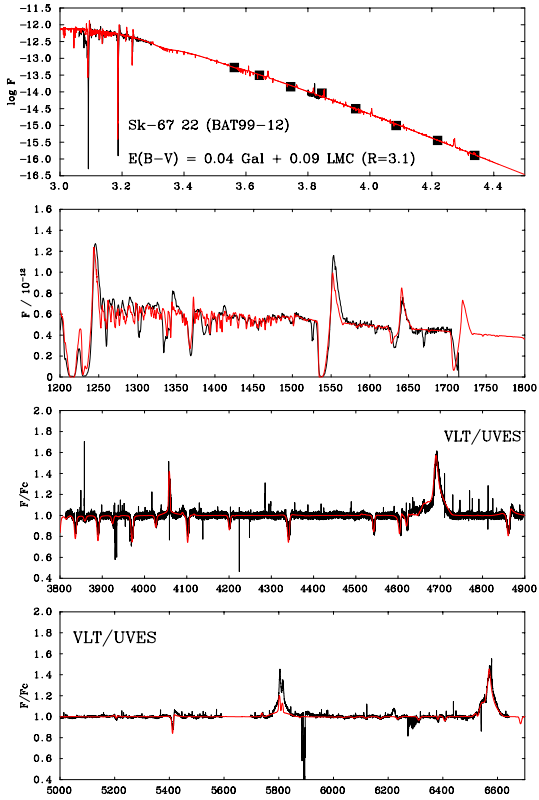}
\caption{Spectral fitting for O2\,If*/WN\,5 template star BAT99-12/Sk -67$^{\circ}$ 22.}
\label{fig:plot_sk6722}
\end{figure}

While the VFTS had made observations of 23 of the W-R and Of/WN stars, analysis was not straight forward as spectra were not flux calibrated. Furthermore, it lacked large sections at the yellow and red wavelengths, which hold important diagnostic lines for W-R star modelling.
The nine stars listed in Table \ref{tab:W-R and Of/WN star parameters}, one for each subtype, were the stars selected as templates for the rest of the census. Fortunately, additional flux calibrated spectra were available for most of these stars, upon which our CMFGEN models were principally based. Figures \ref{fig:plot_sk6722}-\ref{fig:plot_bat90} provide a comparison between the CMFGEN models (red) and the observed spectra (black). Photometric data is also plotted over the CMFGEN spectral energy distributions: broad band photometry (filled squares), narrow band photometry (open squares).

\subsection{W-R Binaries}
\label{sec:W-R Binaries}
\citet{Schnurr:2008}, \citet{Foellmi:2003} and \citet{Bartzakos:2001} all made studies into the binary nature of W-R stars in the LMC. The VFTS supplied new spectra for 19 W-R stars, allowing a fresh look for possible binary candidates with its multi-epoch observations. Updates to previous classifications were given in Table 2 of Paper I.

In order to determine the desired W-R magnitude ($M^{\mathrm{W-R}}_{v}$) from the observed magnitude of a multiple system ($M^{\mathrm{sys}}_{v}$), we used the dilution of different emission lines. For a single undiluted WN\,6 star, the strength of the He\,{\sc ii} $\lambda 4686$ line was taken to be $W_{\lambda}(4686)\approx75\,\AA$ \citep{CrowtherSmith:1997}. Similarly, for a single undiluted WC\,4-5 star, the strength of the C\,{\sc iv} $\lambda 5808$ line was assumed to be $W_{\lambda}(5808)\approx1400\,\AA$ \citep{CrowtherHadfield:2006}. In a binary, the $W_{\lambda}$ would fall as it is diluted by the additional continuum flux of the companion. This reduction in $W_{\lambda}$ relates to the ratio of the continuum fluxes of the two stars, and in turn gives the difference in their apparent magnitudes. As some WN stars do show variation in these line widths, in the case of binary WN/C star (BAT99-92), comparisons were made to the other single WN/C star (BAT99-88) for which $W_{\lambda}(4686)=225\,\AA$ and $W_{\lambda}(5808)=1600\,\AA$

The necessary corrections made to the photometry are given in Table \ref{tab:W-R photometry}. Further spectroscopic comments on the VFTS binary candidates are made below.

ID\# 24/VFTS 019/BAT99-86/Brey 69 - \citet{Foellmi:2003} had classified this star as WN3o+O9: but noted no radial velocity (RV) changes over their epochs. VFTS spectra also showed no RV variation and no absorption either, even in the Balmer lines, supporting a single system. Note, however, upon further inspection of the Pickering lines, a WN\,3(h) classification was favoured over that listed in Table 2 of Paper I.

ID\#375/VFTS 402/BAT99-95/Brey 80 - Absorption was noted in the Balmer and He\,{\sc i} $\lambda4387$ lines. This coincided with RV changes (but no absorption) in the He\,{\sc ii} $\lambda4200$ \& $\lambda4541$ lines. The only exception being He\,{\sc ii} $\lambda6683$ and N\,{\sc iv} $\lambda4057$, both showing variations caused by a possible OB companion.

ID\#543+544/VFTS 507/BAT99-101+102/Brey 87 - The previous \citet{Bartzakos:2001} classification was used to split our narrowband photometry in Table\ref{tab:W-R photometry} into the respective R140a1 and R140a2 components, but we were unable to classify O-type star companion from VFTS spectra.

ID\# 545/VFTS 509/BAT99-103/Brey 87 - The VFTS spectra favoured a WN\,5(h) classification based on the N\,{\sc iv}/N\,{\sc iii} ratio. The two companions noted were suggested by a series of absorption features. The first came from a set of absorption lines at He\,{\sc i} ($\lambda 4026$ \& $\lambda4471$) showing a large velocity shift with respect to the rest frame. This shifted absorption was reflected in the Balmer lines and slightly in the He\,{\sc ii} lines although not always at consistent epochs. This favoured an O-type orbiting companion. Additional central underlying absorption appeared in the Balmer, He\,{\sc i} and He\,{\sc ii} lines, this time showing no velocity shift. This favoured an early O-type star that was separate from the system.

ID\# 727/VFTS 542/BAT99-113  - Details for the new classification can be found in \citet{CrowtherWalborn:2011}. Regarding the binary nature, increased absorption in the Balmer lines was noted coinciding with slightly narrower He\,{\sc i} $\lambda 4026$ absorption. He\,{\sc ii} lines showed small radial velocity changes but no changes to strength to suggest a late O/early B companion.

ID\# 928/VFTS 682 - A new W-R star identified in Paper I. Classified a WN\,5h, the VFTS spectra showed no RV variations to suggest a companion. See also \citet{Bestenlehner:2011}.

ID\# 938/VFTS 695/BAT99-119/Brey 90 - Absorption features present in He\,{\sc ii} $\lambda 4200$ \& $\lambda 4541$ with absorption coinciding in He\,{\sc i} $\lambda 4026$ \& $\lambda 4471$\ on at least one epoch, if not more. However, Balmer lines showed minimal variations, with no changes to H$\gamma$ and only small changes in H$\delta$ peak intensity. A robust classification for the companion remains unknown.

\begin{figure}
\includegraphics[height=0.45\textheight]{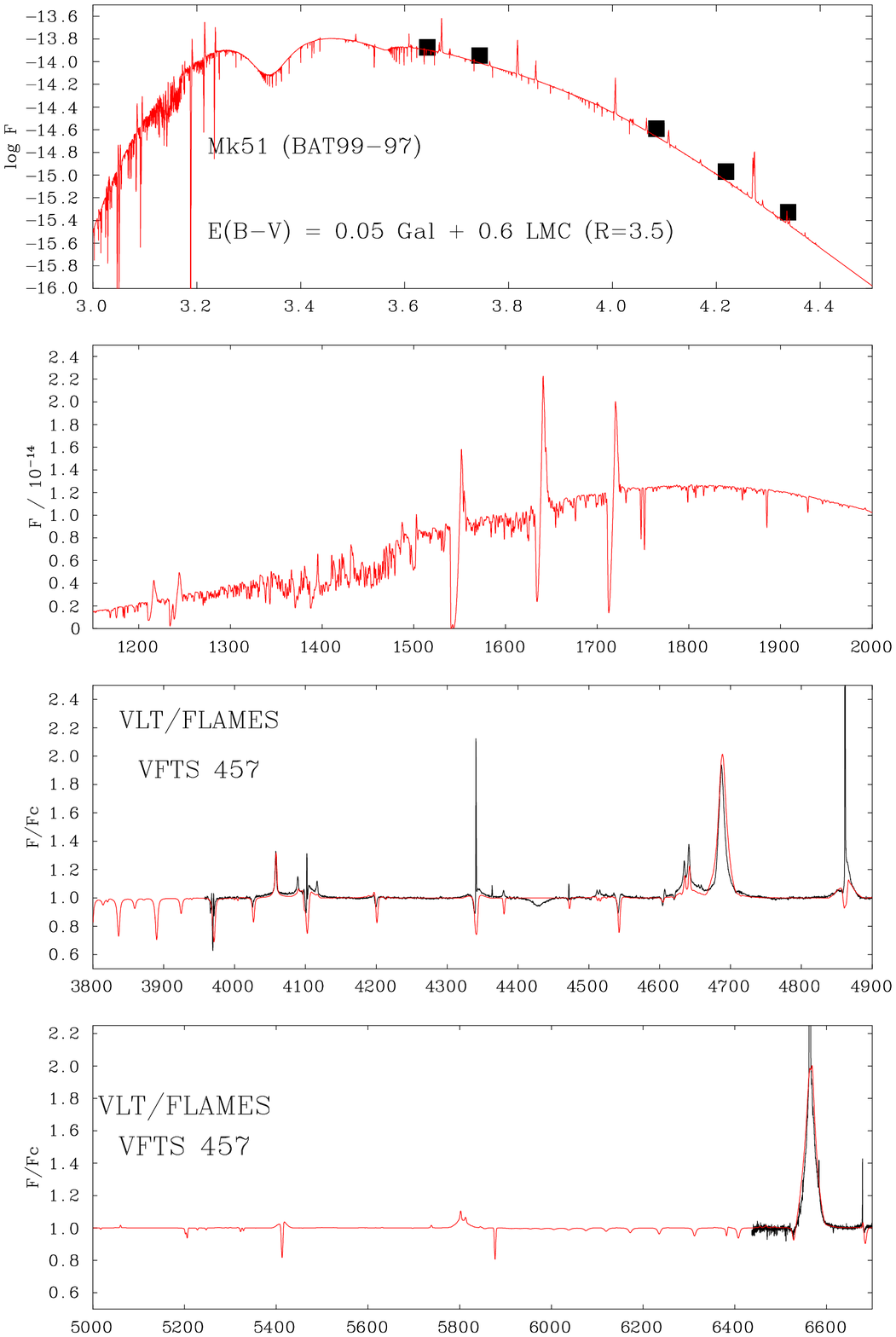}
\caption{Spectral fitting for O3.5\,If*/WN\,7 template star Melnick 51/VFTS 457.}
\end{figure}

\begin{figure}
\includegraphics[height=0.45\textheight]{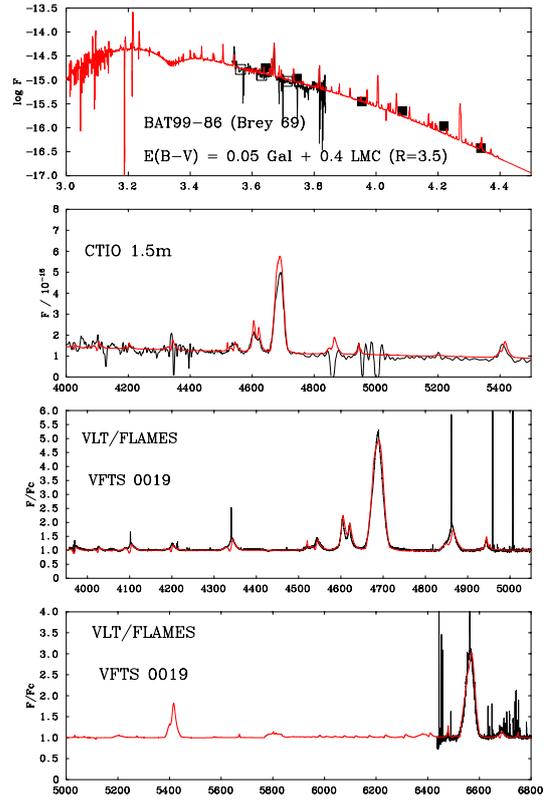}
\caption{Spectral fitting for WN\,3 template star BAT99-86/Brey 69/VFTS 019.}
\end{figure}

\begin{figure}
\includegraphics[height=0.45\textheight]{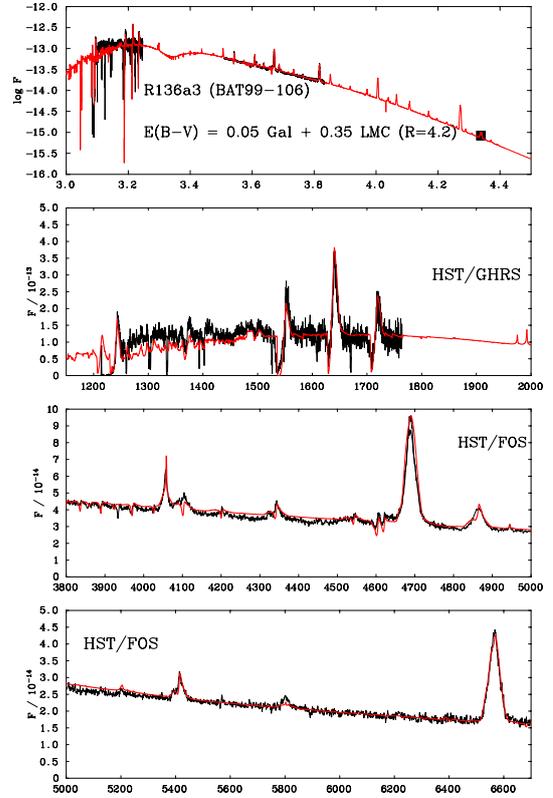}
\caption{Spectral fitting for WN\,5 template star BAT99-106/Brey 82/R136a3.}
\end{figure}

\begin{figure}
\includegraphics[width=0.45\textwidth]{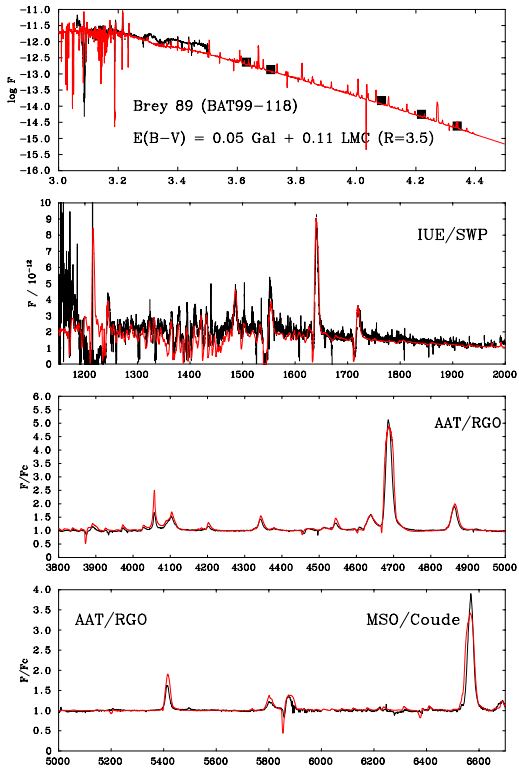}
\caption{Spectral fitting for WN\,6 template star BAT99-118/Brey 89.}
\end{figure}

\begin{figure}
\includegraphics[width=0.45\textwidth]{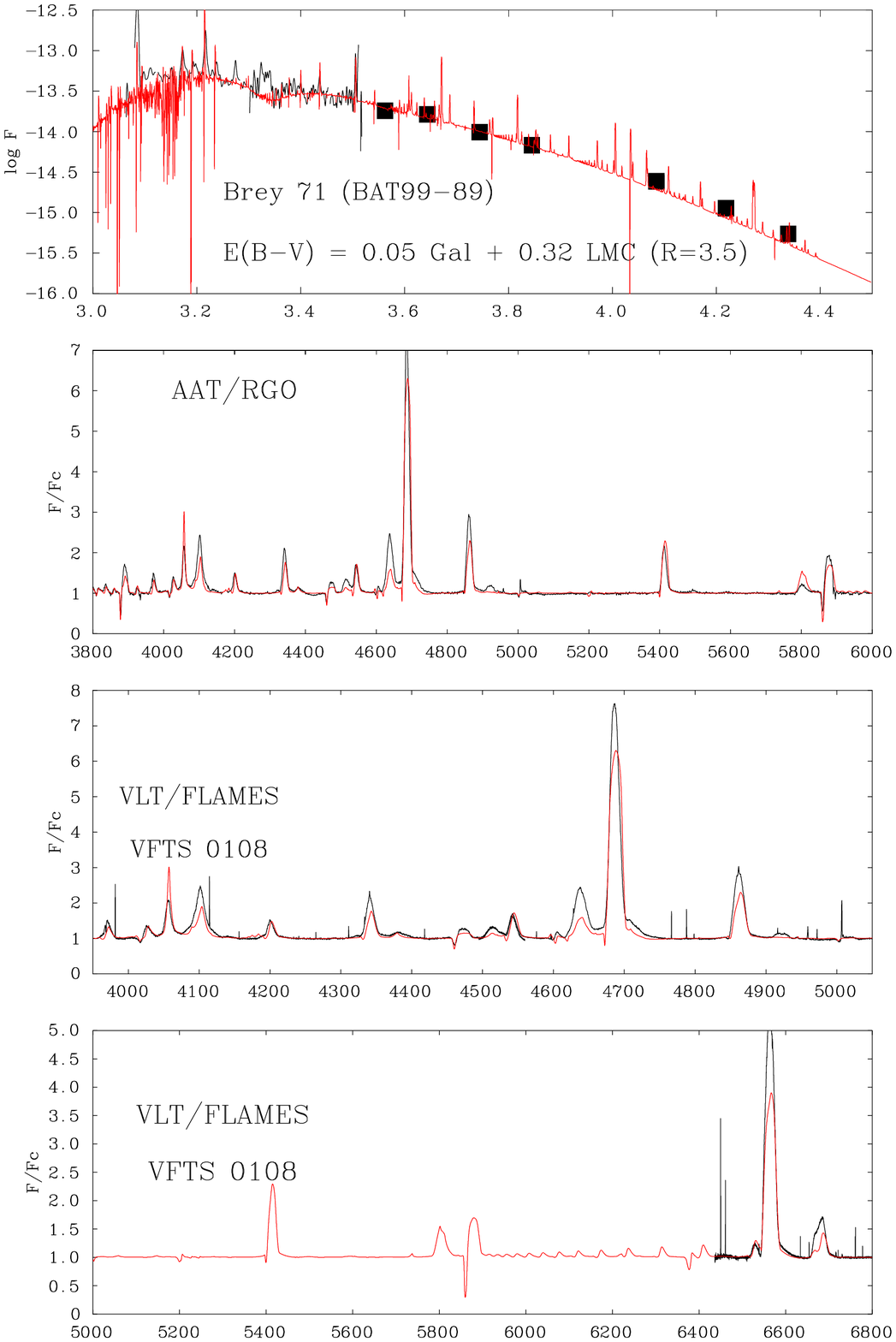}
\caption{Spectral fitting for WN\,7 template star BAT99-89/Brey 71/VFTS 108.}
\end{figure}

\begin{figure}
\includegraphics[height=0.45\textheight]{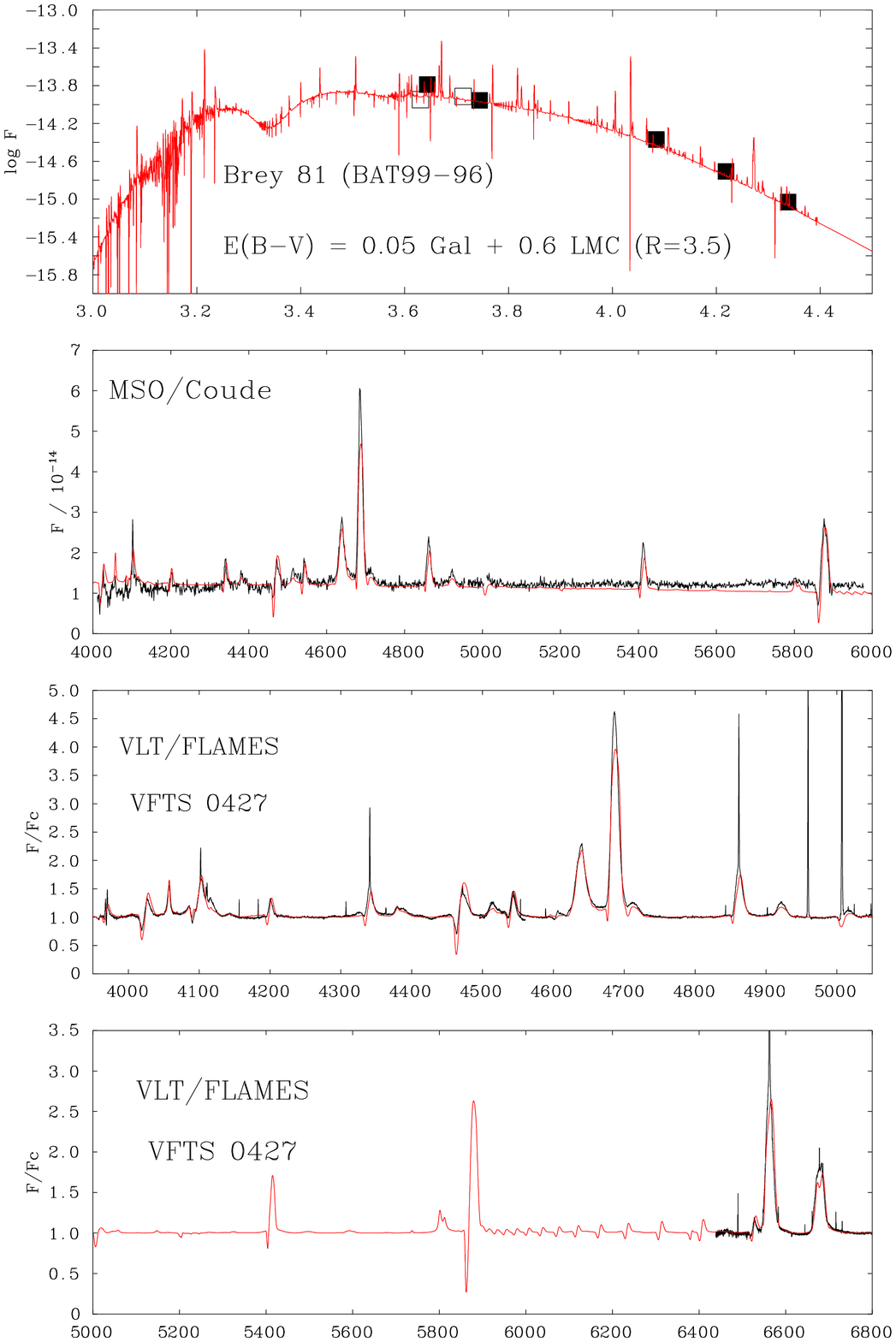} 
\caption{Spectral fitting for WN\,8 template star BAT99-96/Brey 81/VFTS 427.}
\end{figure}

\begin{figure}
\includegraphics[height=0.45\textheight]{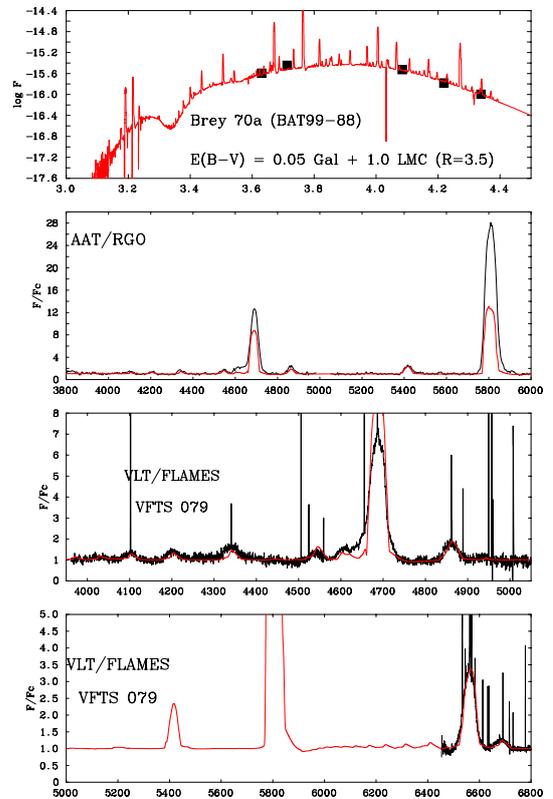}
\caption{Spectral fitting for WN/C template star BAT99-88/Brey 70a/VFTS 079.}
\end{figure}

\begin{figure}
\includegraphics[height=0.45\textheight]{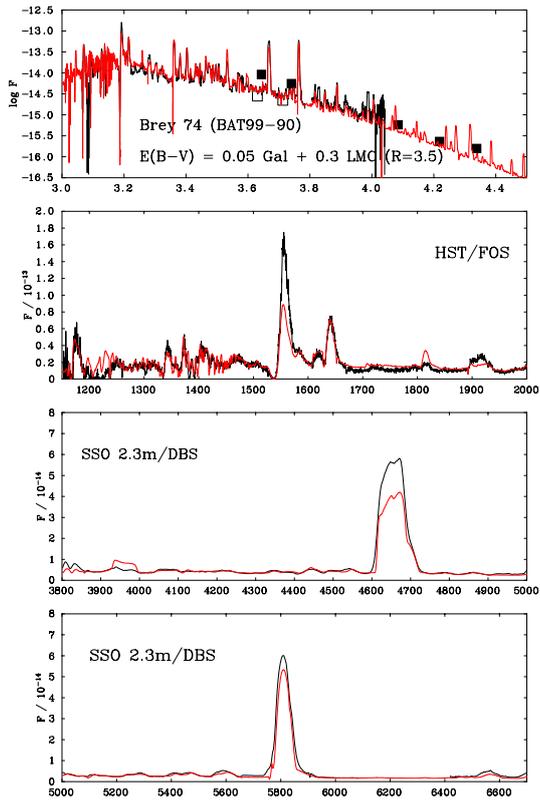}
\caption{Spectral fitting for WC\,4 template star BAT99-90/Brey 74.}
\label{fig:plot_bat90}
\end{figure}

\begin{landscape}
\begin{table}
\begin{center}
\caption{Template W-R and Of/WN star models. Input spectra were both flux calibrated (bold) and non-flux calibrated. Values in parentheses are given in broad band and apply to Of/WN stars. Narrow band is used for all other W-R stars. All properties except for $(m_{V})/m_{v}$ are model derived. Abundance ratios are given in H:He:N for Of/WN and WN stars, He:C:N for WN/C stars and He:C:O for WC stars. See Figures \ref{fig:plot_sk6722}-\ref{fig:plot_bat90} for model fits}
\label{tab:W-R and Of/WN star parameters}
\centering
\begin{tabular} { l @{\hspace{2mm}} l @{\hspace{2mm}} l @{\hspace{2.5mm}} l  c @{\hspace{2mm}} c @{\hspace{2mm}} c  c @{\hspace{2mm}} c @{\hspace{2mm}} c @{\hspace{2mm}} c c @{\hspace{2mm}} c  c  c @{\hspace{2mm}} c @{\hspace{2mm}} c  r}
\multicolumn{2}{l}{Star} & Spectral & Spectra  & $(m_{V})/m_{v}$ & $(A_{V})/A_{v}$ &  $(M_{V})/M_{v}$  & $T_{*}$ &  $T_{\mathrm{eff}}$ & $(BC_{V})/BC_{v}$ & $\log L$ & $\varv_{\infty}$  & $\dot{M}$ & Abundance & $L_{\mathrm{EUV}}$ & $\log q_{\mathrm{0}}$  & $\log Q_{\mathrm{0}}$ & Ref.\\
BAT99 & Alias & Type &  & [mag] & [mag] & [mag] & [kK] & [kK] & [mag] &  [L$_{\odot}$] & [km s$^{-1}$] & [M$_{\odot}$ yr$^{-1}$] & (by mass) & $[L_{\mathrm{Bol}}]$ & [ph cm$^{-2}$ s$^{-1}$] & [ph s$^{-1}$] \\
\hline
\hline
12 & Sk -67$^{\circ}$ 22 & O2\,If*/WN\,5  & {\bf a}, b & (13.49) & (0.4) & $(-5.4)$ & 49.8 & 49.3 & $(-4.4)$ & 5.83 & 2650 & $3.7\times10^{-6}$ & 60:39:0.3 & 0.64 & 24.79 & 49.67 & (1)+(2) \\

97 & Mk 51 & O3.5\,If*/WN\,7  & c & (13.74) & (2.2) & $(-6.9)$ & 41.0 & 40.8 & $(-3.7)$ & 6.14 & 1500 & $4.7\times10^{-6}$ & 60:39:0.5 & 0.49 & 24.40 & 49.83 & (1) \\

\hline
86  & Brey 69 & WN\,3(h)  & c & 16.68 & 1.98 & $-3.75$ & 79.3 & 74.9 & $-5.2$ & 5.43 & 1750 & $3.2\times10^{-5}$& 20:79:0.5 & 0.85 & 25.63 & 49.34 & (1) \\

106  & R136a3 & WN\,5h  &  {\bf d, e, f}  & 13.02 & 1.86 & $-7.29$ & 53.0 & 53.0 & $-4.5$ & 6.58 & 2200 & $3.7\times10^{-5}$ & 40:59:0.4 & 0.69 & 24.90 & 50.47 & (3) \\

118 & Brey 89 & WN\,6h\tablefootmark{\star} & {\bf g, h, i} & 11.20 & 0.58 & $-7.83$ & 45.0 & 39.4 & $-3.5$ & 6.41 & 1450 & $8.0\times10^{-5}$ &20:79:0.5 & 0.54 & 24.62 & 50.26  & (1) \\

89  & Brey 71 & WN\,7h & {\bf g}, c  & 14.13 & 1.53 & $-5.85$ & 49.8 & 44.5 & $-3.6$ & 5.75 & 1000 & $1.9\times10^{-5}$ & 15:84:0.5 & 0.58 & 24.96 & 49.61 & (1)  \\

96  & Brey 81 & WN\,8(h)  & {\bf h}, c & 13.82 & 2.40 & $-7.03$ & 42.5 & 36.9 & $-3.4$ & 6.04 & 850 & $4.0\times10^{-5}$ & 6:93:0.3 & 0.48 & 24.49 & 49.82 & (1) \\

\hline
88  & Brey 70a & WN\,4b/WCE  &  c, g & 17.75 & 4.04 & $-4.74$ & 80.0 & 72.1 & $-5.0$ & 5.78 & 1750 & $1.5\times10^{-5}$ & 99:0.5:0.1 & 0.82 & 25.66 & 49.67 & (1) \\

90  & Brey 74  & WC\,4  &  {\bf e, j} & 15.41 & 1.48 & $-4.52$ & 85.0 & 72.0 & $-4.7$ & 5.57 & 2600 & $1.6\times10^{-5}$ & 45:43:11 & 0.72 & 25.76 & 49.31 & (4) \\

\hline
\end{tabular}
\end{center}

\tablefoot{
\tablefootmark{\star}{Recently revealed as a WN\,5-6+WN\,6-7 binary system by \citet{Sana:2013b} but for our template model we adopt properties in the case of a single star.}\\
Spectra: a - Hubble Space Telescope - Space Telescope Imaging Spectrograph (HST-STIS), b - Very Large Telescope - Fibre Large Array Multi-Element Spectrograph - Ultraviolet and Visual Echelle Spectrograph (VLT-FLAMES-UVES), c - VLT - FLAMES - Multi-object spectrometry mode (VLT-FLAMES-MEDUSA), d - HST-Goddard High-Resolution Spectrograph (HST-GHRS), e - HST-Faint Object Spectrograph (HST-FOS), f - VLT - Spectrograph for Integral Field Observations in the Near Infrared (VLT-SINFONI), g - Anglo Australian Telescope - Royal Greenwich Observatory Spectrograph (AAT-RGO), h - Mount Stromlo Observatory Coud\'{e} Spectrograph (MSO), i - International Ultraviolet Explorer (IUE-HIRES), j - Siding Spring Telescope - Dual-Beam Spectrograph (SSO-DBS).\\
References: (1) - This work, (2) - \citet{DoranCrowther:2011}, (3) - \citet{Crowther:2010}, (4) - \citet{Crowther:2002}.
}
\end{table}
\end{landscape}

\begin{landscape}
\begin{table}
\caption{The photometric properties of all the W-R stars in 30 Dor. Values in \emph{italic} indicate where magnitudes or colours have been assumed. 
The $W_{\lambda}$ of the He\,{\sc ii} $\lambda 4686$ line for WN stars or C\,{\sc iv} $\lambda 5808$ line for WC stars, is listed with references. For multiple systems, a magnitude correction is given, where $\Delta M_{v}=M^{\mathrm{W-R}}_{v} - M^{\mathrm{OB}}_{v}$. Note that the  template W-R stars will have had different $A_{v}$ and $M^{\mathrm{sys}}_{v}$ values derived from their CMFGEN models, as listed in Table \ref{tab:W-R and Of/WN star parameters}, and that the properties given in that table are adopted for the star. Similarly, for comparison, this table gives photometry for stars BAT99-90, 106, 108, 109, 112 but their stellar properties were equivalently adopted from \citet{Crowther:2002} or \citet{Crowther:2010} and were not derived from the values in this table. See text for more details.}
\label{tab:W-R photometry}
\newcounter{foo}
\begin{tabular} {l  l @{\hspace{3mm}} l @{\hspace{3mm}} l  l  l  c @{\hspace{2mm}} c @{\hspace{2mm}} c @{\hspace{2mm}} c  c @{\hspace{2mm}} c @{\hspace{2mm}} c c  @{\hspace{2mm}} c  c  c  c @{\hspace{2mm}} c @{\hspace{2mm}} c }
ID & \multicolumn{3}{c}{Star} & Spectral & Ref. & & $m_{v}$ & $b-v$ & Ref. & & $W_{\lambda}(4686)$ & $W_{\lambda}(5808)$ &  Ref. & $(b-v)_{0}$ & $R_{v}$\tablefootmark{a} & $A_{v}$ & $M^{\mathrm{sys}}_{v}$ & $\Delta M_{v}$ & $M^{\mathrm{W-R}}_{v}$ \\ 
\# & BAT99 & VFTS & Alias & Type & & & [mag] & [mag] & &  & [\AA] & [\AA] & & [mag] & & [mag] & [mag] & [mag] & [mag] \\ 
\hline
24 & 86 & 019 & Br 69          &  WN\,3(h)    &    (\ref{Paper I})\tablefootmark{b}      &  &16.68    &  0.11  & (\ref{SchmutzVacca:1991}) & & 110 & - & (\ref{Paper I}) & $-0.27$  & 4.6 & 1.7 & $-3.6$ & - & -  \\

92 & 88 & 079 & Br 70a        &  WN\,4b/WCE    &    (\ref{Foellmi:2003})   & &  17.75    &  0.46  & (\ref{CrowtherHadfield:2006}) & & 225 & 1660 & (\ref{Paper I},\ref{CrowtherHadfield:2006}) & $-0.23$ & 4.6 & 3.2 & $-3.9$ & - & -    \\

117 & 89 & 108 & Br 71          &  WN\,7h       &     (\ref{CrowtherSmith:1997})   & & 14.13    &  0.22  & (\ref{CrowtherSmith:1997}) & & 96 & - & (\ref{Paper I}) & $-0.28$  & 4.6 & 2.3 & $-6.7$ & - & -    \\

144 & 90 & 136 & Br 74         &  WC\,4  &  (\ref{Smith:1990}) & & 15.42    &  0.12 & (\ref{CrowtherHadfield:2006})  &   & - & 1451 & (\ref{CrowtherHadfield:2006}) & $-0.16$ & 4.6 & 1.3 & $-4.4$ & - & -   \\

155 & 91 & 147 & Br 73-1A             &  WN\,6(h)      &     (\ref{Paper I})   & &  14.98  &  \emph{0.15} & (\ref{Walborn:1999}) &  & 39 & - &  (\ref{Paper I}) & $-0.30$  & 4.6 & 2.1 & $-5.6$ & - & -   \\

185 & 92 & & Br 72, R130                  &  WCE/WN+B1\,I     &   (\ref{ContiMassey:1989})   & &  11.47   &  0.03 & (\ref{SchmutzVacca:1991}) & & 14 & 73 & (\ref{Torres-DodgenMassey:1988},\ref{Schnurr:2008}) & $-0.31$  & 4.6 & 1.6 & $-8.6$ & $+3.2$\tablefootmark{c} & $-5.4$  \\

375 & 95 & 402 & Br 80, R135              &  WN\,7h+OB     &     (\ref{Paper I})    & &  13.04    &  0.09  & (\ref{CrowtherSmith:1997},\ref{Torres-DodgenMassey:1988}) &  & 80 & - & (\ref{Paper I}) & $-0.28$  & 4.6 & 1.7 & $-7.2$ & $-3.5$ & $-7.2$   \\

402 & 96 & 427 & Br 81           &  WN\,8(h)     &    (\ref{CrowtherSmith:1997})     & & 13.82    &  0.49   & (\ref{CrowtherSmith:1997}) & & 40 & - & (\ref{Paper I}) & $-0.30$  & 4.6 & 3.6 & $-8.3$ & - & -    \\

443 & 98 & & Br 79, Mk 49                      &  WN\,6(h)    &    (\ref{CrowtherDessart:1998})     & & 13.37    &  0.10  & (\ref{CrowtherSmith:1997}) & & 19 & - & (\ref{Schnurr:2008}) & $-0.31$   & 4.6 & 1.9 & $-7.0$ & - & -   \\

493 & 100 & 1001 & Br75, R134        &  WN\,7h    &     (\ref{CrowtherSmith:1997})     & & 12.40    &  0.14  & (\ref{CrowtherSmith:1997}) & & 27 & - & (\ref{Schnurr:2008}) & $-0.31$  & 5.4 & 2.4 & $-8.5$ & - & -  \\

543 & 101 & 507 & Br 87, R140a1  &  WC\,4(+WN\,6+O)   &   (\ref{Bartzakos:2001})   & \ldelim \{ {2}{1mm} & \emph{12.50}   &  \emph{0.03}  & (\ref{Breysacher:1986},\ref{DeMarchi:2011}) &  & - & 170\tablefootmark{d} & (\ref{Paper I}) & \emph{-0.32} & 4.6 & 1.6 & $-7.6$ & +2.0 &  $-5.5$ \\

544 & 102 & 507 & Br 87, R140a2  &  WN\,6+(O)   &   (\ref{Moffat:1987})   & & \emph{12.50}   &  \emph{0.03}  & (\ref{Breysacher:1986},\ref{DeMarchi:2011}) &  & 53 & - & (\ref{Paper I}) & $-0.31$ & 4.6 & 1.6 & $-7.6$ & $-1.0$ & $-7.2$ \\ 

545 & 103 & 509 & Br 87, R140b      &  WN\,5(h)+O(+early O) &  (\ref{Paper I})  & & 12.80   &  \emph{0.03} & (\ref{Breysacher:1986},\ref{DeMarchi:2011})  &  & 42 & - & (\ref{Paper I})  & $-0.30$  & 4.6 & 1.5 & $-7.2$ & $-0.3$ & $-6.6$ \\

613 & 106 & & Br 82, R136a3               &  WN\,5h  & (\ref{CrowtherDessart:1998}) & & 13.02    &  0.05   & (\ref{deKoter:1997}) & &  55 & - & (\ref{deKoter:1997}) & $-0.29$ & 5.4 & 1.8 & $-7.3$ & - & -   \\

630 & 108 & & Br 82, R136a1               &  WN\,5h  &  (\ref{CrowtherDessart:1998}) & & 12.23    &  0.03   & (\ref{deKoter:1997}) & & 37 & - & (\ref{deKoter:1997})  & $-0.30$ & 5.4 & 1.8 & $-8.1$ & - & -   \\  

633 & 109 & & Br 82, R136a2               &  WN\,5h &  (\ref{CrowtherDessart:1998}) & & 12.77    &  0.06    & (\ref{deKoter:1997}) & & 41  & - & (\ref{deKoter:1997})  & $-0.30$ & 5.4 & 2.0 & $-7.7$ & - & -   \\

706 & 112 & 1025 & Br 82, R136c       &  WN\,5h  &  (\ref{CrowtherDessart:1998}) & & 13.39  &   \emph{0.20}   & (\ref{MasseyHunter:1998},\ref{DeMarchi:2011}) & &  54 & - & (\ref{deKoter:1997}) & $-0.30$ & 5.4 & 2.7 & $-7.8$ & - & -   \\ 

762 & 115 & & Br 83, Mk33Sb               &  WC\,5      &     (\ref{MasseyHunter:1998})      & & 15.23  &  \emph{0.12}  & (\ref{MasseyHunter:1998}) &  & - & 1100\tablefootmark{d} & (\ref{MasseyHunter:1998}) & \emph{-0.28}  & 5.4 & 2.2 & $-5.4$ & - & -  \\ 

770 & 116 & & Br 84, Mk34                  &  WN\,5h     &    (\ref{CrowtherDessart:1998})     & &  \emph{13.10}    &  \emph{0.14} & (\ref{MasseyHunter:1998},\ref{DeMarchi:2011}) & & 30 & - & (\ref{Schnurr:2008}) & $-0.33$  & 5.4 & 2.6 & $-7.9$ & - & -  \\

862 & 117 & 617 & Br 88, R146           &  WN\,5ha     &     (\ref{Foellmi:2003})     & & 12.95    &  $-0.16$  & (\ref{CrowtherHadfield:2006}) & & 37 & - & (\ref{Paper I}) & $-0.30$   &  4.6 & 0.6 & $-6.2$ & - & -   \\

916 & 118 & & Br 89                      &  WN\,6h\tablefootmark{e}     &      (\ref{Smith:1996})     & & 11.20    &  $-0.17$   & (\ref{SchmutzVacca:1991}) & & 60 & - & (\ref{Schnurr:2008}) & $-0.29$   & 4.6 & 0.6 & $-7.8$ & - & -  \\

928 & 118a &  682   &           &  WN\,5h    &      (\ref{Paper I})      & & \emph{16.30}   &  \emph{0.60}   & (\ref{Paper I},\ref{Breysacher:1986}) &  & 39 & - & (\ref{Paper I})  & $-0.30$   & 4.6 & 4.1 & $-6.3$\tablefootmark{f} & - & -  \\

938 & 119 & 695 & Br 90, R145        &  WN\,6h+?    &     (\ref{Paper I})     & & 12.06    &  0.03   & (\ref{SchmutzVacca:1991}) & & 39 & - & (\ref{Paper I})  & $-0.30$  & 4.6 & 1.5 & $-7.9$ & $-0.1$ & $-7.2$   \\

973 & 121 & 731 & Br 90a           &  WC\,4      &     (\ref{Smith:1990})      & & 17.21    &  0.31   & (\ref{CrowtherHadfield:2006}) & & - & 1258 & (\ref{CrowtherHadfield:2006}) & $-0.28$   & 4.6 & 2.8 & $-4.1$ & - & -   \\

1001 & 122 & 758 & Br 92, R147          &  WN\,5h   &      (\ref{Paper I})     & & 12.75    &  0.03  & (\ref{CrowtherHadfield:2006}) &  & 76 & - & (\ref{Paper I}) & $-0.29$   & 4.6 & 1.5 & $-7.2$ & - & -   \\

\hline
\end{tabular}
\tablefoot{
The photometric data given for stars BAT99-101 and 102 are for the combined R140a system. 
References: (\refstepcounter{foo}\thefoo\label{Paper I}) - Paper I, 
(\refstepcounter{foo}\thefoo\label{Foellmi:2003}) - \citet{Foellmi:2003}, 
(\refstepcounter{foo}\thefoo\label{CrowtherSmith:1997}) - \citet{CrowtherSmith:1997},
(\refstepcounter{foo}\thefoo\label{Smith:1990}) - \citet{Smith:1990},
(\refstepcounter{foo}\thefoo\label{ContiMassey:1989}) - \citet{ContiMassey:1989}
(\refstepcounter{foo}\thefoo\label{CrowtherDessart:1998}) - \citet{CrowtherDessart:1998},
(\refstepcounter{foo}\thefoo\label{Bartzakos:2001}) - \citet{Bartzakos:2001},
(\refstepcounter{foo}\thefoo\label{Moffat:1987}) - \citet{Moffat:1987},
(\refstepcounter{foo}\thefoo\label{MasseyHunter:1998}) - \citet{MasseyHunter:1998},
(\refstepcounter{foo}\thefoo\label{Smith:1996}) - \citet{Smith:1996},
(\refstepcounter{foo}\thefoo\label{SchmutzVacca:1991}) - \citet{SchmutzVacca:1991},
(\refstepcounter{foo}\thefoo\label{CrowtherHadfield:2006}) - \citet{CrowtherHadfield:2006},
(\refstepcounter{foo}\thefoo\label{Walborn:1999}) - \citet{Walborn:1999},
(\refstepcounter{foo}\thefoo\label{Torres-DodgenMassey:1988}) - \citet{Torres-DodgenMassey:1988},
(\refstepcounter{foo}\thefoo\label{deKoter:1997}) - \citet{deKoter:1997},
(\refstepcounter{foo}\thefoo\label{Breysacher:1986}) - \citet{Breysacher:1986},
(\refstepcounter{foo}\thefoo\label{DeMarchi:2011}) - \citet{DeMarchi:2011},
(\refstepcounter{foo}\thefoo\label{Schnurr:2008}) - \citet{Schnurr:2008}.
}
\tablefootmark{a}{Narrow band $R_{v}$ derived from average broadband $R_{V}$ determined for each region (see Appendix \ref{sec:Reddening_appendix}).}\\
\tablefootmark{b}{SpT is amended from WN\,3o to WN\,3(h) following further inspection of the Pickering series.}\\
\tablefootmark{c}{Based on $W_{\lambda}$ dilution of C\,{\sc iv} $\lambda5808$ line.}\\
\tablefootmark{d}{$W_{\lambda}$ of C\,{\sc iv} $\lambda4650$ line (undiluted $W_{\lambda}(4650)\approx1200$) based on BAT99-90.}\\
\tablefootmark{e}{Revealed as a WN\,5-6+WN\,6-7 binary system by \citet{Sana:2013b}}\\
\tablefootmark{f}{$M^{\mathrm{sys}}_{v}$ was estimated using broad band photometry and $v-\mathrm{V} \approx 0.2$\,mag relation from \citet{Breysacher:1986}. However, high reddening allows for a large uncertainty and the star is thought to be more comparable to BAT99-106 (F. Najarro).}\\
\end{table}
\end{landscape}

\section{Reddening}
\label{sec:Reddening_appendix}
A recent tailored analysis, based on the VFTS O-type stars, indicates a varied amount of extinction across the different regions of 30 Dor \citetext{Ma\'{i}z Apell\'{a}niz et al. in prep.}. For now we apply a uniform reddening value of $R_{V}=3.5$ within the MEDUSA region and $R_{V}=4.2$ within the R136 region. These values were derived using additional $K$-band photometry of the census stars, VLT-MAD observations from \citet{Campbell:2010} supplying the R136 region and the InfraRed Survey Facility (IRSF, see Table 6 of Paper I) supplying other stars. \citet{MartinsPlez:2006} offer a $(V-K)_{\mathrm{o}}$ calibration for O-type stars allowing an expected $M_{V}$ to be derived from the $K$-band photometry. This $K$-band derived $M_{V}$ would become consistent with the conventionally derived $M_{V}$ from optical photometry, at an average $R_{V}$ value.

\begin{figure}
\centering
\subfigure[]{\includegraphics[width=0.45\textwidth]{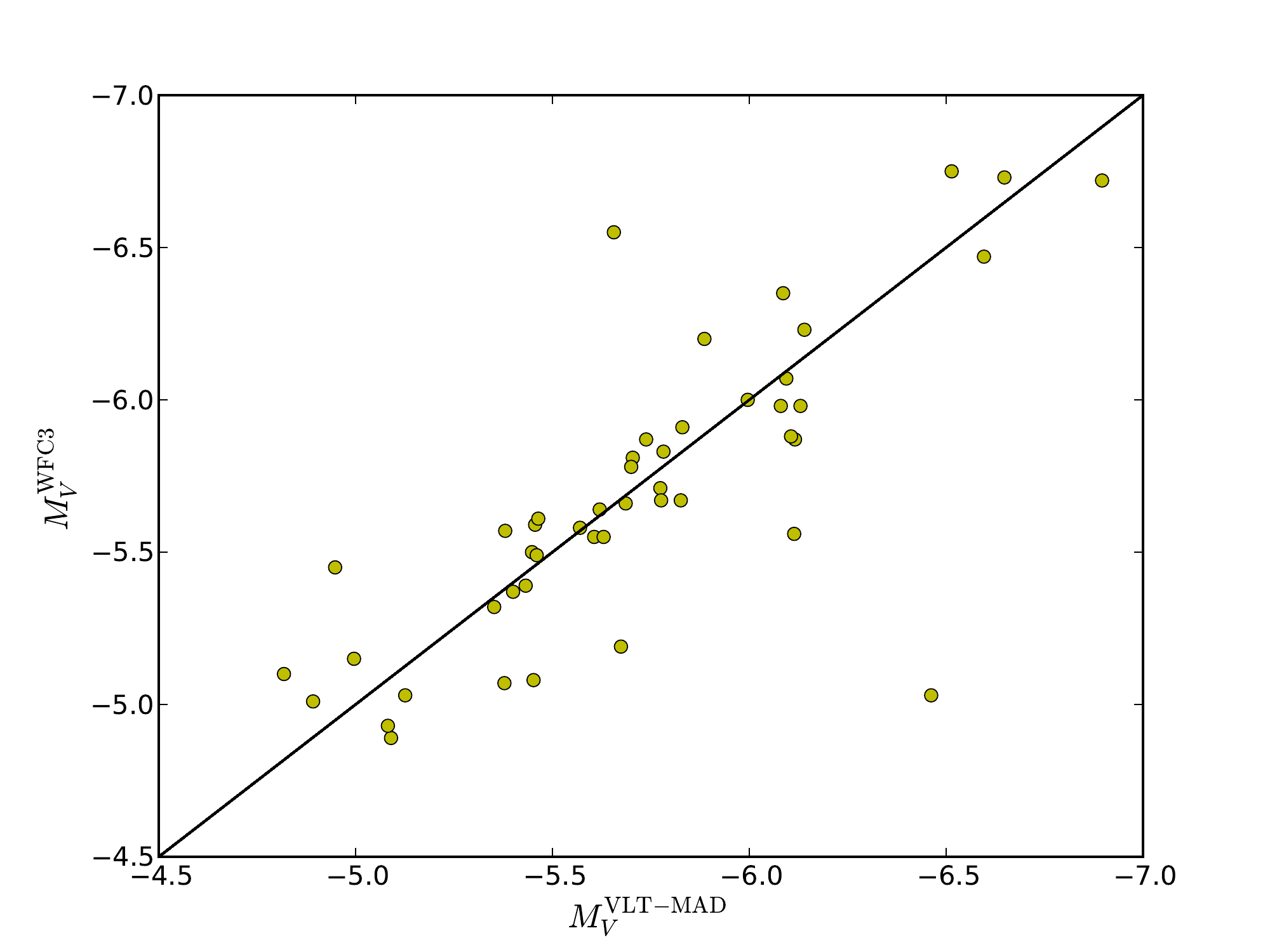}
\label{fig:R136_RV}}
\subfigure[]{\includegraphics[width=0.45\textwidth]{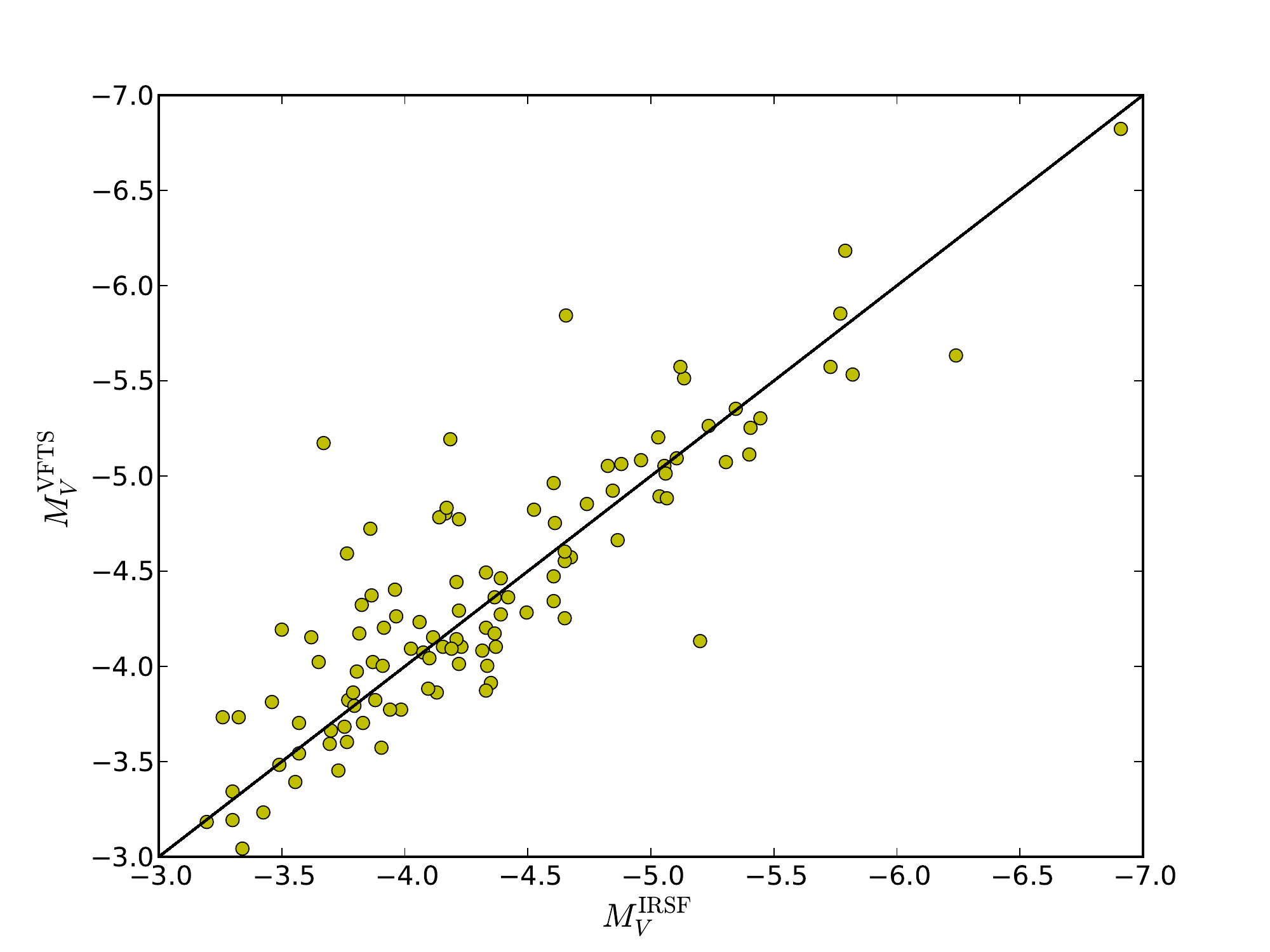}
\label{fig:30Dor_RV}}
\caption{Determination of $R_{V}$ for 30 Dor. In the R136 region (a): The $M_{V}$ of single O-type stars derived from HST/WFC3 photometry is plotted against the expected $M_{V}$ derived from VLT-MAD $K$-band photometry after applying the \citet{MartinsPlez:2006} calibration. In the MEDUSA region (b): The $M_{V}$ derived from VFTS photometry is plotted against the expected $M_{V}$ derived from IRSF $K$-band photometry after applying the \citet{MartinsPlez:2006} calibration. Solid lines indicate the 1:1 relations.}
\label{fig:RV}
\end{figure}

For both regions, a $K$-band extinction of $A_{K}=0.17$\,mag was adopted. Figure \ref{fig:R136_RV} compares the $K$-band derived $M_{V}$ ($M^{\mathrm{VLT-MAD}}_V$) in the R136 region to the $M_{V}$ from the original HST/WFC3 photometry of \citet{DeMarchi:2011} ($M^{\mathrm{WFC3}}_V$). In theory, both values would be equal with all points falling on the 1:1 relationship line. Despite some scattering, most likely where VLT-MAD photometry became less reliable in the core, a value of $R_{V}=4.2$ was favoured within R136. In the case of the MEDUSA region (Figure \ref{fig:30Dor_RV}), only known single VFTS O-type stars were used in the comparison. The $K$-band derived $M_{V}$ ($M^{\mathrm{IRSF}}_V$) was compared to the $M_{V}$ from the original VFTS photometry ($M_{V}^{\mathrm{VFTS}}$), i.e. using Selman, WFI, Parker or CTIO photometry, and favoured a lower value of $R_{V}=3.5$.

An equivalent narrowband $R_{v}$ was required to match the photometry of the W-R stars. We determined this using Equation 24 of \citet{SchmutzVacca:1991}.

\newpage

\section{The Census}
\label{sec:The Census}

\begin{table*}
\begin{center}
\caption{The candidate hot luminous stars in the census as selected by the criteria explained in Section \ref{sec:Candidate selection}. Stars are listed in order of increasing right ascension. The first 20 candidate stars are given here, see online data for full census. References for the photometry are as follows: dM - \citet{DeMarchi:2011}, S - Selman, W - WFI, P - Parker, C - CTIO, Z - \citet{Zaritsky:2004} and Wal - \citet{Walborn:1999}. Popular aliases are also included from the following catalogues: VFTS - Paper I, HSH95 - \citet{Hunter:1995}, S99 - \citet{Selman:1999}, P93 - \citet{Parker:1993}, ST92 - \citet{SchildTestor:1992} and Sk - \citet{Sanduleak:1970} catalogues.}
\label{tab:Candidate stars}
\begin{tabular} { l @{\hspace{2mm}}  l @{\hspace{2mm}} l @{\hspace{2mm}} r @{\hspace{2mm}} r @{\hspace{2mm}} r @{\hspace{2mm}} c @{\hspace{2mm}} ccccc}
ID & $\alpha$ & $\delta$ & $V$ & $B-V$ & Ref. & VFTS  & HSH95 & S99 & P93 & ST92 & Sk \\
\# & (h m s) & $(^{\circ}\,'\,'')$ & (mag) & (mag) & &  & \\
\hline
1 & 05 36 51.12 & $-69$ 05 12.20 & 16.63 & $ 0.67$ & W &     - &      - &     - &      - &       - &           -  \\
2 & 05 36 53.82 & $-69$ 04 30.13 & 17.00 & $ 0.32$ & W &     - &      - &     - &      - &       - &           -  \\
3 & 05 36 55.98 & $-69$ 07 36.70 & 17.36 & $ 0.61$ & W &     - &      - &     - &      - &       - &           -  \\
4 & 05 36 56.85 & $-69$ 05 59.36 & 17.98 & $ 0.68$ & W &     - &      - &     - &      - &       - &           -  \\
5 & 05 36 57.18 & $-69$ 08 47.40 & 15.85 & $ 0.04$ & W &     - &      - &     - &      - &       - &           -  \\
6 & 05 36 57.22 & $-69$ 07 23.33 & 18.33 & $ 0.80$ & W &     - &      - &     - &      - &       - &           -  \\
7 & 05 36 57.26 & $-69$ 05 00.01 & 15.11 & $ 0.69$ & W &     - &      - &     - &      - &       - &           -  \\
8 & 05 36 57.62 & $-69$ 03 44.04 & 14.56 & $ 0.61$ & W &     - &      - &     - &      - &       - &           -  \\
9 & 05 36 58.95 & $-69$ 08 21.66 & 15.33 & $ 0.01$ & W &     - &      - &     - &      - &       - &           -  \\
10 & 05 37 03.59 & $-69$ 10 22.32 & 16.71 & $ 0.32$ & W &     - &      - &     - &      - &       - &           -  \\
11 & 05 37 03.91 & $-69$ 10 03.60 & 13.01 & $ 0.20$ & Z &     - &      - &     - &      - &       - &           -  \\
12 & 05 37 04.01 & $-69$ 08 23.83 & 14.36 & $ 0.80$ & W &     - &      - &     - &      - &       - &           -  \\
13 & 05 37 04.26 & $-69$ 08 05.46 & 16.18 & $ 0.11$ & W &    009 &      - &     - &      - &       - &           -  \\
14 & 05 37 04.47 & $-69$ 04 00.08 & 15.56 & $ 0.20$ & W &     - &      - &     - &      - &       - &           -  \\
15 & 05 37 05.63 & $-69$ 09 12.20 & 15.83 & $ 0.15$ & W &    012 &      - &     - &      - &       - &           -  \\
16 & 05 37 06.29 & $-69$ 04 41.20 & 16.36 & $ 0.13$ & W &    013 &      - &     - &      - &       - &           -  \\
17 & 05 37 08.72 & $-69$ 04 45.51 & 17.36 & $ 0.47$ & W &     - &      - &     - &      - &       - &           -  \\
18 & 05 37 08.88 & $-69$ 07 20.39 & 13.55 & $ 0.04$ & Z &    016 &      - &     - &      - &       - &           -  \\
19 & 05 37 09.68 & $-69$ 02 21.39 & 15.88 & $-0.02$ & W &     - &      - &     - &      - &       - &           -  \\
20 & 05 37 10.17 & $-69$ 05 24.10 & 17.29 & $ 0.37$ & W &     - &      - &     - &      - &       - &           -  \\
\hline
\end{tabular}
\end{center}
\end{table*}

\begin{table*}
\begin{center}
\caption{Stellar parameters of the spectroscopically confirmed hot luminous stars in the census. Absolute magnitudes are given in narrow band for W-R stars and broad band in all other cases. Stars are listed in order of increasing projected distance ($r_{\mathrm{d}}$) from star R136a1. The first 20 stars are given here, see online data for full census. References for the spectral types are are as follows: ST92 - \citet{SchildTestor:1992}, P93 - \citet{Parker:1993}, WB97 - \citet{WalbornBlades:1997}, MH98 - \citet{MasseyHunter:1998}, CD98 - \citet{CrowtherDessart:1998}, B99 - \citet{Bosch:1999}, W99 - \citet{Walborn:1999}, BAT99 - \citet{Breysacher:1999}, Paper I - \citet{Evans:2011}, T11 - \citet{Taylor:2011}, D11 - \citet{Dufton:2011}, CW11 - \citet{CrowtherWalborn:2011}, HB12 - \citet{Henault-Brunet:2012a} , Walborn - Walborn et al. \citetext{in prep.}, McEvoy - McEvoy et al. \citetext{in prep.} and this work. Details of the derivation of the stellar parameters can be found in Section \ref{sec:Stellar Calibrations}.}
\label{tab:Feedback from stars}
\begin{tabular} { l  c  l @{\hspace{2mm}} r  c @{\hspace{2mm}} c @{\hspace{2mm}} c  c @{\hspace{2mm}} c @{\hspace{2mm}} c @{\hspace{2mm}} c @{\hspace{2mm}} c }
ID & $r_{\mathrm{d}}$  & Spectral & Ref. & $T$ & $M_{v}$ or $M_{V}$ & $\log L$ & $Q_{\mathrm{0}}$ & $\varv_{\infty}$ & $\dot{M}$ & $L_{\mathrm{SW}}$ & $D_{\mathrm{mom}}$ \\
\# & (pc) & Type & & (kK) & (mag) & $(\mathrm{L}_{\odot})$ & (ph s$^{-1}$) &  (km s$^{-1}$) & ($\mathrm{M}_{\odot}$ yr$^{-1}$)  & (erg s$^{-1}$) & (g cm$^{\frac{3}{2}}$ s$^{-2}$) \\
\hline
 630 & 0.00 & WN\,5h & CW11 & 53 & $-8.3$ & 6.9 & 6.6e+50 & 2600 & 5.0e-05 & 1.1e+38 & 1.3e+36  \\
 633 & 0.02 & WN\,5h & CW11 & 53 & $-7.8$ & 6.8 & 4.8e+50 & 2450 & 5.0e-05 & 8.7e+37 & 1.0e+36  \\
 642 & 0.07 & O2\,If* & CW11 & 46 & $-6.3$ & 6.1 & 7.4e+49 & 3000 & 5.0e-06 & 1.3e+37 & 9.6e+34  \\
 638 & 0.09 & O3\,III(f*) & MH98 & 47 & $-5.7$ & 5.8 & 4.6e+49 & 3200 & 2.5e-06 & 8.4e+36 & 5.0e+34  \\
 640 & 0.12 & O3\,V & MH98 & 48 & $-5.5$ & 5.8 & 3.9e+49 & 3000 & 2.0e-06 & 5.1e+36 & 3.0e+34  \\
 613 & 0.12 & WN\,5h & CW11 & 53 & $-7.4$ & 6.6 & 3.0e+50 & 2200 & 4.0e-05 & 5.6e+37 & 6.5e+35  \\
 651 & 0.14 & O3\,III(f*) & MH98 & 47 & $-5.5$ & 5.8 & 3.9e+49 & 3200 & 2.0e-06 & 6.8e+36 & 3.8e+34  \\
 614 & 0.15 & O7\,V & MH98 & 38 & $-6.2$ & 5.8 & 2.5e+49 & 2300 & 1.6e-06 & 2.8e+36 & 2.7e+34  \\
 656 & 0.18 & O3\,V & CD98 & 48 & $-5.6$ & 5.9 & 4.5e+49 & 3000 & 2.0e-06 & 6.2e+36 & 3.8e+34  \\
 595 & 0.19 & O5::\,V & MH98 & 42 & $-4.9$ & 5.4 & 1.3e+49 & 2700 & 4.0e-07 & 1.0e+36 & 6.1e+33  \\
 602 & 0.22 & O3\,V & MH98 & 48 & $-5.9$ & 6.0 & 5.6e+49 & 3000 & 3.2e-06 & 8.3e+36 & 5.4e+34  \\
 662 & 0.25 & O9\,V & MH98 & 34 & $-5.1$ & 5.2 & 3.9e+48 & 1500 & 3.2e-07 & 2.3e+35 & 2.8e+33  \\
 661 & 0.26 & O3-6\,V & MH98 & 44 & $-5.5$ & 5.7 & 2.9e+49 & 3000 & 1.3e-06 & 3.2e+36 & 2.0e+34  \\
 644 & 0.28 & O3\,V & MH98 & 48 & $-5.5$ & 5.8 & 4.0e+49 & 3000 & 2.0e-06 & 5.1e+36 & 3.0e+34  \\
 670 & 0.29 & O3\,III(f*) & MH98 & 47 & $-5.5$ & 5.8 & 3.9e+49 & 3200 & 2.0e-06 & 6.8e+36 & 3.8e+34  \\
 682 & 0.38 & O3\,If* & MH98 & 42 & $-6.1$ & 5.9 & 4.7e+49 & 3700 & 2.0e-06 & 8.3e+36 & 4.8e+34  \\
 624 & 0.41 & O3\,V & MH98 & 48 & $-5.7$ & 5.9 & 4.8e+49 & 3000 & 2.5e-06 & 6.8e+36 & 4.3e+34  \\
 678 & 0.43 & O3\,III(f*) & MH98 & 47 & $-5.7$ & 5.9 & 4.9e+49 & 3200 & 2.5e-06 & 8.9e+36 & 5.3e+34  \\
 687 & 0.43 & O3III\,(f*) & CD98 & 47 & $-6.2$ & 6.0 & 7.1e+49 & 3200 & 4.0e-06 & 1.4e+37 & 9.1e+34  \\
 648 & 0.44 & O3\,V & MH98 & 48 & $-5.6$ & 5.8 & 4.4e+49 & 3000 & 2.0e-06 & 5.9e+36 & 3.6e+34  \\
\hline
\end{tabular}
\end{center}
\end{table*}

\end{document}